\documentclass[twocolumn]{aastex631}

\usepackage{natbib,aas_macros,amsmath}
\usepackage{multirow,color}

\newcommand{\rrangle}{\right\rangle}
\newcommand{\llangle}{\left\langle}

\begin{document}

\shorttitle{ALMA CO Observations for Luminous Star-forming Galaxies at $\lowercase{z} = 6$}
\shortauthors{Ono et al.}
\submitjournal{ApJ in press}

\title{%
ALMA Observations for CO Emission from Luminous Lyman-break Galaxies at $\lowercase{z}=6.0293$--$6.2037$
}

\author[0000-0001-9011-7605]{Yoshiaki Ono}
\affiliation{Institute for Cosmic Ray Research, The University of Tokyo, 5-1-5 Kashiwanoha, Kashiwa, Chiba 277-8582, Japan}

\author[0000-0001-7201-5066]{Seiji Fujimoto}
\affiliation{Cosmic Dawn Center (DAWN), Jagtvej 128, DK2200 Copenhagen N, Denmark}
\affiliation{Niels Bohr Institute, University of Copenhagen, Lyngbyvej 2, DK2100 Copenhagen \O, Denmark}

\author[0000-0002-6047-430X]{Yuichi Harikane}
\affiliation{Institute for Cosmic Ray Research, The University of Tokyo, 5-1-5 Kashiwanoha, Kashiwa, Chiba 277-8582, Japan}
\affiliation{Department of Physics and Astronomy, University College London, Gower Street, London WC1E 6BT, UK}

\author[0000-0002-1049-6658]{Masami Ouchi}
\affiliation{National Astronomical Observatory of Japan, 2-21-1 Osawa, Mitaka, Tokyo 181-8588, Japan}
\affiliation{Institute for Cosmic Ray Research, The University of Tokyo, 5-1-5 Kashiwanoha, Kashiwa, Chiba 277-8582, Japan}
\affiliation{Kavli Institute for the Physics and Mathematics of the Universe (WPI), The University of Tokyo, 5-1-5 Kashiwanoha, Kashiwa-shi, Chiba, 277-8583, Japan}

\author[0000-0002-3258-3672]{Livia Vallini}
\affiliation{Scuola Normale Superiore, Piazza dei Cavalieri 7, 56126 Pisa, Italy}

\author[0000-0002-9400-7312]{Andrea Ferrara}
\affiliation{Scuola Normale Superiore, Piazza dei Cavalieri 7, 56126 Pisa, Italy}

\author{Takatoshi Shibuya}
\affiliation{Kitami Institute of Technology, 165 Koen-cho, Kitami, Hokkaido 090-8507, Japan}

\author[0000-0002-7129-5761]{Andrea Pallottini}
\affiliation{Scuola Normale Superiore, Piazza dei Cavalieri 7, 56126 Pisa, Italy}

\author[0000-0002-7779-8677]{Akio K. Inoue}
\affiliation{Department of Physics, School of Advanced Science and Engineering, Faculty of Science and Engineering, Waseda University, 3-4-1, Okubo, Shinjuku, Tokyo 169-8555, Japan}
\affiliation{Waseda Research Institute for Science and Engineering, Faculty of Science and Engineering, Waseda University, 3-4-1, Okubo, Shinjuku, Tokyo 169-8555, Japan}

\author[0000-0001-6186-8792]{Masatoshi Imanishi}
\affil{National Astronomical Observatory of Japan, 2-21-1 Osawa, Mitaka, Tokyo 181-8588, Japan}
\affil{Department of Astronomical Science, The Graduate University for Advanced Studies, SOKENDAI, Mitaka, Tokyo 181-8588, Japan}

\author[0000-0002-2597-2231]{Kazuhiro Shimasaku}
\affiliation{Department of Astronomy, Graduate School of Science, The University of Tokyo, 7-3-1 Hongo, Bunkyo-ku, Tokyo 113-0033, Japan}
\affiliation{Research Center for the Early Universe, Graduate School of Science, The University of Tokyo, 7-3-1 Hongo, Bunkyo-ku, Tokyo 113-0033, Japan}

\author[0000-0002-0898-4038]{Takuya Hashimoto}
\affiliation{Tomonaga Center for the History of the Universe (TCHoU), Faculty of Pure and Applied Sciences, University of Tsukuba, Tsukuba, Ibaraki 305-8571, Japan}

\author[0000-0003-1700-5740]{Chien-Hsiu Lee}
\affiliation{NSF's National Optical-Infrared Astronomy Research Laboratory, Tucson, AZ 85719, USA}

\author[0000-0001-6958-7856]{Yuma Sugahara}
\affil{National Astronomical Observatory of Japan, 2-21-1 Osawa, Mitaka, Tokyo 181-8588, Japan}
\affil{Waseda Research Institute for Science and Engineering, Faculty of Science and Engineering, Waseda University, 3-4-1, Okubo, Shinjuku, Tokyo 169-8555, Japan}

\author[0000-0003-4807-8117]{Yoichi Tamura}
\affil{Division of Particle and Astrophysical Science, Graduate School of Science, Nagoya University, Nagoya 464-8602, Japan}

\author[0000-0002-4052-2394]{Kotaro Kohno}
\affiliation{Institute of Astronomy, Graduate School of Science, The University of Tokyo, 2-21-1 Osawa, Mitaka, Tokyo 181-0015, Japan}
\affiliation{Research Center for the Early Universe, Graduate School of Science, The University of Tokyo, 7-3-1 Hongo, Bunkyo-ku, Tokyo 113-0033, Japan}

\author[0000-0001-7825-0075]{Malte Schramm}
\affiliation{Graduate School of Science and Engineering, Saitama University, 255 Shimo-Okubo, Sakura-ku, Saitama City, Saitama 338-8570, Japan}


\begin{abstract}
We present our new Atacama Large Millimeter/submillimeter Array (ALMA) observations 
targeting CO(6--5) emission from three luminous  
Lyman break galaxies (LBGs) 
at $z_{\rm spec} = 6.0293$--$6.2037$ 
found in the Subaru/Hyper Suprime-Cam survey,  
whose [{\sc Oiii}]$88\mu$m and [{\sc Cii}]$158\mu$m emission have been detected with ALMA.
We find a marginal detection of the CO(6--5) line from one of our LBGs, J0235--0532, 
at the $\simeq 4 \sigma$ significance level 
and obtain upper limits for the other two LBGs, J1211--0118 and J0217--0208.  
Our $z=6$ luminous LBGs are consistent with the previously found 
correlation between the CO luminosity and the infrared luminosity.
The unique ensemble of the multiple far-infrared emission lines 
and underlying continuum fed to a photodissociation region model 
reveal that J0235--0532 has a relatively high hydrogen nucleus density 
that is comparable to those of low-$z$ (U)LIRGs, quasars, and Galactic star-forming regions 
with high $n_{\rm H}$ values, 
while the other two LBGs have lower $n_{\rm H}$ 
consistent with local star-forming galaxies. 
By carefully taking account of various uncertainties 
we obtain total gas mass and gas surface density constraints from their CO luminosity measurements.
We find that J0235--0532 locates below the Kennicutt-Schmidt (KS) relation, 
comparable to the previously CO(2--1) detected $z=5.7$ LBG, HZ10.   
Combined with previous results for dusty starbursts at similar redshifts, 
the KS relation at $z=5$--$6$ is on average consistent with the local one.

\end{abstract}

\keywords{%
\href{http://astrothesaurus.org/uat/734}{High-redshift galaxies (734)}; 
\href{http://astrothesaurus.org/uat/979}{Lyman-break galaxies (979)}; 
\href{http://astrothesaurus.org/uat/262}{CO line emission (262)}; 
\href{http://astrothesaurus.org/uat/844}{Interstellar line emission (844)}; 
\href{http://astrothesaurus.org/uat/1073}{Molecular gas (1073)}; 
\href{http://astrothesaurus.org/uat/847}{Interstellar medium (847)}; 
\href{http://astrothesaurus.org/uat/1223}{Photodissociation regions (1223)}; 
\href{http://astrothesaurus.org/uat/594}{Galaxy evolution (594)}; 
\href{http://astrothesaurus.org/uat/595}{Galaxy formation (595)}; 
\href{http://astrothesaurus.org/uat/563}{Galactic and extragalactic astronomy (563)}; 
\href{http://astrothesaurus.org/uat/529}{Far infrared astronomy (529)}; 
\href{http://astrothesaurus.org/uat/1647}{Submillimeter astronomy (1647)}; 
\href{http://astrothesaurus.org/uat/1061}{Millimeter astronomy (1061)} 
}

\section{Introduction} \label{sec:introduction}

\begin{deluxetable*}{cccc} 
\tablecolumns{4} 
\tablewidth{0pt} 
\tablecaption{Summary of the Properties of Our Targets 
\label{tab:targets}}
\tablehead{
    \colhead{} 
    &  \colhead{J1211--0118}
    &  \colhead{J0235--0532}
    &  \colhead{J0217--0208}
}
\startdata 
R.A.									& 12:11:37.112 			& 02:35:42.412 				& 02:17:21.603 \\
Decl.									& $-$01:18:16.500 		& $-$05:32:41.623 				& $-$02:08:52.778 \\
$M_{\rm UV}$ (mag) 					& $-22.8$ 				& $-22.8$ 						& $-23.3$ \\
$L_{\rm UV}$ ($10^{11} L_\odot$) 			& 2.7 				& 2.9 						& 4.3 \\
SFR$_{\rm UV}$ ($M_\odot$ yr$^{-1}$) 		& $48 \pm 3$ 			& $48 \pm 4$ 					& $76 \pm 4$ \\
$r_{\rm e}$ (kpc)$^{\textcolor{red}{\dagger 1}}$	& 1.20				& 0.97$^{\textcolor{red}{\dagger 2}}$	& $0.57$ \\
EW$_0^{{\rm Ly}\alpha}$ ({\AA})			& $6.9 \pm 0.8$ 		& $41 \pm 2$ 					& $15 \pm 1$ \\
$\beta_{\rm UV}$ 						& $-2.0 \pm 0.5$ 		& $-2.6 \pm 0.6$ 				& $-0.1 \pm 0.5$ \\ 
$z_{\rm sys}$ 							& $6.0293 \pm 0.0002$ 	& $6.0901 \pm 0.0006$ 			& $6.2037 \pm 0.0005$  \\ 
$L_{\rm [OIII]} / L_{\rm [CII]}$ 				& $3.4 \pm 0.6$ 		& $8.9 \pm 1.7$ 				& $6.0 \pm 1.7$ 
\enddata 
\tablecomments{Most of the values presented in this Table 
have been obtained in the previous studies 
(\citealt{2018PASJ...70S..35M}; \citealt{2020ApJ...896...93H}).
}
\tablenotetext{\textcolor{red}{$^{\dagger 1}$}}{Half-light radius measured with the Subaru/HSC $z$-band images, 
which trace the rest UV continuum emission (Section \ref{subsec:SKrelation}).
}
\tablenotetext{\textcolor{red}{$^{\dagger 2}$}}{This $r_{\rm e}$ value is measured with SExtractor, 
while the $r_{\rm e}$ values for the other targets are measured with GALFIT. 
This is because a numerical convergence problem may have occurred 
in the profile fitting with GALFIT for J0235--0532 (Section \ref{subsec:SKrelation}).
}
\end{deluxetable*} 

Constraining the properties of molecular gas in galaxies across cosmic time 
is important to understand galaxy formation and evolution 
(see the reviews of \citealt{2013ARA&A..51..105C}; \citealt{2020ARA&A..58..157T}). 
Although star formation proceeds through the conversion of molecular hydrogen, H$_2$, into stars,  
it is difficult to directly detect emission from H$_2$ in molecular clouds 
due to the lack of a permanent dipole moment 
and the high temperatures necessary to excite even the lowest transitions.\footnote{Specifically, 
the two lowest H$_2$ rotational transitions 
have upper level energies of $h \nu / k_{\rm B} = 510$ K and $1015$ K above ground 
(\citealt{1984CaJPh..62.1639D}), 
and the lowest H$_2$ vibrational transition is even more difficult to excite, 
corresponding to $h \nu / k_{\rm B} = 6471$ K 
(\citealt{2013ARA&A..51..207B}).}
Instead, emission lines from rotational transitions of carbon monoxide, CO, are 
often employed to trace cold molecular gas in galaxies 
that is responsible for star formation activities.

Searches for CO line emission at $z \gtrsim 3$ 
have mainly focused on the most luminous sources like quasars 
(e.g., 
\citealt{2003Natur.424..406W}; 
\citealt{2003A&A...409L..47B}; 
\citealt{2007A&A...467..955W}; 
\citealt{2007A&A...472L..33M}; 
\citealt{2013ApJ...773...44W})
and dusty starburst galaxies 
(e.g., 
\citealt{2003ApJ...597L.113N}; 
\citealt{2005MNRAS.359.1165G}; 
\citealt{2013MNRAS.429.3047B}; 
\citealt{2013Natur.496..329R}; 
\citealt{2013ApJ...767...88W}; 
\citealt{2016MNRAS.457.4406A}; 
\citealt{2018NatAs...2...56Z}). 
For example, 
\cite{2010ApJ...720L.131R} have detected CO(2--1), CO(5--4), and CO(6--5) emission 
in a $z=5.3$ dusty star-forming galaxy (SFG), AzTEC-3, 
and revealed a large molecular gas reservoir, 
maintaining its intense starburst with $\gtrsim 1000 \, M_\odot$ yr$^{-1}$. 
Another high-$z$ example is a dusty SFG at $z=5.7$, CRLE, 
whose CO(2--1) as well as [{\sc Cii}]$158\mu$m and [{\sc Nii}]$205\mu$m have been detected 
in \cite{2018ApJ...861...43P}, 
showing a large amount of molecular gas reservoir 
with an intense starburst with $\simeq 1500  \, M_\odot$ yr$^{-1}$.

On the contrary, little progress has been made for high-$z$ normal SFGs  
such as Lyman break galaxies (LBGs), 
which are more representative of the high-$z$ galaxy population.
Although a handful of CO detections have been reported in mostly lensed LBGs at $z \sim 3$  
\citep[e.g.,][]{2004ApJ...604..125B,2017MNRAS.468.3468G}, 
to date only a few CO detections from normal SFGs at $z>3$ have been reported:  
i.e., luminous LBGs, LBG-1\footnote{LBG-1 is also named HZ6 (\citealt{2015Natur.522..455C}).} 
and HZ10,\footnote{CRLE 
is serendipitously discovered in ALMA observations targeting HZ10 (\citealt{2018ApJ...861...43P}).}
 at $z=5.3$--$5.7$ 
(\citealt{2019ApJ...882..168P}; 
see also, \citealt{2014ApJ...796...84R})
and a likely damped Ly$\alpha$ absorber host, Serenity-18, at $z=5.9$ 
(\citealt{2018ApJ...863L..29D}). 
CO lines from normal SFGs at $z > 3$ are typically too faint to allow for an investigation of 
the galaxy properties related to molecular gas components at high redshifts 
(e.g., \citealt{2022arXiv220301345H}) 
such as the gas surface density, the gas mass fraction, and the gas depletion time, 
and comparison with the Kennicutt-Schmidt (KS) relation 
(\citealt{1959ApJ...129..243S}; \citealt{1998ApJ...498..541K}), 
which are critically important to understand the star formation process 
(e.g., \citealt{2016ApJ...833..112S}; \citealt{2021ApJ...908...61K}) 
but have not yet been constrained well compared to those at lower redshifts.

In this study, 
we present our Atacama Large Millimeter/submillimeter Array (ALMA) 
observations targeting CO(6--5) emission 
at $\nu^{\rm (rest)}_{\rm CO(6-5)} = 691.47$ GHz in the rest-frame, 
corresponding to the rest-frame wavelength of 
$\lambda^{\rm (rest)}_{\rm CO(6-5)}  = 433.6$ $\mu$m, 
as well as dust continuum emission 
in three LBGs at $z=6$ 
that have been identified in the Subaru/Hyper Suprime-Cam (HSC) survey 
(\citealt{2018PASJ...70S...4A}). 
Previous optical spectroscopic observations have detected Ly$\alpha$ emission 
from the three LBGs 
(\citealt{2018PASJ...70S..35M}),
and subsequent ALMA observations have detected 
[{\sc Oiii}]$88\mu$m and [{\sc Cii}]$158\mu$m emission lines 
in these galaxies (\citealt{2020ApJ...896...93H}).

\begin{deluxetable*}{ccccccccc} 
\tablecolumns{9} 
\tablewidth{0pt} 
\tablecaption{Summary of Our ALMA Observations and Data
\label{tab:ALMA_observations}}
\tablehead{
    \colhead{Target} 
    &  \colhead{Date}
    &  \colhead{Configuration}
    &  \colhead{Central frequencies of SPWs}
    &  \colhead{$t_{\rm int}$}
    &  \colhead{PWV}
    &  \colhead{$\sigma_{\rm cont}$}
    &  \colhead{Beam FWHM}
    &  \colhead{PA}
\\
    \colhead{} 
    &  \colhead{(YYYY-MM-DD)}
    &  \colhead{}
    &  \colhead{(GHz)}
    &  \colhead{(min)}
    &  \colhead{(mm)}
    &  \colhead{($\mu$Jy beam$^{-1}$)}
    &  \colhead{ }
    &  \colhead{(deg)}
\\
    \colhead{(1)} 
    &  \colhead{(2)}
    &  \colhead{(3)}
    &  \colhead{(4)}
    &  \colhead{(5)}
    &  \colhead{(6)}
    &  \colhead{(7)}
    &  \colhead{(8)}
    &  \colhead{(9)}
}
\startdata 
J1211--0118 & 2019-10-01 & C43-4 & $97.801$, $99.488$, $109.625$, $111.500$ & $59.5$ & $3.4$ & $17.4$ & $2\farcs36 \times 2\farcs15$ & $59.56$ \\
J0235--0532 & 2019-11-12 & C43-3 & $96.964$, $98.652$, $108.825$, $110.700$ & $91.7$ & $5.5$ & $9.7$ & $3\farcs17 \times 2\farcs72$ & $-72.15$ \\
J0217--0208 & 2019-11-12 & C43-3 & $95.434$, $97.122$, $107.325$, $109.200$ & $72.6$ & $5.2$ & $7.4$ & $3\farcs18 \times 2\farcs77$ & $70.38$ 
\enddata 
\tablecomments{
(1) Target ID. 
(2) Observation date. 
(3) Antenna configuration. 
(4) Central frequencies of the four SPWs. 
(5) On-source integration time. 
(6) Precipitable water vapor. 
(7) The $1\sigma$ level of the continuum image. 
(8) The synthesized beam FWHM in units of arcsec $\times$ arcsec. 
(9) The position angle of the synthesized beam. 
}
\end{deluxetable*} 

This paper is outlined as follows. 
After introducing our three $z = 6$ luminous LBGs in Section \ref{sec:targets}, 
we describe our new ALMA observations and data reduction processes in Section \ref{sec:observations}. 
Our results for the CO emission and dust continuum emission 
from the three $z = 6$ luminous LBGs are presented in Section \ref{sec:results}.  
We discuss their gaseous properties in Section \ref{sec:discussion} 
and present a summary in Section \ref{sec:summary}. 
Throughout this paper, 
we use magnitudes in the AB system (\citealt{1983ApJ...266..713O})
and assume a flat universe with 
$\Omega_{\rm m} = 0.3$, $\Omega_\Lambda = 0.7$, and $H_0 = 70$ km s$^{-1}$ Mpc$^{-1}$. 
In this cosmological model, 
an angular dimension of $1.0$ arcsec corresponds to 
a physical dimension of $5.710$ kpc at $z=6.0$ 
(e.g., Equation 18 of \citealt{1999astro.ph..5116H}).
We adopt the \cite{2003PASP..115..763C} initial mass function (IMF) 
with lower and upper mass cutoffs of $0.1 M_\odot$ and $100 M_\odot$, respectively. 
Where necessary to convert star formation rates (SFRs) in the literature 
from the \cite{1955ApJ...121..161S} IMF 
and the \cite{2001MNRAS.322..231K} IMF 
to the Chabrier IMF, 
we multiply constant factors of 
$\alpha_{\rm SC} = 0.63$ 
and 
$\alpha_{\rm KC} = 0.94 \, (= 0.63/0.67)$, respectively 
(\citealt{2014ARA&A..52..415M}).

\section{Targets} \label{sec:targets}

To constrain the properties of molecular gas in $z=6$ normal SFGs, 
we target three luminous LBGs at $z_{\rm spec} = 6.029$--$6.204$:  
J1211--0118, J0235--0532, and J0217--0208. 
Their basic properties reported in previous work are summarized in Table \ref{tab:targets}. 
These LBGs have been spectroscopically identified with Ly$\alpha$ emission 
(\citealt{2018PASJ...70S..35M}) 
and their [{\sc Oiii}]$88\mu$m, [{\sc Cii}]$158\mu$m, and dust continuum emission 
have been observed with ALMA
(\citealt{2020ApJ...896...93H}). 
Their total SFRs, SFR$_{\rm tot}$, have been estimated to be 
$\sim 100 M_\odot$ yr$^{-1}$ 
as the sum of the dust-unobscured and dust-obscured SFRs 
based on the rest-frame ultraviolet (UV) and infrared (IR) continuum emission, 
SFR$_{\rm UV}$ and SFR$_{\rm IR}$, respectively.
These SFRs are estimated 
by using Equation (1) and Equation (4) of \cite{1998ARA&A..36..189K} 
and considering the conversion factor from the Salpeter IMF to the Chabrier IMF. 
For details, see Appendix \ref{sec:standard_equations}.
Because of their moderately high total SFRs, 
the CO emission line fluxes of our targets are expected to be high, 
if their molecular gas is not already depleted by recent star formation.

Although their UV absolute magnitudes are $M_{\rm UV} \simeq -23.0$ mag, 
around which the luminosity functions of galaxies and AGNs are almost comparable 
\citep[e.g.,][]{2018PASJ...70S..10O},  
their rest UV spectra exhibit no clear signatures of AGNs 
such as broad Ly$\alpha$ or {\sc Nv} 1240{\AA}, 
suggesting that they are normal SFGs. 
Note that, because they are not located in a foreground galaxy cluster field 
or close to a foreground massive red galaxy, 
they are unlikely to be affected by strong lensing. 
Thus, they are great laboratories to investigate typical properties of high-$z$ normal SFGs 
with no systematic uncertainties of lensing models.

\begin{figure}[h]
\begin{center}
   \includegraphics[width=0.47\textwidth]{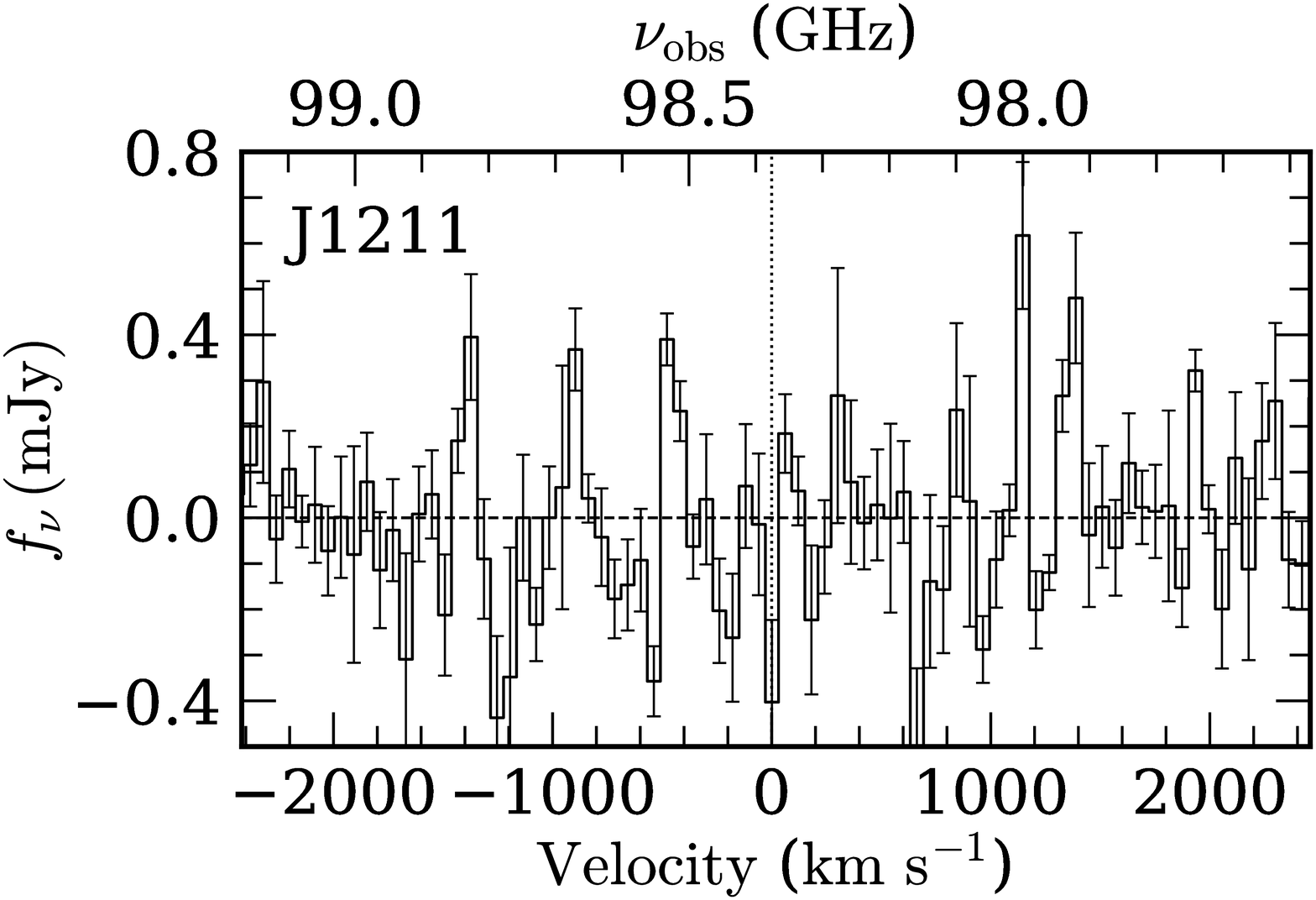}
   \includegraphics[width=0.47\textwidth]{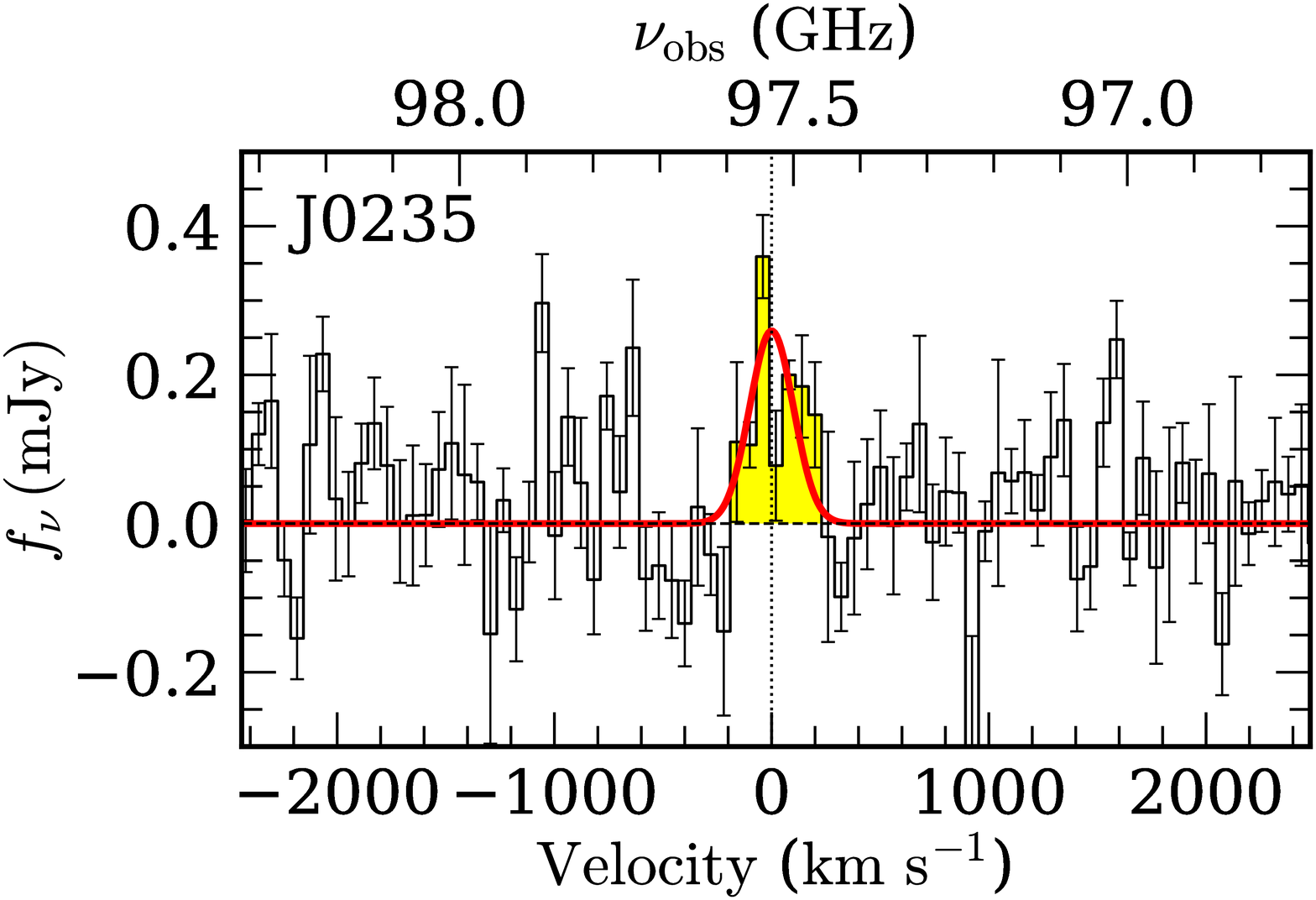}
   \includegraphics[width=0.47\textwidth]{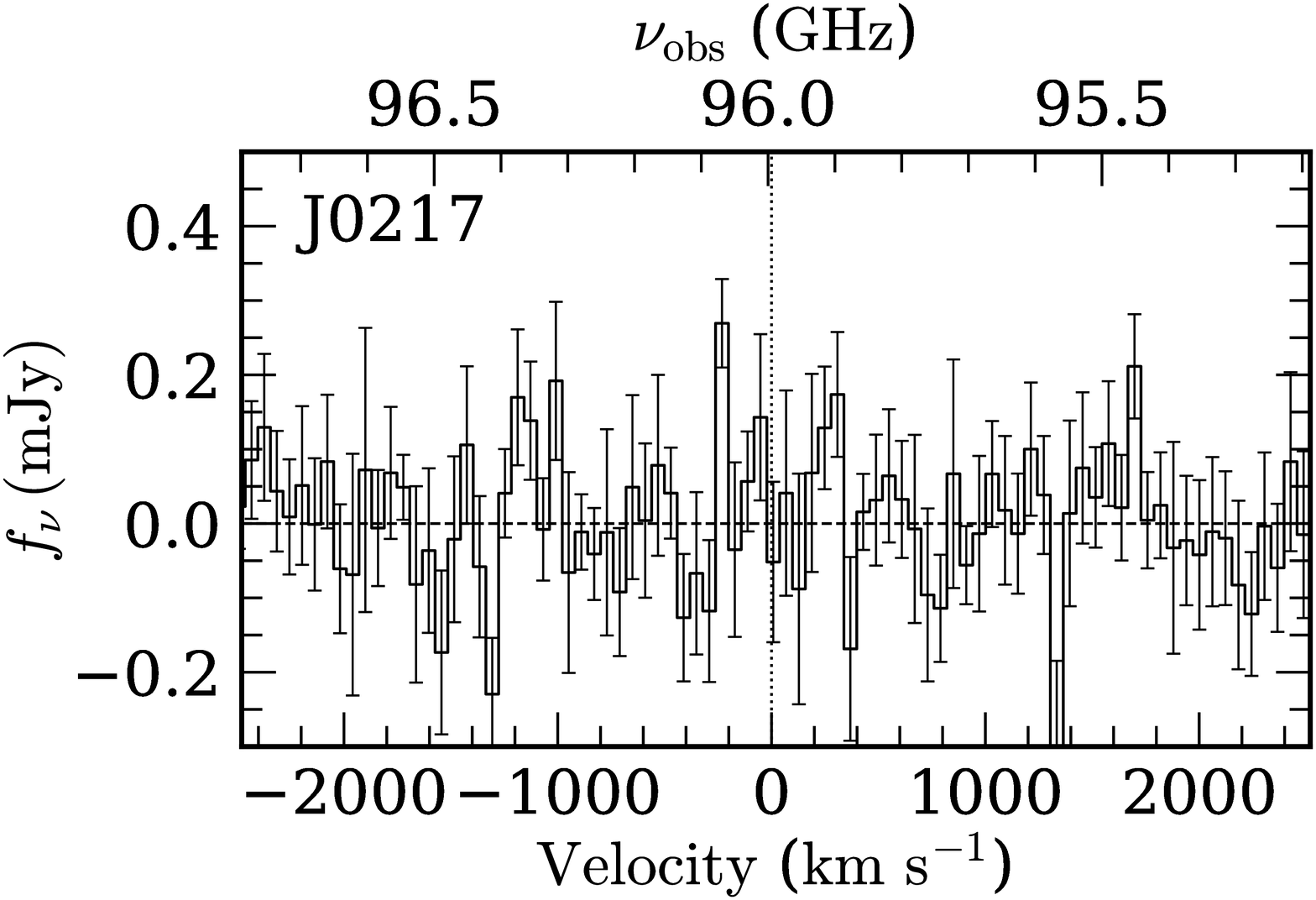}
\caption{
ALMA spectra of J1211--0118 (top), J0235--0532 (middle), and J0217--0208 (bottom) 
around the redshifted frequency of the CO(6--5) emission line (black histogram) 
extracted by placing a beam aperture 
(For details, see the text in Section \ref{subsec:results_co}).
The dotted vertical line corresponds to the systemic redshift 
determined with the FIR emission lines of [{\sc Oiii}] and [{\sc Cii}] 
(\citealt{2020ApJ...896...93H}). 
The red curve in the middle panel represents the best-fit Gaussian function 
to the CO(6--5) emission line. 
}
\label{fig:CO_spectrum}
\end{center}
\end{figure}

\section{ALMA Observations and Data Reduction} \label{sec:observations}

\begin{figure}[h]
\begin{center}
   \includegraphics[height=0.27\textheight]{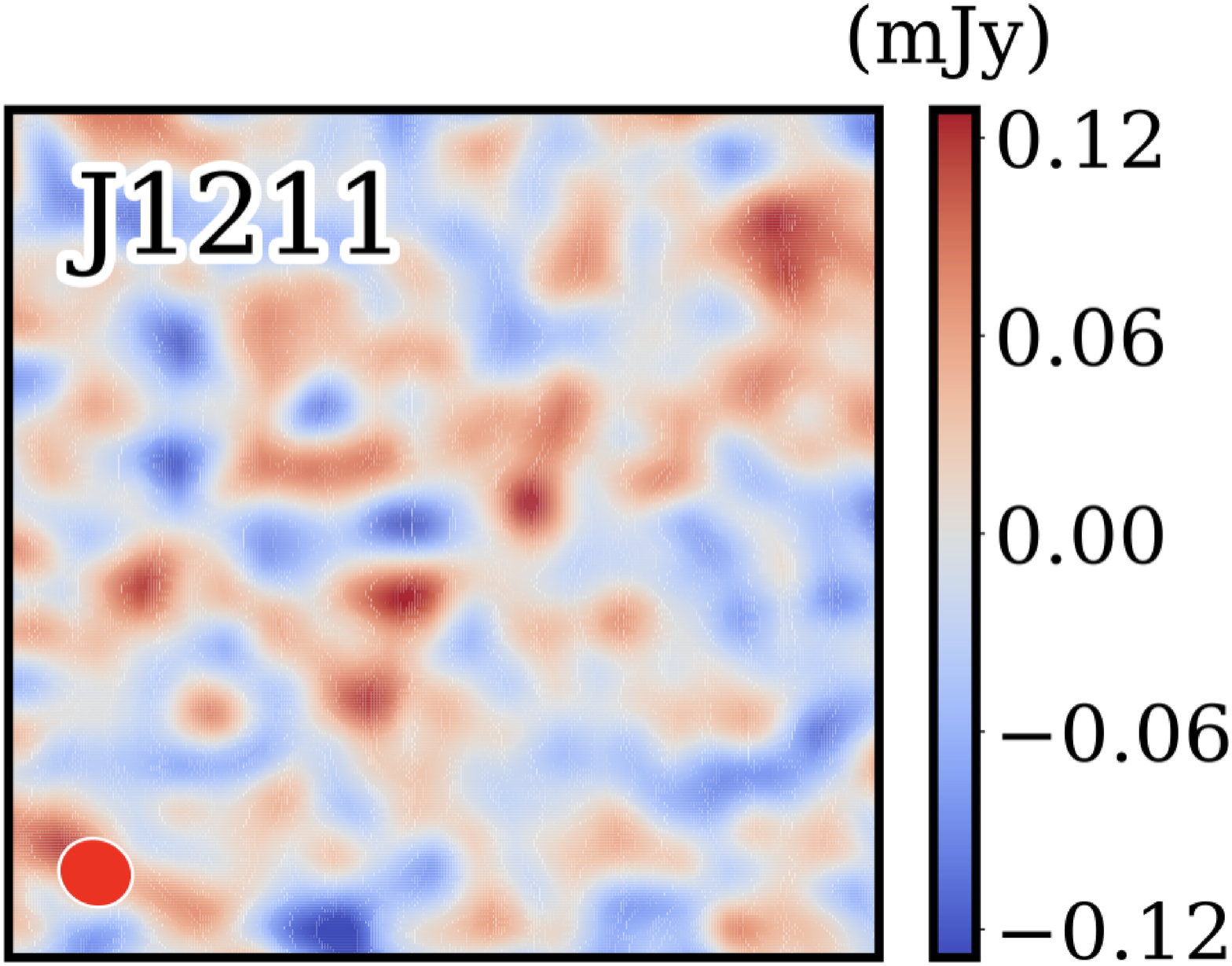} 
   \includegraphics[height=0.27\textheight]{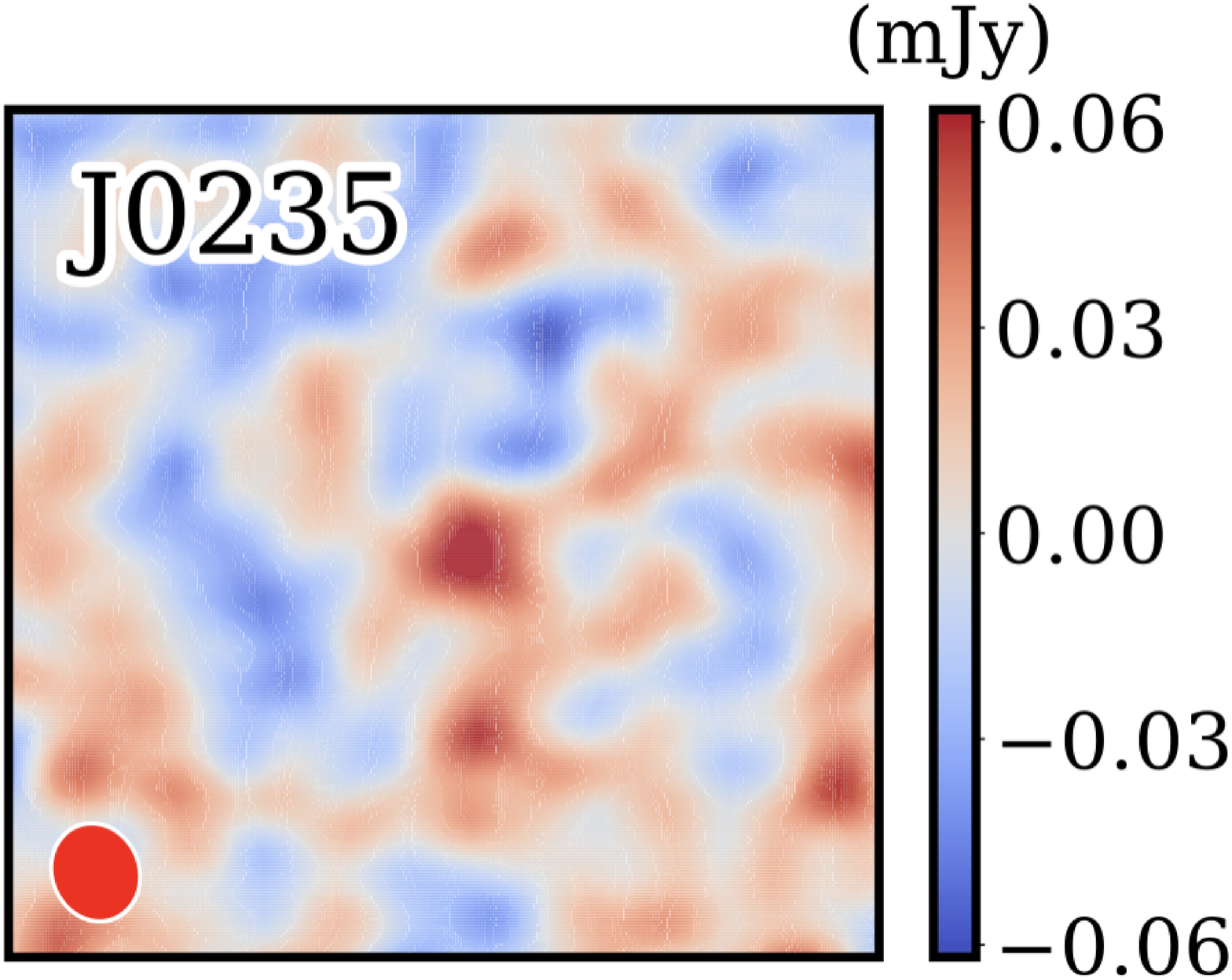} 
   \includegraphics[height=0.27\textheight]{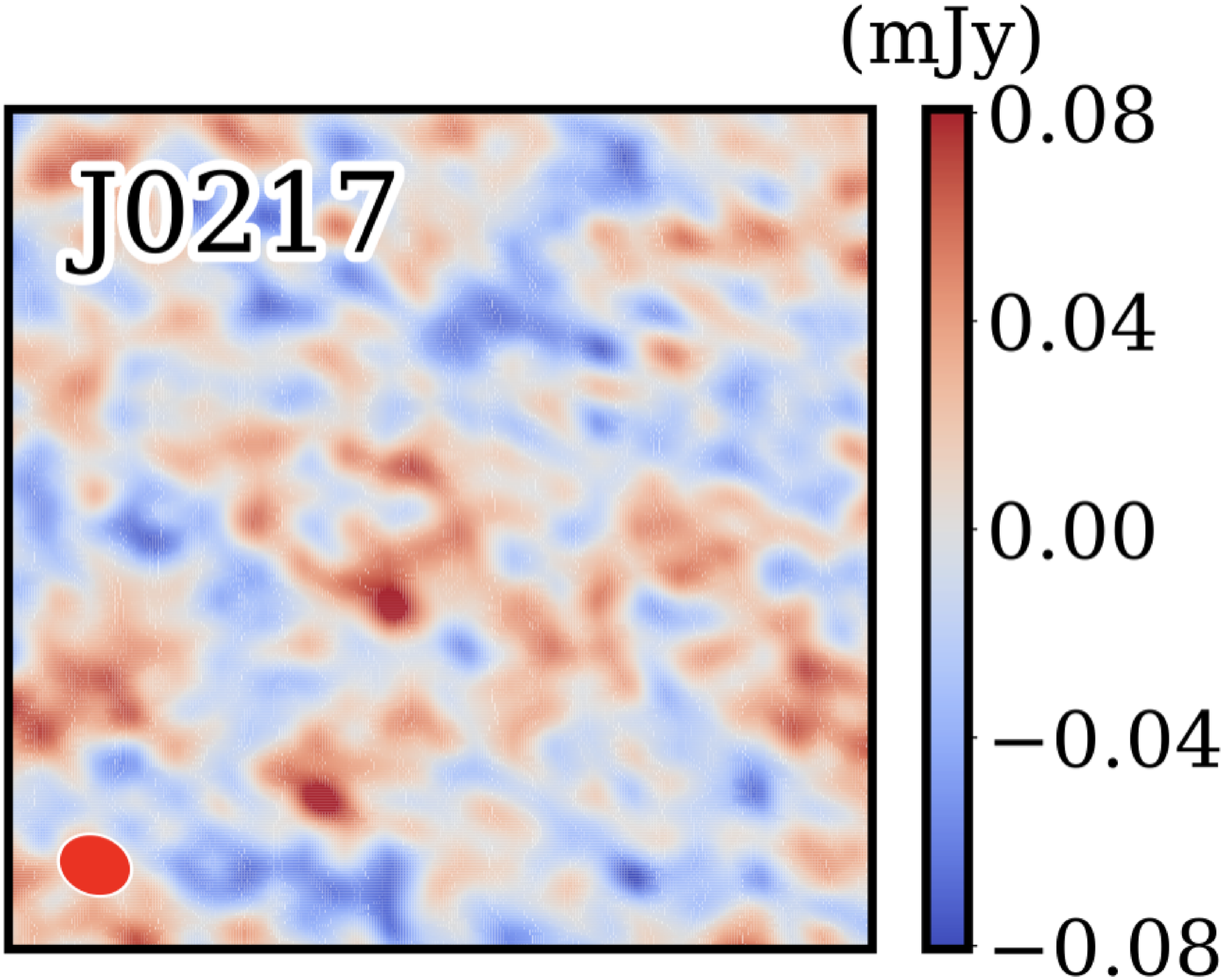} 
\caption{
Zeroth moment images showing the integrated CO(6--5) flux densities 
of J1211--0118 (top), J0235--0532 (middle), and J0217--0208 (bottom).  
The synthesized beam is shown in the bottom left corner in each image. 
The size of each image is $30'' \times 30''$.
The color bar range for the flux densities corresponds to $\pm 3$ times the standard deviation. 
}
\label{fig:CO_moment_zero}
\end{center}
\end{figure}

\begin{figure*}[h]
\begin{center}
   \includegraphics[width=0.32\textwidth]{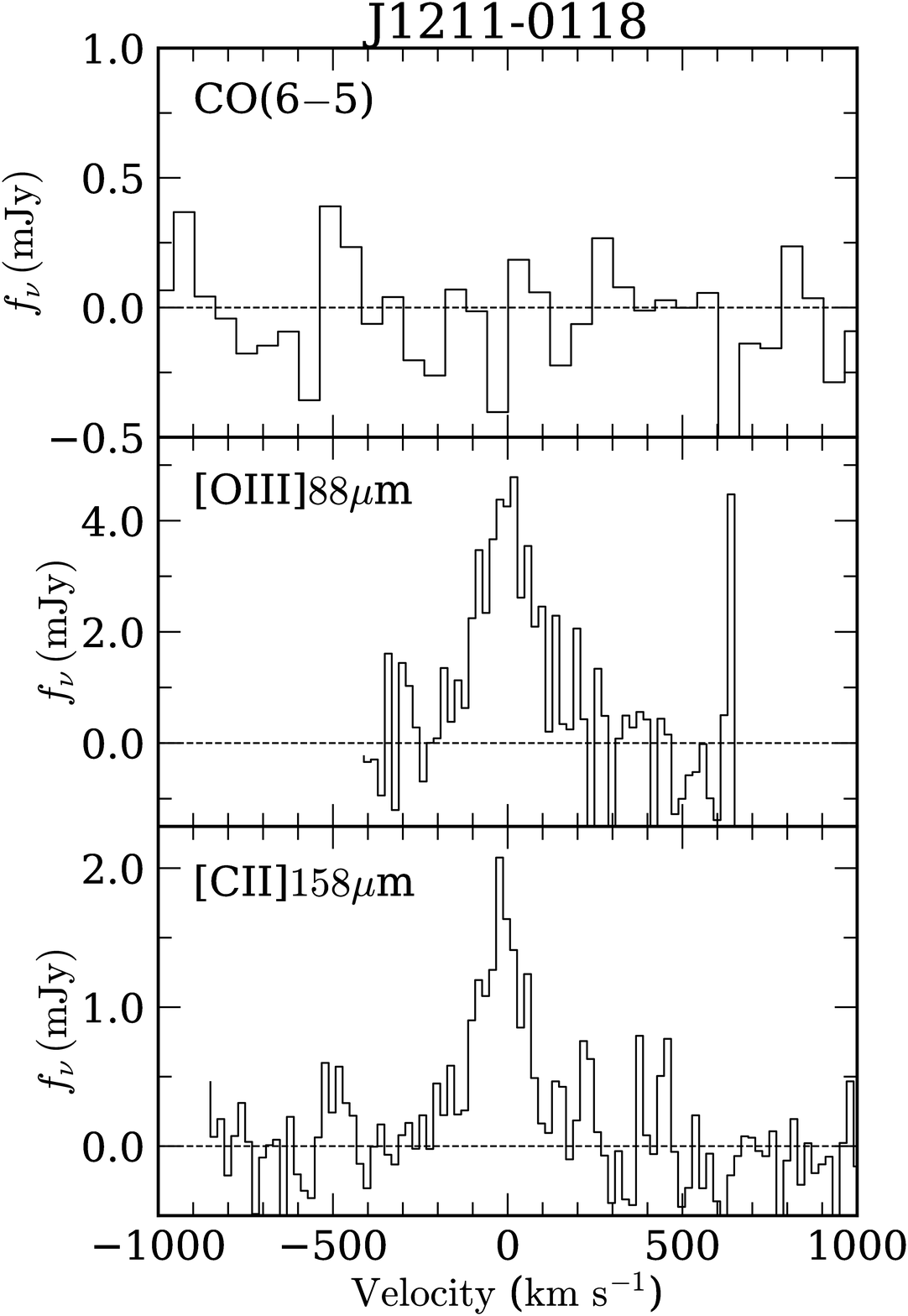}
   \includegraphics[width=0.32\textwidth]{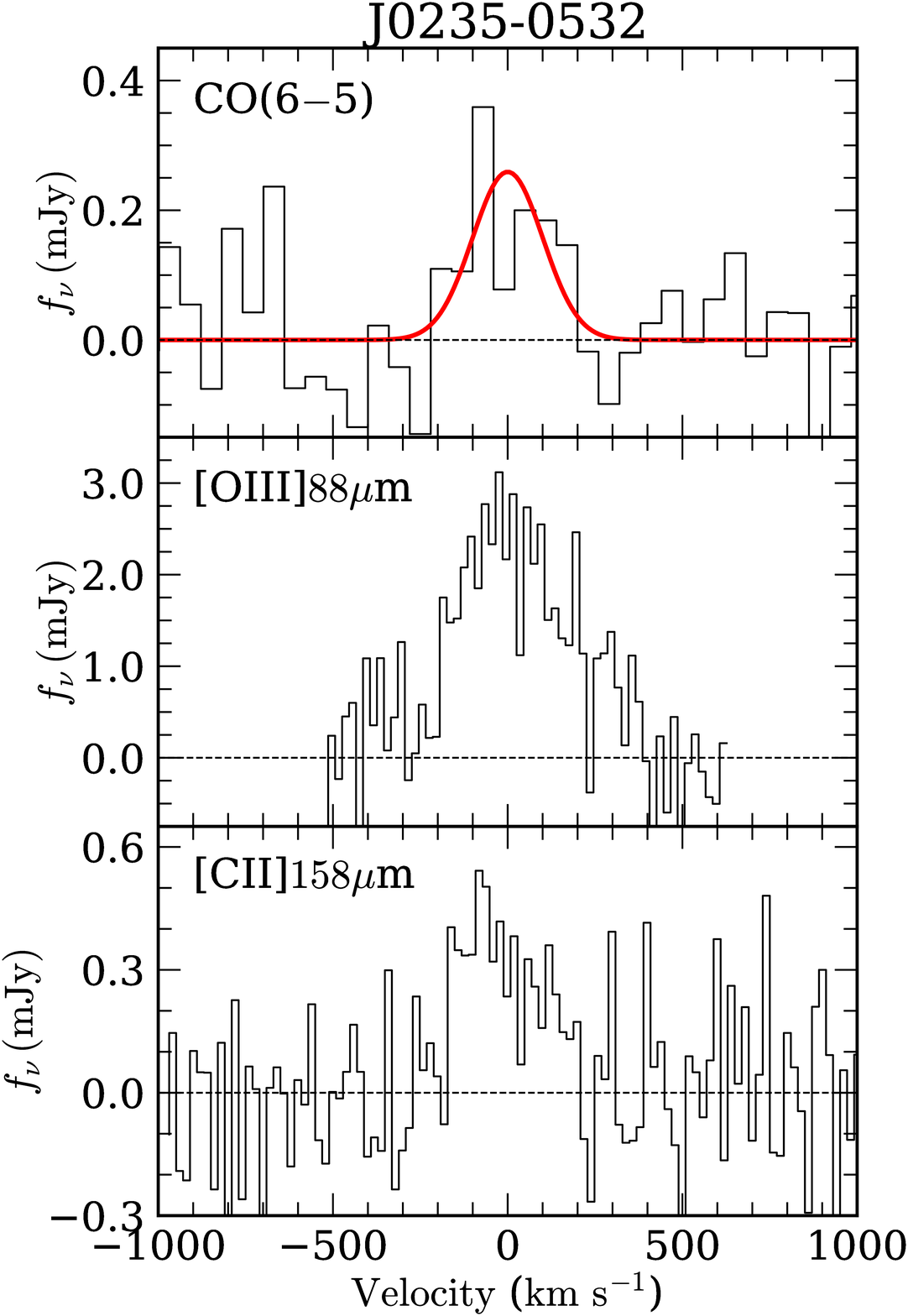}
   \includegraphics[width=0.32\textwidth]{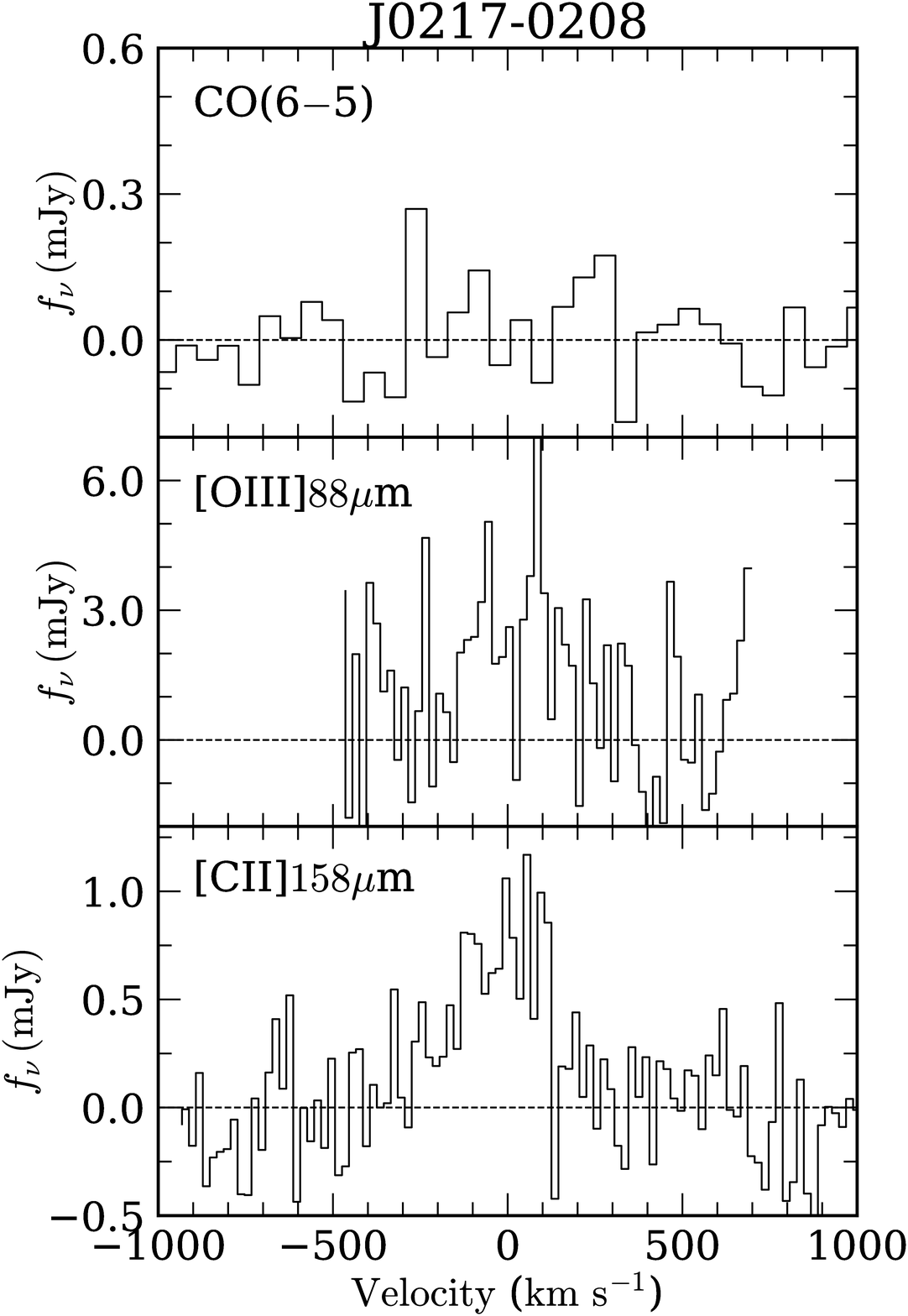}
\caption{
ALMA spectra 
of J1211--0118 (left), J0235--0532 (center), and J0217--0208 (right) 
around CO(6--5), [{\sc Oiii}]$88\mu$m, and [{\sc Cii}]$158\mu$m from top to bottom. 
The CO spectra are the same as the ones shown in Figure \ref{fig:CO_spectrum}, 
but the velocity range is limited to [$-1000$ km s$^{-1}$, $1000$ km s$^{-1}$].  
The spectra for [{\sc Oiii}]$88\mu$m and [{\sc Cii}]$158\mu$m 
have been obtained in \cite{2020ApJ...896...93H}.
}
\label{fig:CO_OIII_CII_J0235}
\end{center}
\end{figure*}

\begin{figure}[h]
\begin{center}
   \includegraphics[height=0.28\textheight]{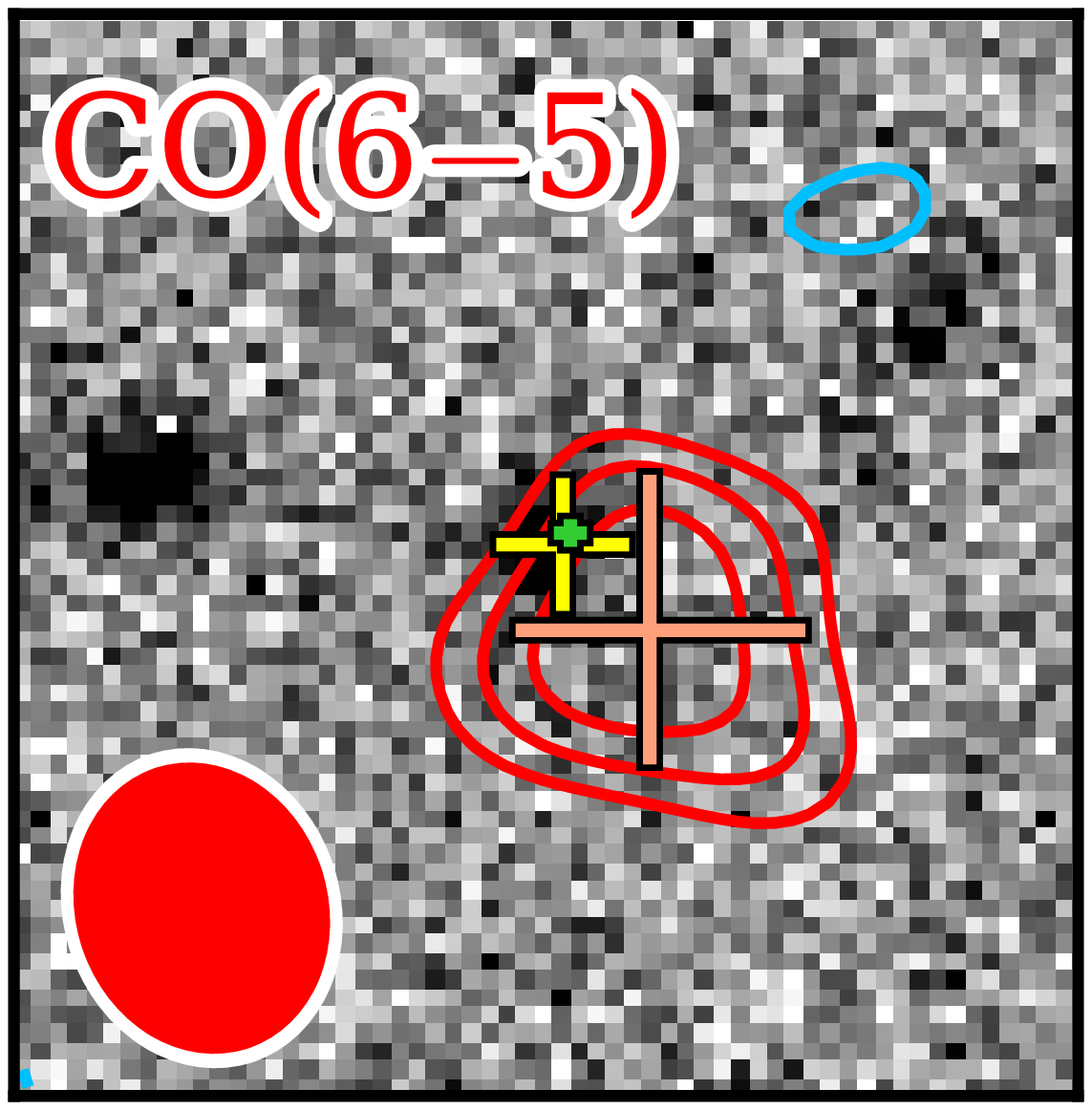}
\caption{
CO(6--5) contours for J0235--0532. 
The red (blue) contours are multiples of $0.5\sigma$ ($-0.5\sigma$) starting at $2 \sigma$ ($-2 \sigma$). 
The synthesized beam is shown in the bottom left corner. 
The gray background is the Subaru HSC $z$-band image of J0235--0532 
that captures the rest-frame UV continuum emission. 
The positions of the CO(6--5) and UV continuum emission are consistent 
within the large uncertainties (light-red cross) estimated from the Monte Carlo simulations
(for details, see the text in Section \ref{subsec:results_co}). 
The yellow and green crosses denote 
the positions of the [{\sc Cii}] and [{\sc Oiii}] emission, respectively, 
and the sizes of the crosses are their uncertainties 
(\citealt{2020ApJ...896...93H}), 
which are also consistent with that of CO(6--5). 
The size of the image is $10'' \times 10''$. 
}
\label{fig:CO_image_J0235}
\end{center}
\end{figure}

Our targets were observed during ALMA Cycle 7 with Band 3 
between 2019 October 01 and 2019 November 12 
(Project code: 2019.1.00156.S; PI: Y. Ono).
The number of antennas used in the observations is $45$. 
The antenna configurations were C43-3 
for J0235--0532 and J0217--0208, 
and C43-4 for J1211--0118. 
The maximum baselines of C43-3 and C43-4 are $500.2$ m and $2617.4$ m, respectively. 
The minimum baseline of these configurations is $15.1$ m.
We used four spectral windows (SPWs) with $1.875$ GHz bandwidths 
in the Frequency Division Mode, 
yielding the total frequency coverage of $7.5$ GHz. 
The velocity resolution was set to $3.9$ MHz, which corresponds to about $10$ km s$^{-1}$. 
One of the SPWs was used for the CO(6--5) line 
and the other SPWs were used for the dust continuum.  
Note that CO(6--5) is the lowest CO excitation 
that can be observed for $z=6$ galaxies with ALMA Band 3--10. 
The details of the observations are presented in Table \ref{tab:ALMA_observations}.

We reduce the ALMA data by using the Common Astronomy Software Applications 
(CASA; \citealt{2007ASPC..376..127M}) package\footnote{\url{https://casa.nrao.edu/}}
version 5.6.1. 
Using the CLEAN task, 
we produce continuum images and data cubes for our targets 
with the natural weighting. 
We apply a Gaussian taper with FWHM$=$2\farcs0 
to improve the signal-to-noise ratio (S/N) 
for potentially existing low surface brightness emission. 
We adopt a pixel scale of 0\farcs1 
and a common spectral channel bin of about $60$ km s$^{-1}$.
Table \ref{tab:ALMA_observations} 
presents the $1\sigma$ flux density levels 
and the spatial resolutions, and the synthesized beam position angles 
for the continuum images. 
Note that, although two of our targets were also observed with 
Northern Extended Millimeter Array (NOEMA), 
both show no detection, which is consistent with the ALMA results (Appendix \ref{sec:noema_observations}).

\begin{deluxetable*}{cccc} 
\tablecolumns{4} 
\tablewidth{0pt} 
\tablecaption{Summary of Our Observational Results
\label{tab:observational_results}}
\tablehead{
    \colhead{} 
    &  \colhead{J1211--0118}
    &  \colhead{J0235--0532}
    &  \colhead{J0217--0208}
}
\startdata 
CO(6--5) integrated flux (Jy km s$^{-1}$) 				& $<0.0713$ 			& $0.0652 \pm 0.0175$		 					& $<0.0609$ \\
CO(6--5) FWHM (km s$^{-1}$) 						& --- 					& $237 \pm 51$ 								& --- \\
$L_{\rm CO(6-5)}$ ($10^7 L_\odot$) 					& $<3.41$ 			& $2.88 \pm 0.773$ 								& $<3.95$ \\
$L'_{\rm CO(6-5)}$ ($10^9$ K km s$^{-1}$ pc$^2$) 		& $<3.22$ 			& $2.72 \pm 0.73$ 								& $<3.74$ \\
$f_{\nu,430\mu{\rm m}}$ ($\mu$Jy)					& $<52.3$ 			& $<29.1$ 									& $<22.1$ \\
$L_{\rm IR}$ ($10^{11} L_\odot$) 					& $3.6^{+34.4}_{-1.9}$ 	& $5.8^{+19.4}_{-5.8}$$^{\textcolor{red}{\dagger 1}}$	& $2.0^{+5.9}_{-0.3}$ \\
$T_{\rm dust}$ (K) 								& $40^{+44}_{-14}$ 		& 50--80 (fixed) 								& $31^{+26}_{-9}$ \\ 
$f_{\rm CMB}^{\rm CO(6-5)}$ 						& 0.72 				& 0.79--0.89									& 0.55 \\
SFR$_{\rm IR}$ ($M_\odot$ yr$^{-1}$) 				& $39^{+375}_{-21}$ 	& $63^{+211}_{-63}$$^{\textcolor{red}{\dagger 1}}$		& $22^{+64}_{-3}$ \\
SFR$_{\rm tot}$ ($M_\odot$ yr$^{-1}$) 				& $88^{+375}_{-21}$ 	& $112^{+211}_{-64}$	 						& $98^{+65}_{-5}$ \\
$M_{\rm gas}$ ($10^{10} M_\odot$) 					& $< 8.99$ 			& $7.59 \pm 4.74$ 								& $< 10.4$ \\
$\Sigma_{\rm SFR}$ ($M_\odot$ yr$^{-1}$ kpc$^{-2}$) 	& $9.6^{+41.5}_{-2.3}$	& $18.9^{+35.6}_{-10.7}$							& $48.2^{+31.6}_{-2.5}$ \\
$\Sigma_{\rm gas}$ ($10^3 M_\odot$ pc$^{-2}$) 		& $< 9.9$				& $12.8 \pm 8.0$								& $< 5.1$ \\
$f_{\rm gas}$$^{\textcolor{red}{\dagger 2}}$ 			& $< 0.59$ 			& $0.55^{+0.12}_{-0.23}$ 							& $< 0.46$ \\
$t_{\rm dep}$ (Gyr) 								& $< 1.02$			& $0.68^{+0.90}_{-0.47}$							& $< 1.06$ 
\enddata 
\tablecomments{
The upper limits are $3\sigma$.
}
\tablenotetext{\textcolor{red}{$^{\dagger 1}$}}{%
These values are the $3 \sigma$ upper limits when $T_{\rm dust} = 50$ K, 
and the upper error takes into account the case when $T_{\rm dust} = 80$ K.  
For details, see the text in Section \ref{subsec:results_dust}.
}
\tablenotetext{\textcolor{red}{$^{\dagger 2}$}}{%
The quoted uncertainties in the gas fraction do not include 
the systematic uncertainty associated with the stellar mass estimates. 
}
\end{deluxetable*} 

\section{Results} \label{sec:results}

\subsection{CO(6--5)} \label{subsec:results_co}

\begin{figure*}[h]
\begin{center}
   \includegraphics[scale=0.32]{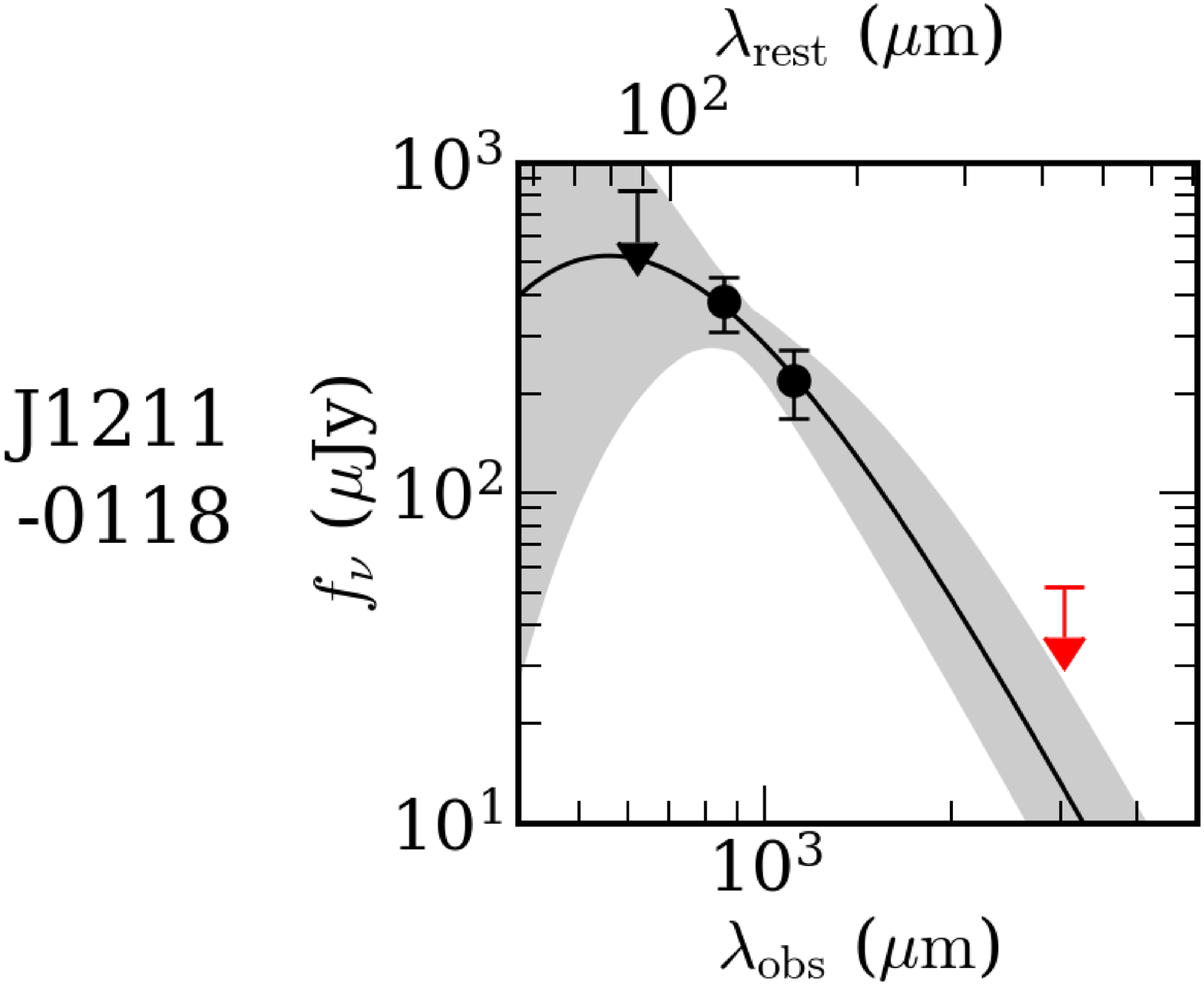}
   \includegraphics[scale=0.32]{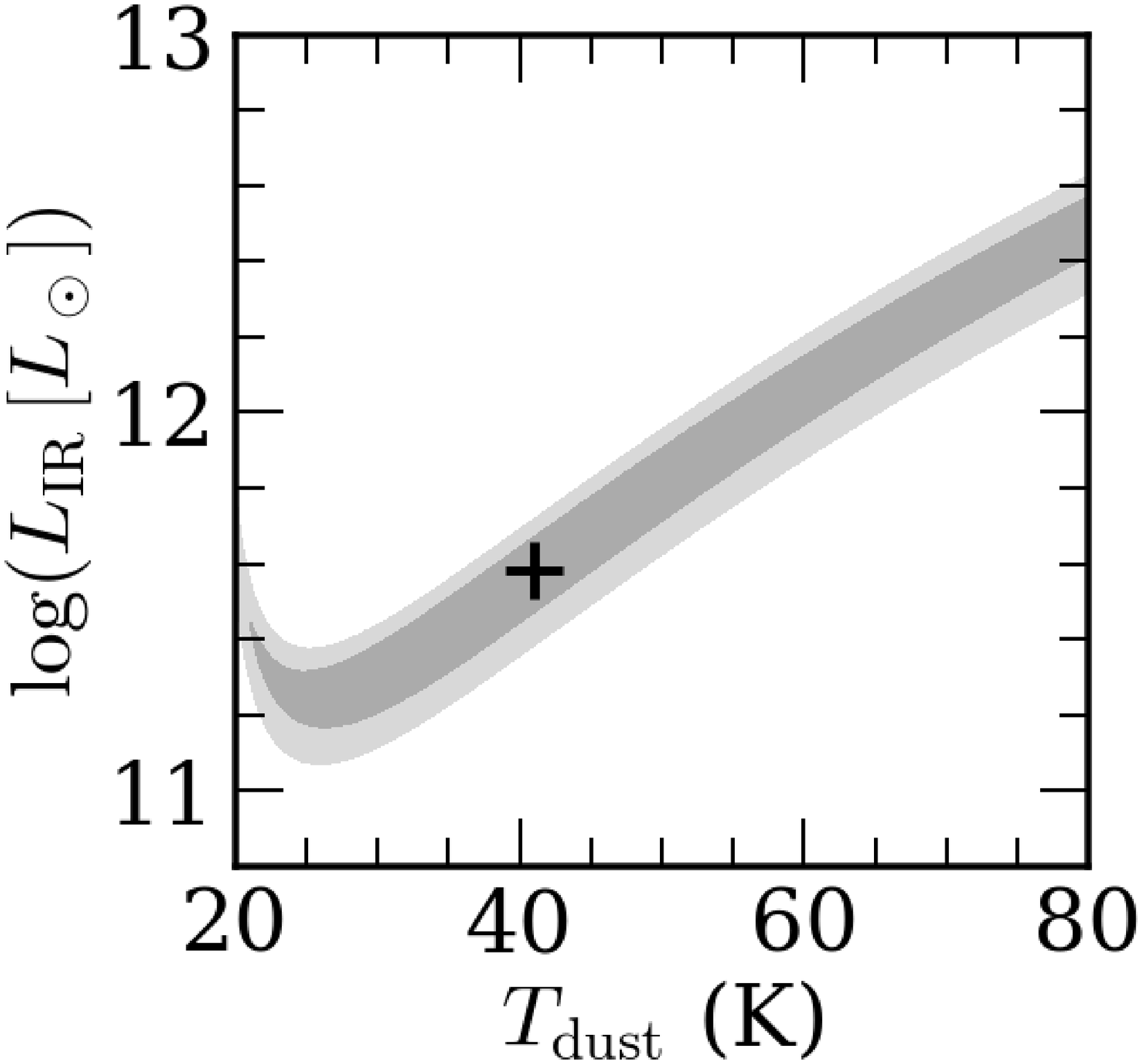}
   \includegraphics[scale=0.32]{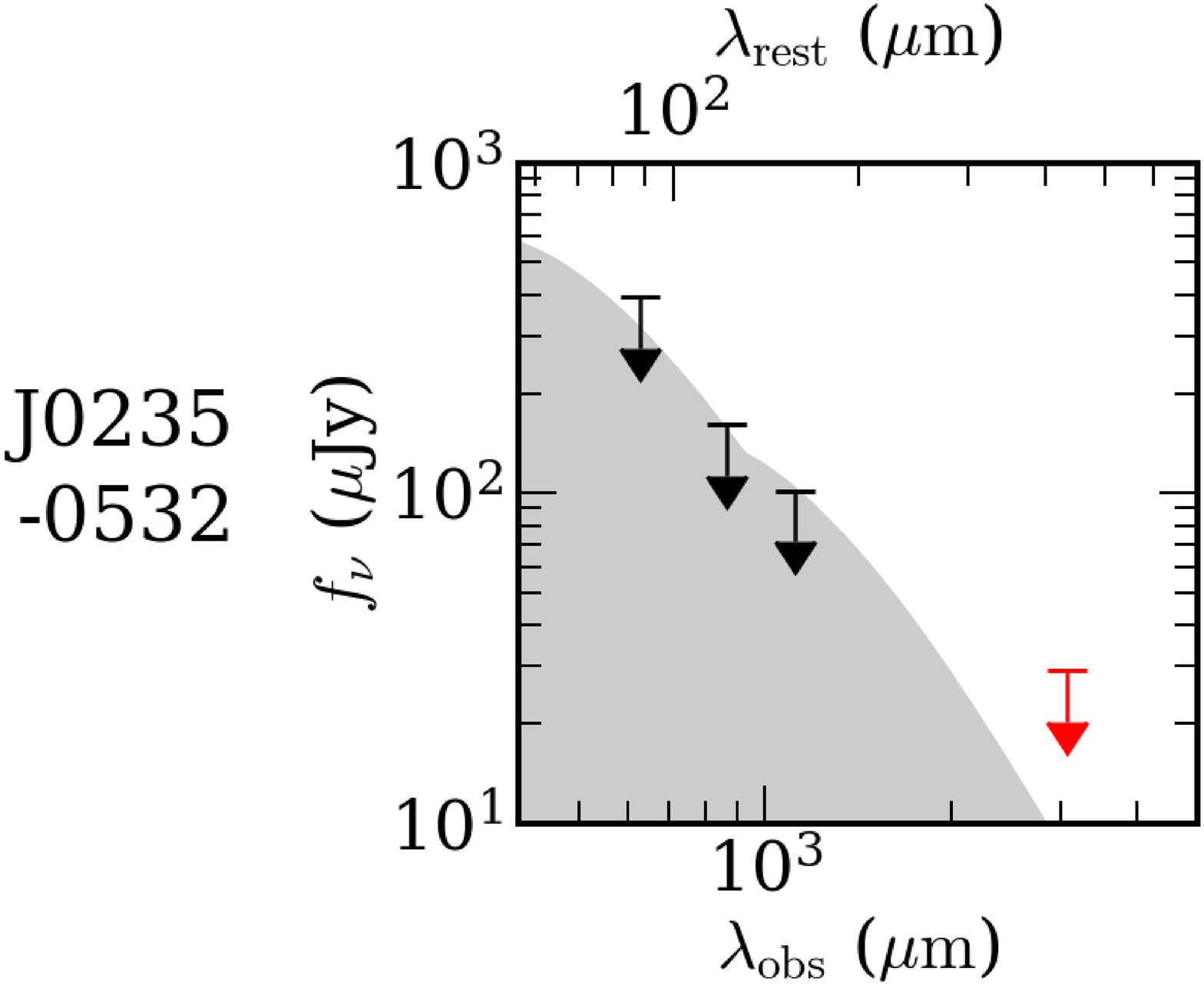}
   \includegraphics[scale=0.32]{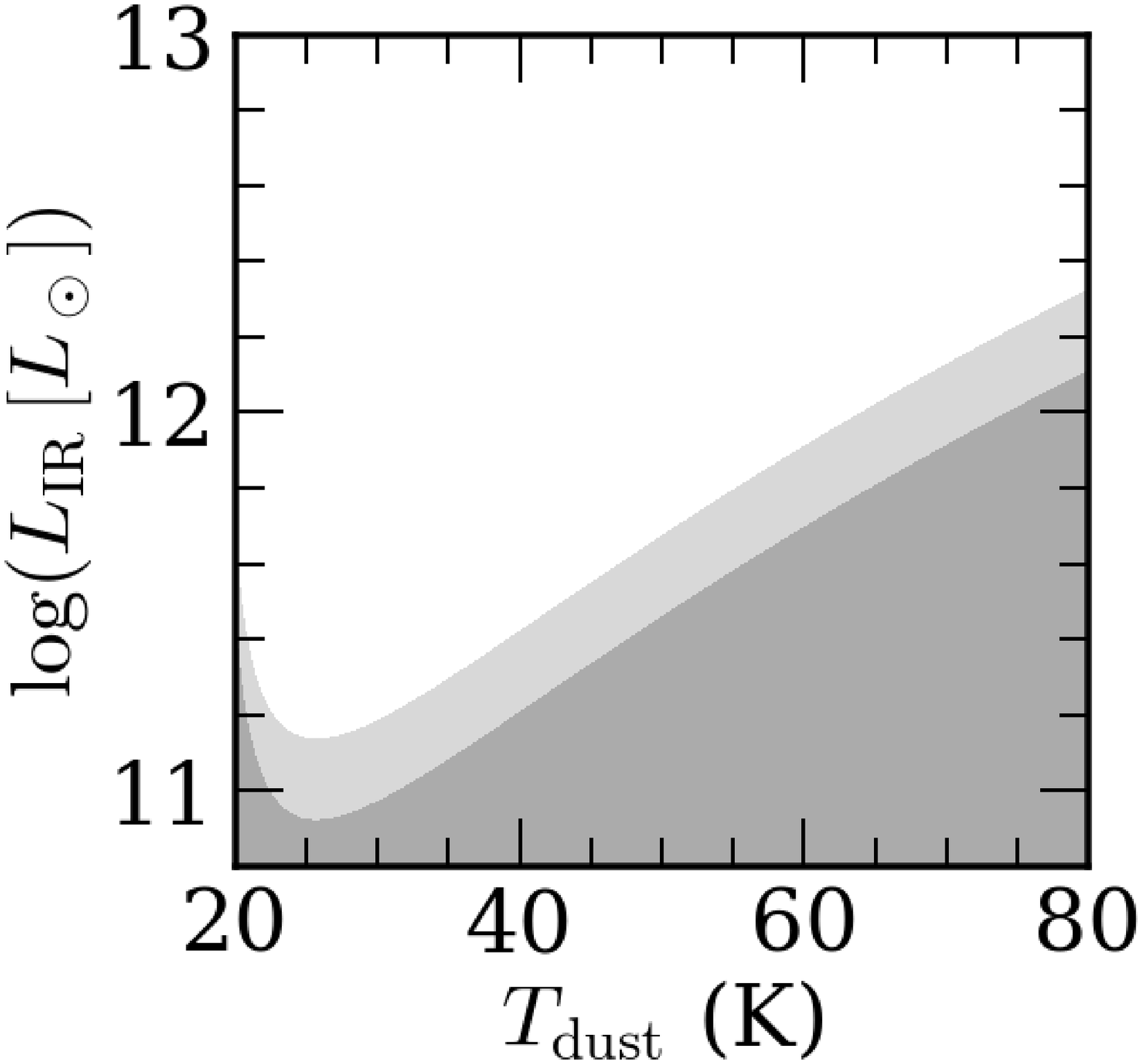}
   \includegraphics[scale=0.32]{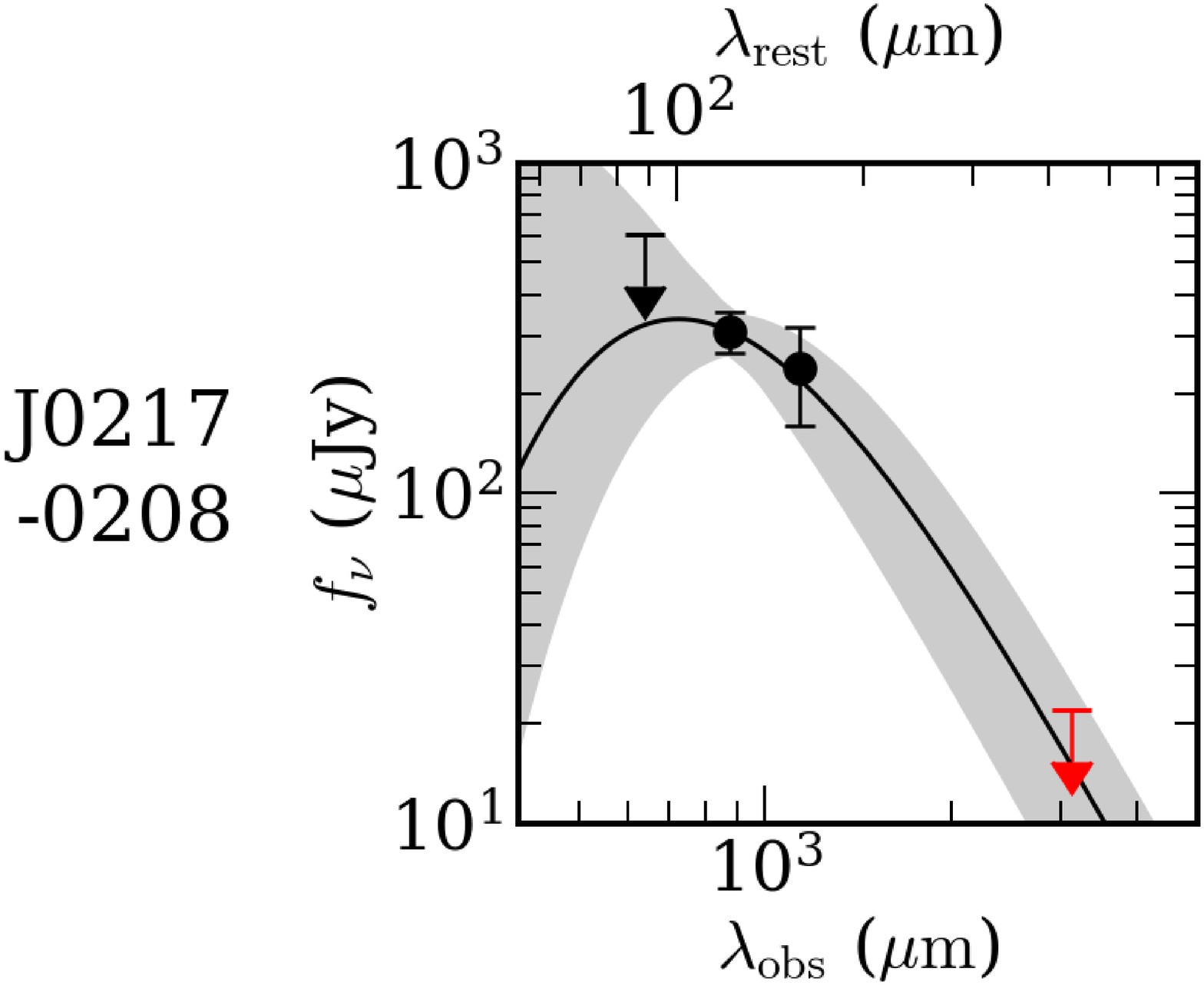}
   \includegraphics[scale=0.32]{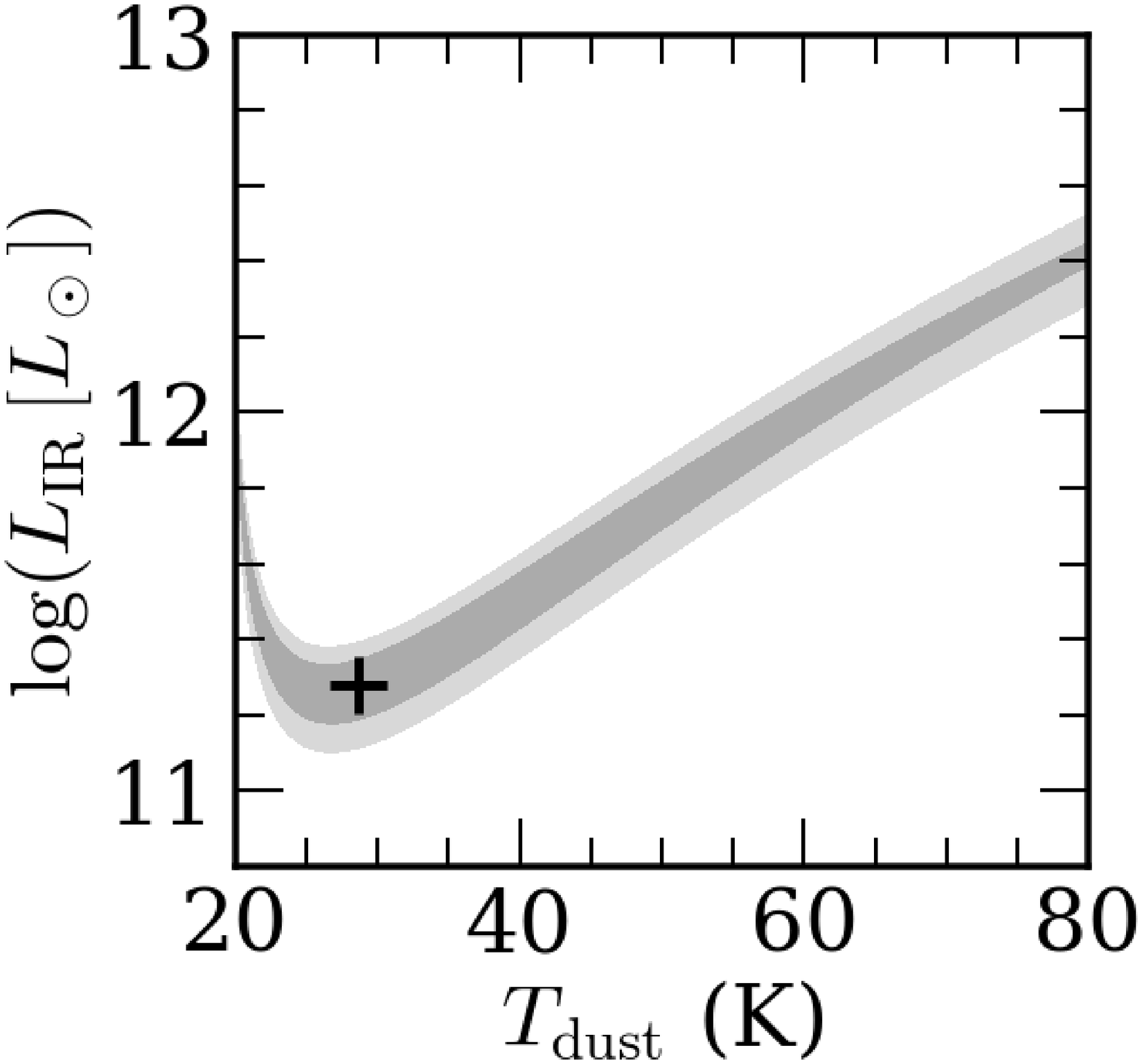}
\caption{
\textbf{Left}: 
Dust continuum SEDs of our $z=6$ luminous LBGs, 
J1211--0118, J0235--0532, and J0217--0208 from top to bottom. 
The red arrows are the $3\sigma$ upper limits 
on the flux densities obtained in our ALMA observations.
The black circles and downward arrows denote 
the observed flux densities and $3\sigma$ upper limits, respectively, 
obtained in \cite{2020ApJ...896...93H}. 
The black solid curve indicates the best-fit modified blackbody 
and the gray shade corresponds to the $1\sigma$ uncertainties. 
\textbf{Right}: 
Error contours for the two parameters of $L_{\rm IR}$ and $T_{\rm dust}$ 
in the modified blackbody fitting. 
The dark and light shades denote the $1\sigma$ and $2\sigma$ confidence regions, respectively. 
The black cross corresponds to the best-fit parameters. 
}
\label{fig:continuumSED}
\end{center}
\end{figure*}

\begin{figure*}[h]
\begin{center}
   \includegraphics[width=1.0\textwidth]{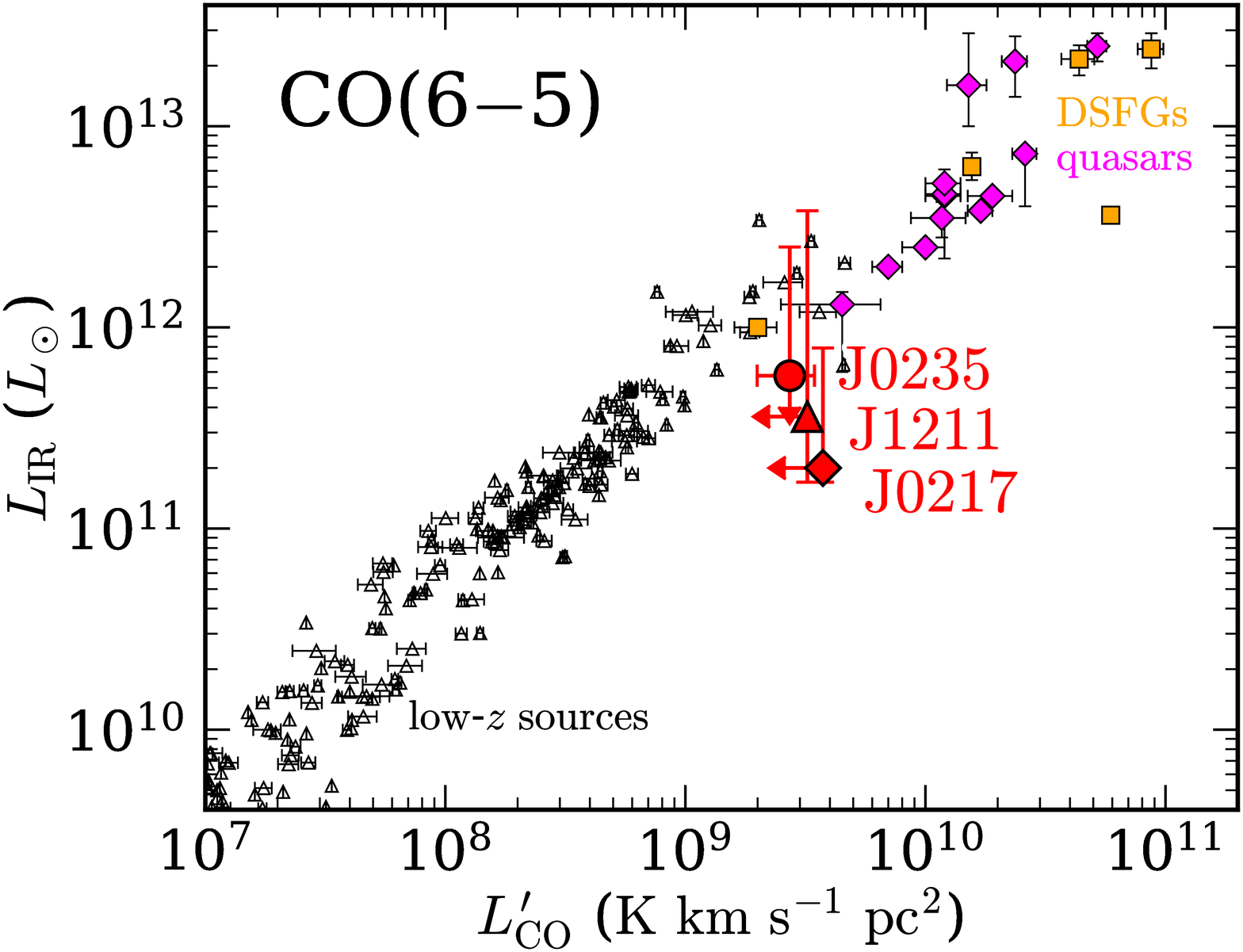}
\caption{
IR luminosity integrated over the wavelength range of $8$--$1000 \mu$m, $L_{\rm IR}$, 
vs. CO(6--5) luminosity in units of K km s$^{-1}$ pc$^2$, $L'_{\rm CO}$. 
The red circle is our ALMA result for a luminous LBG at $z=6$, J0235--0532, 
whose CO(6--5) emission shows the $4\sigma$ significance level, 
with $L_{\rm IR}$ and $L'_{\rm CO}$ in the case of $T_{\rm dust} = 50$ K 
(\citealt{2019PASJ...71...71H}; \citealt{2020ApJ...896...93H}).
The upper error bar along the $y$-axis for J0235--0532 
considers a higher dust temperature case of $T_{\rm dust} = 80$ K. 
The red triangle and diamond are also our ALMA results for 
the other luminous LBGs at $z=6$, J1211--0118 and J0217--0208, respectively, 
which shows no significant CO(6--5) detection. 
The red arrows correspond to the $3\sigma$ upper limits.
The orange squares are high-$z$ DSFGs 
(\citealt{2019A&A...628A..23A}; \citealt{2019ApJ...887...55C}; \citealt{2018ApJ...863L..29D}; 
\citealt{2010ApJ...720L.131R}; \citealt{2017ApJ...842L..15S}) 
and the magenta diamonds are high-$z$ quasars 
(\citealt{2019MNRAS.489.3939C}; \citealt{2017ApJ...845..154V}; \citealt{2010ApJ...714..699W}; 
\citealt{2011AJ....142..101W}; \citealt{2016ApJ...830...53W}) 
at $z\sim5$--$7$. 
The black triangles are nearby galaxies, Seyfert galaxies, and (U)LIRGs 
at low redshifts ($z < 0.1$) 
compiled by \cite{2015ApJ...810L..14L}.
}
\label{fig:CO65_LIR}
\end{center}
\end{figure*}

For J0235--0532, the CO(6--5) emission line is 
marginally detected at the expected frequency from the systemic redshift, 
while for the other two targets, the CO(6--5) emission is not significantly detected. 
Figure \ref{fig:CO_spectrum} shows the ALMA spectra of our $z=6$ luminous LBGs   
around their CO(6--5) emission line as expected from their systemic redshift 
measured by \cite{2020ApJ...896...93H} 
with the far-infrared (FIR) emission lines of [{\sc Oiii}] and [{\sc Cii}].
The spectrum of J0235--0532 is extracted 
by placing a single beam aperture around the peak position of the CO emission 
in the CO(6--5) moment zero map (velocity integrated map; Figure \ref{fig:CO_moment_zero}), 
because the CO emission is not spatially resolved in the ALMA data. 
We fit Gaussian functions to the observed spectrum of J0235--0532  
from $97.3$ GHz to $97.8$ GHz 
and obtain the best-fit Gaussian function as presented in Figure \ref{fig:CO_spectrum}. 
The integrated flux of this line calculated from the best-fit function is $0.0652 \pm 0.0175$ Jy km s$^{-1}$, 
indicating that the CO(6--5) emission line of J0235--0532 
shows a marginal detection at the $\simeq 4\sigma$ significance level. 
Reassuringly the velocity width of the CO line is comparable to 
those of the previously detected [{\sc Cii}] and [{\sc Oiii}] lines (\citealt{2020ApJ...896...93H}). 
Because the CO(6--5) is not significantly detected 
for J1211--0118 and J0217--0208 
(Figures \ref{fig:CO_spectrum} and \ref{fig:CO_moment_zero}), 
we extract their spectra by placing a beam aperture based on the coordinates of their rest UV continuum emission. 
The upper limits of their CO(6--5) line fluxes 
are calculated from the square root of the sum of the squared flux density errors 
in the range of $\pm 250$ km s$^{-1}$ 
around the expected CO(6--5) frequency from their systemic redshift.
The range of $\pm 250$ km s$^{-1}$ 
is comparable to the $2 \times$ FWHM 
of their [{\sc Oiii}] and [{\sc Cii}] emission lines 
(their FWHMs are about 
$170$--$370$ km s$^{-1}$; Table 1 of \citealt{2020ApJ...896...93H}). 
The integrated emission line flux or the upper limit for each target 
and the observed FWHM of the detected line 
are presented in Table \ref{tab:observational_results}.

Figure \ref{fig:CO_OIII_CII_J0235} compares the ALMA spectra of our $z=6$ luminous LBGs 
around the CO(6--5) emission line 
with those around the [{\sc Oiii}] and [{\sc Cii}] emission lines.
Although the S/N of the CO(6--5) line of J0235--0532 is not high, 
the redshifts based on the CO, [{\sc Oiii}], and [{\sc Cii}] lines are broadly consistent with each other. 
Some previous studies have shown significant velocity shifts 
between [{\sc Oiii}] and [{\sc Cii}] 
(e.g., BDF-3299 in \citealt{2017A&A...605A..42C}; 
cf. \citealt{2019PASJ...71...71H}; \citealt{2020MNRAS.493.4294B} as counter-examples); 
our results suggest that the velocity shift in J0235--0532, if any, is smaller 
than the previous results.

Figure \ref{fig:CO_image_J0235}  
presents the CO(6--5) emission contours of J0235--0532 
with the Subaru HSC $z$-band image probing the rest-frame UV continuum emission. 
Although the positions of CO(6--5) and UV continuum appear to be slightly offset, 
this may be caused by 
the relatively low S/N of the CO emission. 
We estimate the uncertainties of the CO peak position 
by running a suite of Monte Carlo simulations in the same way as \cite{2020ApJ...896...93H}. 
We add artificial noises to the actual data 
according to a Gaussian random distribution 
with a standard deviation equal to the $1\sigma$ noise of the data, 
and remeasure the peak positions one thousand times 
to estimate the uncertainties of the CO peak position. 
We find that the CO peak position is consistent with that of the UV continuum  
within the $2 \sigma$ uncertainties. 
In Figure \ref{fig:CO_image_J0235}, 
we also present the [{\sc Cii}] and [{\sc Oiii}] positions obtained in \cite{2020ApJ...896...93H}, 
confirming that the CO peak position is also consistent with those of [{\sc Cii}] and [{\sc Oiii}].

For a sanity check of the position of J0235--0532 in the HSC astrometry, 
which has been calibrated 
against the Pan-STARRS first data release (DR1) catalog (\citealt{2016arXiv161205560C}), 
we use nearby ($<1'$) bright stars 
whose positions are accurately measured 
in the Gaia early data release 3 (EDR3) catalog
(\citealt{2016A&A...595A...1G,2021A&A...649A...1G}).\footnote{\url{https://gea.esac.esa.int/archive/}} 
We confirm that the positional differences of the nearby bright stars 
between the HSC data and the Gaia EDR3 catalog 
are only within $<0\farcs01$ with no systematic offsets, 
which is consistent with similar comparison results in previous work 
on a much larger scale 
(Section 6.3 of \citealt{2019PASJ...71..114A}).

As mentioned above, 
the significance of the CO(6--5) line from J0235--0532 is only about $4 \sigma$. 
However, the observed CO peak position on the sky is consistent with that of the UV continuum, 
and the observed CO frequency is also in good agreement with 
those of the previously detected FIR emission lines;  
the probability of these events occurring simultaneously by chance is lower 
than the estimate above.
We calculate the combined probability 
that these three events occur simultaneously by chance 
based on Fisher's method (\citealt{1970smrw.book.....F}; 
see also, \citealt{2015ApJ...814...20F}; 
\citealt{2018AJ....156...24M}; 
\citealt{2021arXiv210809288K}; 
cf. \citealt{10.1093/biomet/asx076}). 
First, the probability that the CO detection is a false positive 
can be calculated from the significance of the CO line from J0235--0532. 
We obtain a false positive probability ($p$-value) of $p_1 \simeq 9.7 \times 10^{-5}$, 
assuming that the flux measurement errors follow a Gaussian distribution.
Second, the probability that the CO peak position is 
consistent with the previously detected source position 
can be calculated from the fraction of the area corresponding to the $2\sigma$ range 
to that of the obtained ALMA data. 
The $p$-value for this event is estimated to be $p_2 \simeq 7.3 \times 10^{-4}$.
Third, the probability that the detected line frequency 
coincides with those of the previously detected FIR lines can be calculated from 
the fraction of the $2 \times$ FWHM frequency range to that of the four SPWs. 
The $p$-value for this event is estimated to be $p_3 \simeq 2.1 \times 10^{-2}$.
From these individual $p$-values, 
we calculate a test statistic (TS) for the combined probability, 
\begin{equation}
{\rm TS}
	= -2 \sum_{i=1}^k \ln p_i, 
\end{equation}
where $k=3$ in this case. 
By comparing this TS with 
the $\chi^2$ distribution with $2k$ degrees of freedom, $\chi^2_{2k}$, 
we obtain a combined $p$-value 
$p_{\rm com} \simeq 3.4 \times 10^{-7}$ 
from 
\begin{equation}
p_{\rm com}
	= \int^\infty_{\rm TS} \chi^2_{2k} (x) dx. 
\end{equation}
We then solve the following equation, 
\begin{equation}
p_{\rm com}
	= \int^\infty_{{\rm S/N}_{\rm com}} \dfrac{1}{\sqrt{2\pi}} \exp \left( - \dfrac{x^2}{2} \right) dx,
\end{equation}
to obtain an equivalent Gaussian standard deviation of 
S/N$_{\rm com}$ $\simeq 5.0$ as the combined significance. 
In this paper, this signal for J0235--0532 is regarded as the CO(6--5) line.  
However, because the combined significance is still not very high, 
it is necessary to secure a firm detection of this CO line with follow-up observations.

Although the CO peak position 
is consistent with that of the UV continuum, 
the apparent offset (with large uncertainties) 
might indicate a hint of 
photoevaporation of photodissociation regions (PDRs; 
e.g., \citealt{2017MNRAS.467.1300V}; \citealt{2017MNRAS.471.4476D}; \citealt{2017A&A...605A..42C}). 
A partial displacement between the UV continuum tracing {\sc Hii} regions 
and CO(6--5) emission tracing dense clumps within giant molecular clouds 
(GMCs; e.g., \citealt{2007ARA&A..45..565M}) is expected 
when PDRs are photoevaporated (For details, see discussion in \citealt{2017MNRAS.467.1300V}). 
This scenario can be verified 
by observing the CO emission from J0235--0532 
with a higher S/N and better resolution.

From the integrated CO(6--5) emission line flux, 
we obtain the CO(6--5) luminosity in units of $L_\odot$ 
by using Equation (18) of \cite{2014PhR...541...45C}.
We also calculate 
the CO(6--5) luminosity in units of K km s$^{-1}$ pc$^2$ defined as 
Equation (19) of \cite{2014PhR...541...45C}. 
These equations are presented in Section \ref{sec:standard_equations}.
In these calculations, the effect of the cosmic microwave background (CMB) is taken into account 
by dividing the observed integrated flux by a factor of $f_{\rm CMB}$ 
(\citealt{2013ApJ...766...13D}),
\begin{equation}
f_{\rm CMB}
	= 1 - \dfrac{B_\nu [T_{\rm CMB} (z)]}{B_\nu [T_{\rm exc}]}, 
\end{equation}
where $B_\nu$ is the Planck function, 
$T_{\rm CMB} (z) = 2.73 (1+z)$ K is the temperature of the CMB, 
and 
$T_{\rm exc}$ is the excitation temperature of the CO(6--5) transition. 
Assuming the local thermal equilibrium (LTE), 
$T_{\rm exc}$ is equal to the kinetic temperature of the gas, $T_{\rm kin}$, 
and then to the dust temperature, $T_{\rm dust}$, 
i.e., $T_{\rm exc} = T_{\rm kin} = T_{\rm dust}$.\footnote{Note that, 
even in the case of non-LTE, 
if the gas temperature and density are relatively high, 
the CMB effect for the CO(6--5) line is comparable to that in the LTE case 
(see the non-LTE example with $T_{\rm kin} = 40$ K 
and the number density of H$_2$ molecules of $n_\mathrm{H_2} = 10^{4.2}$ cm$^{-3}$ 
in Figure 10 of \citealt{2013ApJ...766...13D}; 
cf. Figure 6 of \citealt{2013ApJ...766...13D}).}
Here we use $T_{\rm dust}$ estimated in Section \ref{subsec:results_dust}. 
The obtained CO luminosities or upper limits, 
as well as the $f_{\rm CMB}$ values, 
are presented in Table \ref{tab:observational_results}.\footnote{Because 
J0235--0532 shows no dust continuum detection 
in our data and the previous data (\citealt{2020ApJ...896...93H}), 
the obtained constraint on $T_{\rm dust}$ is not stringent. 
We thus consider the $T_{\rm dust}$ uncertainty 
when we obtain the uncertainties of the physical quantities of J0235--0532 
that are related to $T_{\rm dust}$ 
such as the IR luminosity and the CO luminosity. 
For details, see Section \ref{subsec:results_dust}.}
Note, as a caveat, that this prescription assumes a uniform kinetic temperature 
for CO and dust continuum emitting regions. 
However, in reality, PDRs have a kinetic temperature profile 
that depends on the radiation field and the gas density. 
If the kinetic temperature of the CO emitting regions is higher than adopted here, 
$T_{\rm exc}$ would be higher and thus $f_{\rm CMB}$ would be higher 
(Section 2.4 of \citealt{2015ApJ...813...36V}; 
see also, Section 4.3 of \citealt{2018MNRAS.473..271V}). 
In this sense, the \cite{2013ApJ...766...13D} prescription may provide a pessimistic estimate 
of the fraction of the intrinsic flux observed.

\subsection{Dust Continuum Emission} \label{subsec:results_dust}

The dust continuum emission 
from our $z=6$ luminous LBGs 
at $\lambda_{\rm obs} \simeq 3$ mm ($\lambda_{\rm rest} \simeq 430 \, \mu$m) 
is not significantly detected. 
The $3\sigma$ upper limits of 
their dust continuum flux densities are 
$52.3 \, \mu$Jy for J1211--0118, 
$29.1 \, \mu$Jy for J0235--0532,
and  
$22.1 \, \mu$Jy for J0217--0208.
Their dust continuum emission maps are presented in Appendix \ref{sec:dust_continuum_emission_maps}.

In order to characterize their dust continuum emission properties, 
we combine our ALMA results at $\lambda_{\rm rest} \simeq 430 \, \mu$m 
with the results of \cite{2020ApJ...896...93H} at shorter wavelengths of 
$\lambda_{\rm rest} \simeq 90$--$160 \, \mu$m, 
and fit modified blackbody spectral energy distributions (SEDs) 
to the observed flux densities 
by varying $T_{\rm dust}$ and $L_{\rm IR}$. 
We calculate the intrinsic dust continuum flux densities of a modified blackbody SED 
by using Equation (\ref{eq:modified_blackbody_SED}), and then obtain 
the expected dust continuum flux densities of the modified blackbody, 
$f_\nu^{\rm (exp)}$, from $f_\nu^{\rm (int)}$ 
by considering the CMB heating and attenuation effects 
based on the prescription of \cite{2013ApJ...766...13D} 
in the same way as described in Section \ref{subsec:results_co}. 
In the dust continuum SED fitting, 
we require that $T_{\rm dust}$ be higher than 
the CMB temperature at the redshift of the galaxy 
($\sim 20$ K at $z \sim 6$).

Figure \ref{fig:continuumSED} shows the modified blackbody SED fitting results to the observed SEDs.  
For J1211--0118 and J0217--0208, 
modified blackbody SEDs fit well with the observed SEDs. 
The best-fit IR luminosities and dust temperatures are 
($L_{\rm IR}$, $T_{\rm dust}$) 
$=$ ($3.6^{+34.4}_{-1.9} \times 10^{11} L_\odot$, $40^{+44}_{-14}$ K) 
for J1211--0118, 
and 
($2.0^{+5.9}_{-0.3} \times 10^{11} L_\odot$, $31^{+26}_{-9}$ K) 
for J0217--0208, 
which are consistent with the results of \cite{2020ApJ...896...93H}. 
Considering the large $T_{\rm dust}$ uncertainties, 
the reason why the CO emission lines are not detected for these targets 
may be that the CMB attenuation effect for these targets are relatively large 
($f_{\rm CMB}$ is small) due to low $T_{\rm dust}$.
Because our observations only add an upper limit to the observed SEDs 
on the longer wavelength side of the SED peak, 
the parameter constraints do not become stronger compared to the previous work. 
The two parameters of $L_{\rm IR}$ and $T_{\rm dust}$ are still degenerate,  
which will be greatly improved if deep observations for the dust continuum emission 
at shorter wavelengths than the SED peak are conducted.  
Note that another method has been proposed recently 
to determine $T_{\rm dust}$ and the dust mass assuming dust to be in radiative equilibrium 
if the source size of dust continuum emission is obtained 
(\citealt{2020MNRAS.495.1577I}). 
Alternatively, it would be possible to have independent estimates of 
$L_{\rm IR}$ and $T_{\rm dust}$, as well as the dust mass, 
based on the dust continuum and [{\sc Cii}] line luminosities 
by adopting the method recently presented in \cite{2021MNRAS.503.4878S}.

For J0235--0532, although the allowed parameter ranges are determined 
based on the upper limits of the flux densities, 
the obtained constraints on $L_{\rm IR}$ and $T_{\rm dust}$ are not stringent.
Following \cite{2020ApJ...896...93H}, 
we adopt $T_{\rm dust} = 50$ K for this galaxy without continuum detection 
as a fiducial value (see also, \citealt{2019PASJ...71...71H})
for comparisons with previous studies,  
which yields a $3\sigma$ upper limit of $L_{\rm IR} < 5.8 \times 10^{11} L_\odot$. 
We also consider a higher dust temperature case of $T_{\rm dust} = 80$ K 
as a systematic uncertainty. 
This is because J0235--0532 has the highest [{\sc Oiii}]/[{\sc Cii}] luminosity ratio (Table \ref{tab:targets}), 
possibly suggesting a relatively high dust temperature. 
In fact, previous observations for nearby galaxies 
have shown that SFGs with higher [{\sc Oiii}]/[{\sc Cii}] ratios 
tend to have higher $T_{\rm dust}$ values (\citealt{2018ApJ...869L..22W}), 
although the [{\sc Oiii}]/[{\sc Cii}] ratios of their SFGs 
are not as high as those of J0235--0532. 
More recently, 
dust continuum observations for 
a $z=8.31$ galaxy with a similarly high [{\sc Oiii}]/[{\sc Cii}] ratio, MACS0416-Y1,
have suggested a possibility that its dust temperature may be extremely high, 
exceeding $80$ K, although its physical origin is still under discussion 
(\citealt{2020MNRAS.493.4294B}). 
One possible physical explanation for very high dust temperatures is that 
part of their dust is locked in molecular clouds and/or young star clusters that host active star formation. 
Based on hydrodynamic simulations, 
\cite{2018MNRAS.477..552B} have shown that, 
in such a situation, dust is heated by the strong interstellar radiation fields and 
can show a very high dust temperature, efficiently emitting FIR continuum,  
which can explain the high FIR luminosity 
without invoking mechanisms for massive dust production at high redshifts 
(See also, e.g., \citealt{2019MNRAS.488.2629A}; \citealt{2020MNRAS.497..956S}).
The higher $T_{\rm dust}$ case for J0235--0532 yields a more conservative upper limit of 
$L_{\rm IR} < 2.5 \times 10^{12} L_\odot$ ($3\sigma$).

In Figure \ref{fig:CO65_LIR}, 
we compare the CO(6--5) and IR luminosities of our luminous LBGs at $z=6$ 
with those of nearby sources at $z < 0.1$ 
(\citealt{2015ApJ...810L..14L}) 
as well as dusty star-forming galaxies 
(DSFGs; \citealt{2019A&A...628A..23A}; \citealt{2019ApJ...887...55C}; \citealt{2018ApJ...863L..29D}; 
\citealt{2010ApJ...720L.131R}; \citealt{2017ApJ...842L..15S}) 
and quasars 
(\citealt{2019MNRAS.489.3939C}; \citealt{2017ApJ...845..154V}; \citealt{2010ApJ...714..699W}; 
\citealt{2011AJ....142..101W}; \citealt{2016ApJ...830...53W}) 
at comparable redshifts of $z \sim 5$--$7$ to our targets. 
For nearby sources, 
the correlation between CO and IR luminosities has been found over a wide luminosity range
(\citealt{2015ApJ...810L..14L}), 
which can be interpreted as an integrated KS relation 
because the CO and IR luminosities are correlated with 
gas mass and SFR, respectively
(e.g., \citealt{2017A&A...603A..93M}).
We find that 
our result for J0235--0532, which is the only one of our targets 
showing CO(6--5) detection at the $\simeq 4\sigma$ significance level, 
is broadly consistent with previous results 
owing to the relatively large uncertainty on the IR luminosity. 
For J1211--0118 and J0217--0208, whose CO emission is not detected, 
our results are also consistent with previous results. 
In other words, 
the obtained CO luminosity upper limits for these two sources 
are not deep enough to know 
whether they deviate from the correlation between $L'_{\rm CO}$ and $L_{\rm IR}$ 
seen in low-$z$ sources or not, 
which can be distinguished by much deeper CO observations.

Note that 
the CO spectral line energy distribution (SLED) excitation varies as a function of gas density, 
radiation field, mach number within GMCs, and presence of shocks 
(e.g., \citealt{2018MNRAS.473..271V}; \citealt{2021A&A...652A..66P}),  
and thus the CO(6--5) emission line, 
which traces dense gas with critical density of $n_{\rm crit} = 2.9 \times 10^5$ cm$^{-3}$, 
would trace a fraction of the total molecular gas, i.e., dense clumps within GMCs.  
Thus, the relation between $L'_{\rm CO(6-5)}$ and $L_{\rm IR}$ 
would not be entirely related to the $M_{\rm gas}$-SFR relation, 
and the interpretation as an integrated version of the KS relation could be partially hampered.
In Sections \ref{subsec:total_gas_mass} and \ref{subsec:SKrelation}, 
we convert $L'_{\rm CO(6-5)}$ to $L'_{\rm CO(1-0)}$ 
by adopting the average CO SLED for SFGs at lower redshifts 
to obtain the gas mass and gas surface density estimates from $L'_{\rm CO(1-0)}$, 
and compare them with the KS relation found in the local Universe, 
although the systematic uncertainties in such conversions are not small 
(Section \ref{subsec:systematic_uncertainties}).

\section{Discussion} \label{sec:discussion}

In this section, 
first we discuss physical origins 
for the relatively strong CO emission of J0235--0532 
compared to the other two targets 
based on comparisons with a PDR model and previous results.   
Next, 
we derive the total gas mass constraints from our CO results for the $z=6$ luminous LBGs, 
and present comparisons of their gas surface densities with previous results. 
Finally, we caution that the obtained gas mass constraints 
still have substantial systematic uncertainties.
Note that we also present other gaseous properties of gas fraction and gas depletion timescale 
and compare them with previous results in Appendix \ref{sec:extra_results}.

\subsection{Physical Reasons for the Luminous CO(6--5) Emission in J0235--0532} \label{subsec:reason_for_luminous_CO}

In this study, we have observed CO(6--5) emission for the three luminous LBGs at $z = 6$ 
with the comparable total SFRs of $\sim 100 M_\odot$ yr$^{-1}$.
As a result, CO(6--5) is marginally detected in J0235--0532 at the $\simeq 4 \sigma$ significance level, 
but not in the other two LBGs.
In this section, we discuss physical reasons for this difference.

Because the [{\sc Cii}] emission has also been detected for these LBGs
in \cite{2020ApJ...896...93H}, 
we calculate the line ratio of CO(6--5) to [{\sc Cii}] 
as well as the ratio of the [{\sc Cii}] to IR luminosity, 
which are useful for obtaining constraints 
on the physical properties of PDRs in galaxies 
such as the density of hydrogen nuclei, $n_{\rm H}$, 
and incident far-ultraviolet (FUV) radiation field, $U_{\rm UV}$, with $6$--$13.6$ eV 
based on comparisons with theoretical models for PDRs. 
For the PDR modeling, 
we use the Photodissociation Region Toolbox 
(PDRT; 
\citealt{1999ApJ...527..795K}; 
\citealt{2006ApJ...644..283K}; 
\citealt{2008ASPC..394..654P}),\footnote{\url{http://dustem.astro.umd.edu/index.html}} 
which calculates 
various line and continuum intensity ratios 
for combinations of $n_{\rm H}$ and $U_{\rm UV}$ 
by solving for 
the equilibrium chemistry, thermal balance, and radiation transfer through a PDR layer 
in a self-consistent way. 
Specifically, we use the wk2006 model of the PDRT 
with solar metallicity for comparisons with previous results.

Because the [{\sc Cii}] emission comes from not only PDRs but also {\sc Hii} regions, 
we need to subtract the contribution of [{\sc Cii}] emission from {\sc Hii} regions 
for comparisons with the PDRT calculation results. 
For this purpose, we refer to Figure 9 of \cite{2019A&A...626A..23C},
which presents the dependency of 
the fraction of [{\sc Cii}] emission from {\sc Hii} regions, $f_{\rm [CII]}^{\rm (ion)}$, 
on gas-phase metallicity 
(See also Figure 4 of \citealt{2017ApJ...845...96C}; 
\citealt{2019ApJ...886...60S}; \citealt{2021ApJ...909..130R}; 
see also theoretical results such as 
\citealt{2017MNRAS.468.4831K}; \citealt{2017ApJ...846..105O}; 
\citealt{2019MNRAS.489....1F}; \citealt{2019MNRAS.487.1689P}). 
Based on interstellar medium (ISM) absorption line analyses detected in the stacked spectrum of 
$z \sim 6$ luminous LBGs with $M_{\rm UV} \simeq -23$ mag 
including one of our targets, J1211--0118, 
\cite{2020ApJ...902..117H} have found that 
their gas-phase metallicity is close to solar.\footnote{This result 
is consistent with our use of the wk2006 model of the PDRT with solar metallicity.} 
By combining these two previous results, 
$f_{\rm [CII]}^{\rm (ion)}$ of our $z=6$ luminous LBGs would be about 
$f_{\rm [CII]}^{\rm (ion)} \simeq 0.3$.
Because the observed $f_{\rm [CII]}^{\rm (ion)}$ values have a scatter of $\simeq 0.1$--$0.2$, 
here we consider it as a systematic uncertainty. 
We also apply a factor of two correction for the observed CO flux 
considering line luminosity from both sides of each optically thick cloud 
for comparisons with the PDRT calculation results, 
as suggested by \cite{1999ApJ...527..795K} 
(see also, \citealt{2016ApJ...830...53W}; \citealt{2019ApJ...876...99S}; \citealt{2019ApJ...876..112R}).

Because the $f_{\rm [CII]}^{\rm (ion)}$ values are correlated 
with the [{\sc Cii}]/[{\sc Nii}]$122\mu$m luminosity ratio 
as presented in Figure 10 of \cite{2019A&A...626A..23C}, 
we can also evaluate $f_{\rm [CII]}^{\rm (ion)}$ from [{\sc Cii}]/[{\sc Nii}]. 
However, the [{\sc Nii}] emission has not been detected in any of our targets \citep{2020ApJ...896...93H}, 
and the lower limits on the [{\sc Cii}]/[{\sc Nii}] ratios are not so stringent. 
Calculating the $3 \sigma$ lower limits on the [{\sc Cii}]/[{\sc Nii}] ratios based on Table 1 of \cite{2020ApJ...896...93H}, 
we obtain $L_{\rm [CII]}/L_{\rm [NII]} > 0.36$--$2.3$. 
We confirm that the expected range of the $f_{\rm [CII]}^{\rm (ion)}$ values 
from the lower limits of the [{\sc Cii}]/[{\sc Nii}] ratios 
are consistent with those expected from the gas metallicity.

\begin{figure*}[h]
\begin{center}
   \includegraphics[width=0.49\textwidth]{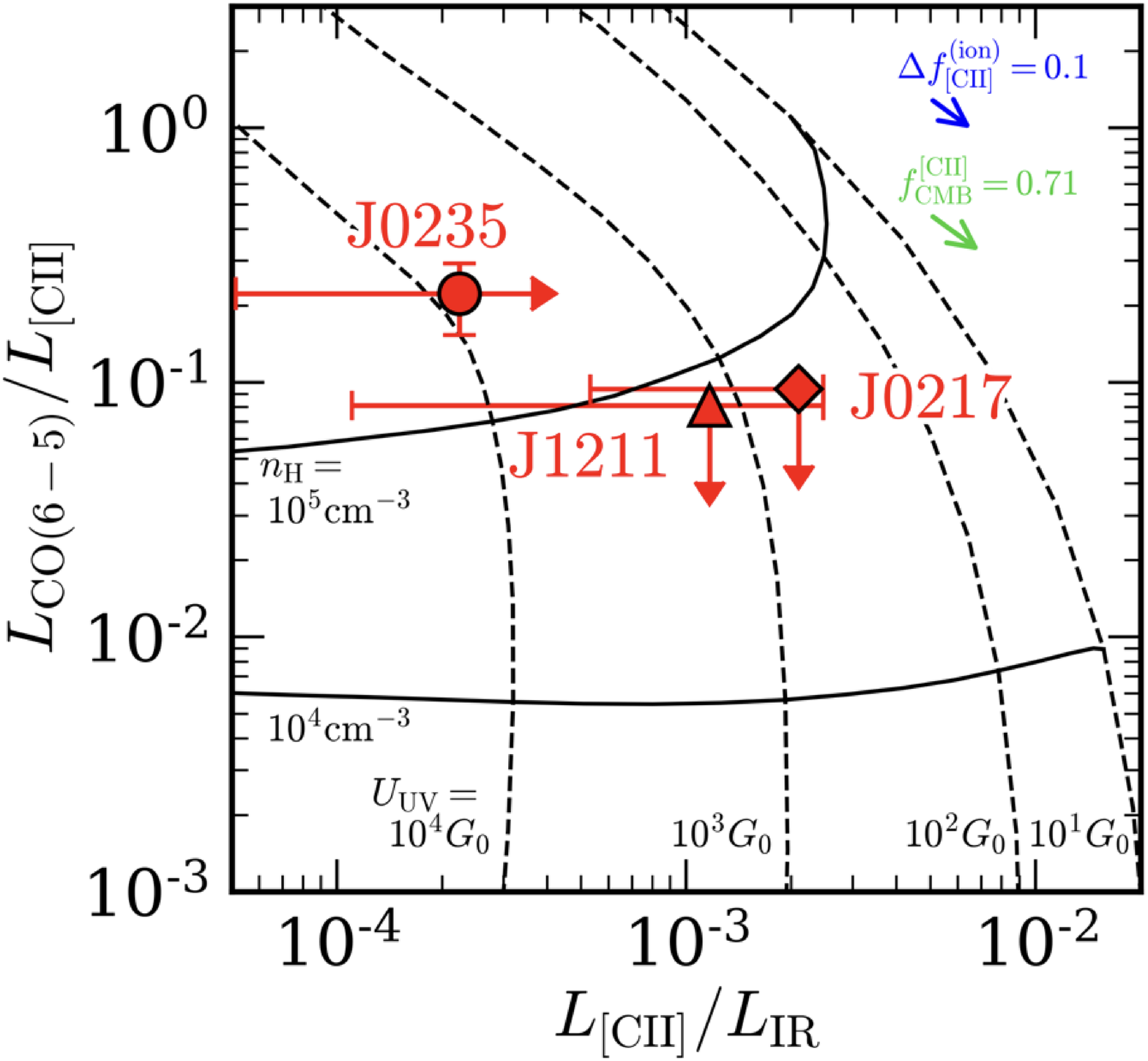}
   \includegraphics[width=0.49\textwidth]{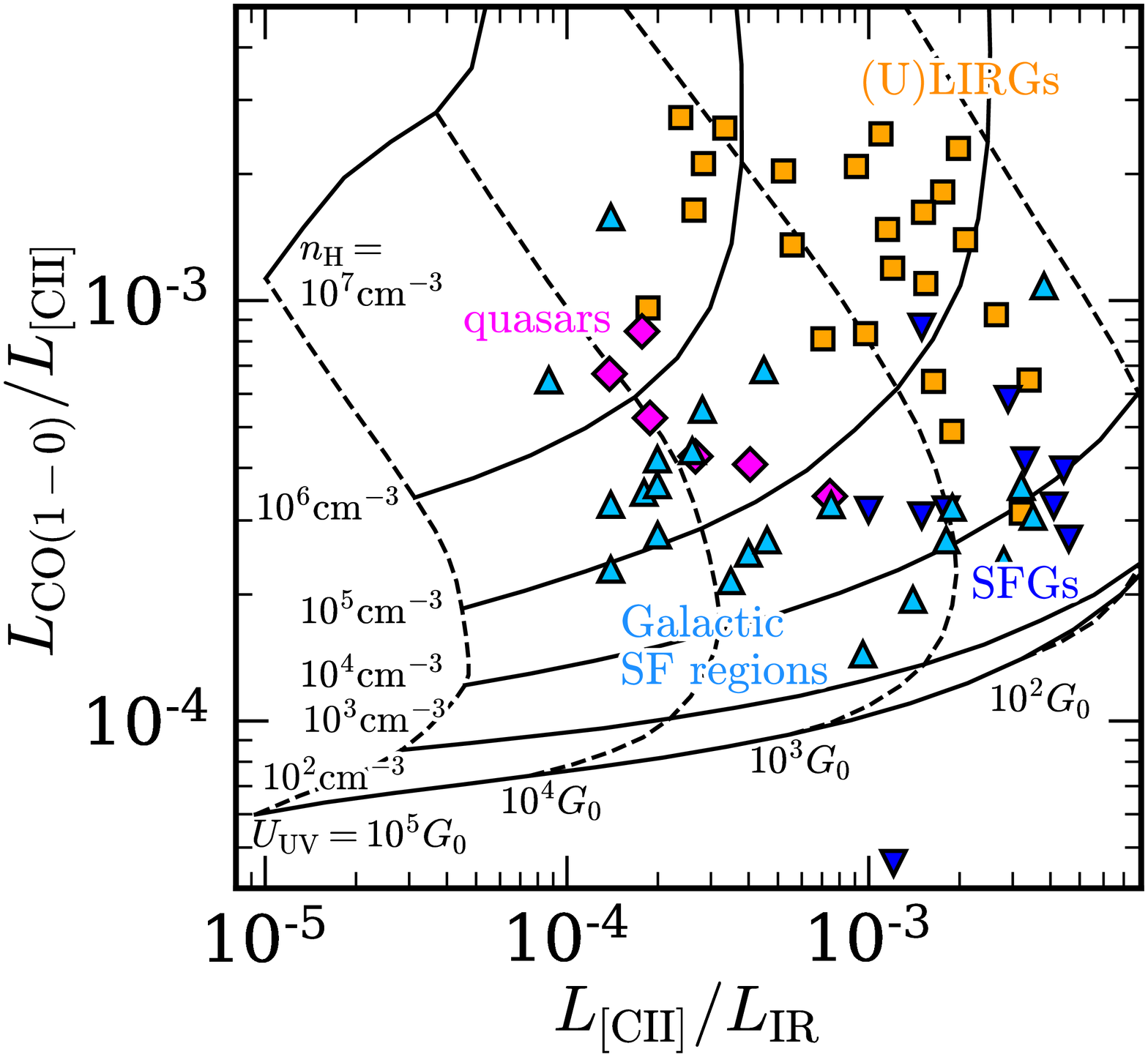}
   \includegraphics[width=0.49\textwidth]{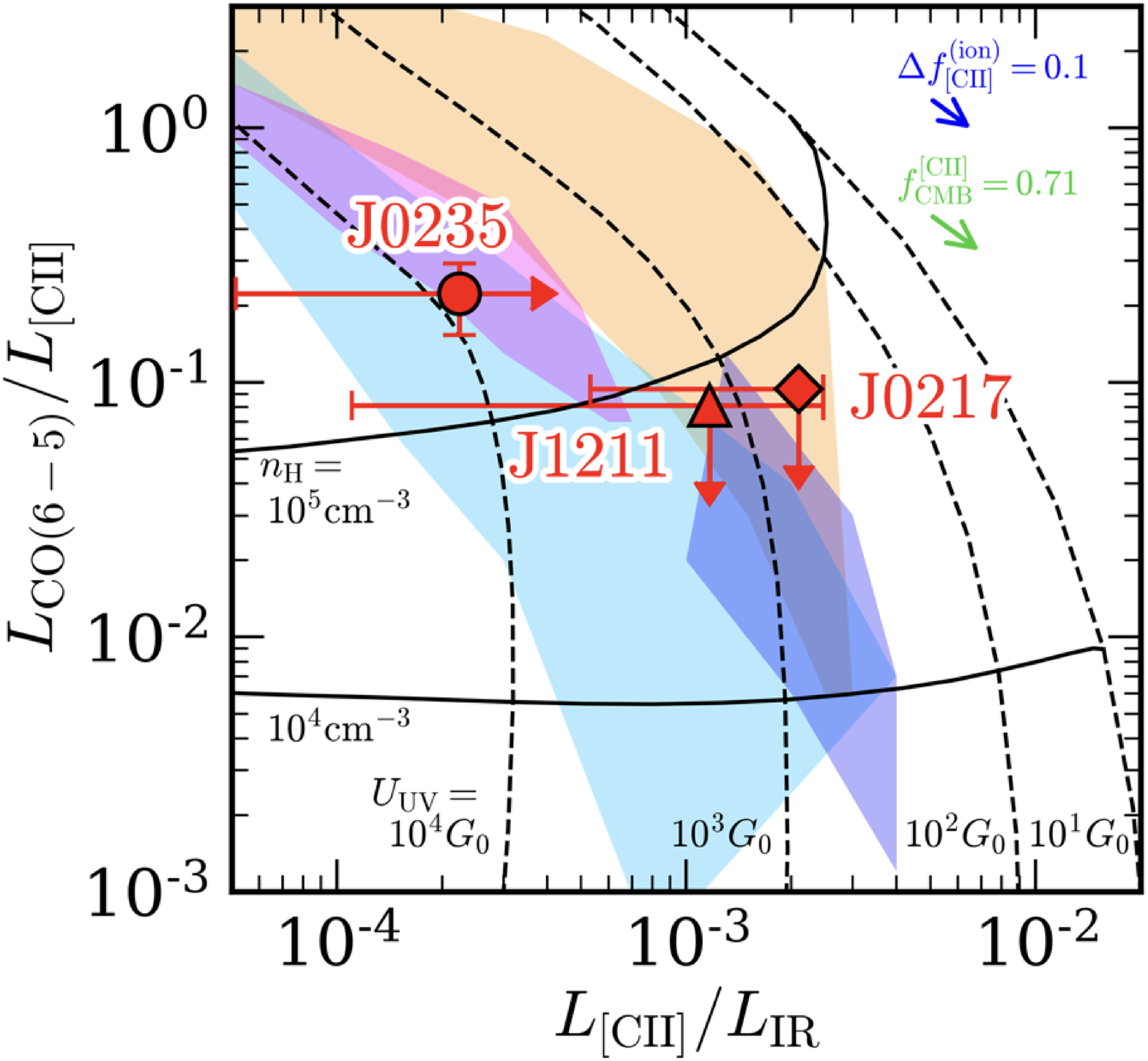}
\caption
{\textbf{Top Left}: 
$L_{\rm CO(6-5)}$/$L_{\rm [CII]}$ vs. $L_{\rm [CII]}$/$L_{\rm IR}$. 
The red circle is our ALMA result for a luminous LBG at $z=6$ with $\simeq 4 \sigma$ CO(6--5) detection, 
J0235--0532, in the case of  $T_{\rm dust} = 50$ K, 
and the lower error bar along the $x$-axis considers the higher dust temperature case of $T_{\rm dust} = 80$ K. 
The red triangle and diamond are also our ALMA results for 
the other luminous LBGs at $z=6$, J1211--0118 and J0217--0208, respectively, 
which shows no significant CO(6--5) detection. 
The red arrows correspond to the $3\sigma$ limits. 
The solid and dashed curves represent the theoretical calculations 
with the PDRT with 
constant hydrogen nucleus densities, $n_{\rm H}$, in units of cm$^{-3}$, 
and FUV (6--13.6 eV) radiation fields, $U_{\rm UV}$, 
in units of the average interstellar radiation field 
in the vicinity of the Sun, $G_0 = 1.6 \times 10^{-3}$ erg s$^{-1}$ cm$^{-2}$ 
(\citealt{1968BAN....19..421H}), 
respectively. 
The blue arrow in the upper right corner 
represents the amount of shift when $f_{\rm [CII]}^{\rm (ion)}$ increases by $0.1$. 
The green arrow in the upper right corner 
show the systematic uncertainties of the CMB effect on the [{\sc Cii}] emission 
\citep{2019MNRAS.487.3007K}. 
\textbf{Top Right}: 
$L_{\rm CO(1-0)}$/$L_{\rm [CII]}$ vs. $L_{\rm [CII]}$/$L_{\rm IR}$. 
The orange squares denote LIRGs and ULIRGs 
(\citealt{2015ApJ...801...72R}).
The blue downward triangles and cyan triangles represent 
local SFGs and Galactic star-forming regions, respectively 
(\citealt{1991ApJ...373..423S}). 
The magenta diamonds present high-$z$ quasars 
(\citealt{1999ApJ...518L..65B}; \citealt{2005A&A...440L..51M}; \citealt{2006ApJ...645L..97I}; 
\citealt{2012ApJ...752L..30W}; \citealt{2013ApJ...772..103L}; \citealt{2014ApJ...783...71W}; 
\citealt{2015MNRAS.451.1713S}; \citealt{2016ApJ...830...53W}; \citealt{2017ApJ...837..146V}; 
compiled by \citealt{2019ApJ...876...99S}). 
The solid and dashed curves are the same as in the left panel. 
In both panels, the CO line luminosities are multiplied by a factor of two 
as recommended by \cite{1999ApJ...527..795K} 
(for details, see the text in Section \ref{subsec:reason_for_luminous_CO}; 
see also, \citealt{2019ApJ...876...99S}). 
\textbf{Bottom}: 
Same as the top left panel, but with the colored shaded regions 
that roughly correspond to the low-$z$ CO(1--0) results presented in the top right panel. 
}
\label{fig:CIIFIR_CO65CII}
\end{center}
\end{figure*}

By taking account of these points, 
in the top left panel of Figure \ref{fig:CIIFIR_CO65CII}, 
we compare our ALMA results 
for $L_{\rm CO(6-5)}$/$L_{\rm [CII]}$ vs. $L_{\rm [CII]}$/$L_{\rm IR}$ 
with the PDRT calculation results. 
Because the PDRT does not include the CMB temperature, 
the observed CO and IR luminosities are corrected for the CMB effect (Section \ref{sec:results}).\footnote{Although 
we compare the observed results corrected for the CMB effect with the PDRT calculation results,  
the PDRT calculation with the consideration of the CMB effect may change the shape of the diagnostic 
(M. Wolfire, private communication). 
We need to check this point when the theoretical models are updated in the future.}
In the same way as in Figure \ref{fig:CO65_LIR}, 
we adopt $T_{\rm dust} = 50$ K for J0235--0532 as a fiducial value 
and consider up to $T_{\rm dust} = 80$ K as a systematic uncertainty 
yielding a conservative lower limit of $L_{\rm [CII]}$/$L_{\rm IR}$. 
We find that the $n_{\rm H}$ value of J0235--0532 is 
higher than those of J1211--0118 and J0217--0208.
Because we only obtain the lower limit for the $L_{\rm [CII]}$/$L_{\rm IR}$ ratio of J0235--0532 
due to the non-detection of the dust continuum emission, 
it is unclear whether the incident FUV radiation in J0235--0532 is stronger than the others or not.

For comparisons of the $n_{\rm H}$ and $U_{\rm UV}$ values 
of our $z=6$ luminous LBGs 
with those of other sources at lower redshifts, 
in the top right panel of Figure \ref{fig:CIIFIR_CO65CII}, 
we show previous observation results for 
$L_{\rm CO(1-0)}$/$L_{\rm [CII]}$ vs. $L_{\rm [CII]}$/$L_{\rm IR}$ 
of luminous infrared galaxies (LIRGs), 
ultra-luminous infrared galaxies (ULIRGs; \citealt{2015ApJ...801...72R}), 
quasars 
(\citealt{1999ApJ...518L..65B}; \citealt{2005A&A...440L..51M}; \citealt{2006ApJ...645L..97I}; 
\citealt{2012ApJ...752L..30W}; \citealt{2013ApJ...772..103L}; \citealt{2014ApJ...783...71W}; 
\citealt{2015MNRAS.451.1713S}; \citealt{2016ApJ...830...53W}; \citealt{2017ApJ...837..146V}; 
compiled by \citealt{2019ApJ...876...99S}),  
local SFGs such as spiral galaxies, and Galactic star-forming regions 
(\citealt{1991ApJ...373..423S}), 
as well as the PDRT model calculation results.  
For easier comparison, 
the bottom panel of Figure \ref{fig:CIIFIR_CO65CII} is the same as the top left panel 
but with the color shaded regions 
that roughly corresponds to the locations of low-$z$ sources in the previous work 
with CO(1--0) observations shown in the top right panel. 
We find that 
the relatively high $n_{\rm H}$ value of J0235--0532, 
$n_{\rm H} \gtrsim 10^5$ cm$^{-3}$ depending on $U_{\rm UV}$, 
is consistent with those of LIRGs and ULIRGs 
with relatively low $n_{\rm H}$ values in that population, 
as well as  
those of quasars and Galactic star-forming regions 
with high $n_{\rm H}$ and $U_{\rm UV}$ values 
considering the high $T_{\rm dust}$ case. 
We also find that 
J1211--0118 and J0217--0208, 
likely showing moderate $U_{\rm UV}$ values 
with $n_{\rm H}$ upper limits around $10^5$ cm$^{-3}$  
are consistent with nuclear regions of local SFGs 
and Galactic star-forming regions 
with relatively low $n_{\rm H}$ values.

It should be noted that there are two systematic uncertainties 
in this comparison.
One is related to $f_{\rm [CII]}^{\rm (ion)}$. 
As mentioned above, the relation between $f_{\rm [CII]}^{\rm (ion)}$ and metallicity 
has a scatter of about $0.1$--$0.2$ 
(e.g., Figure 9 of \citealt{2019A&A...626A..23C}).  
In our $L_{\rm CO(6-5)}$/$L_{\rm [CII]}$ vs. $L_{\rm [CII]}$/$L_{\rm IR}$ figures, 
we show the blue arrow in the upper right corner 
that corresponds to the amount of shift 
when $f_{\rm [CII]}^{\rm (ion)}$ is increased by $0.1$. 
We confirm that this systematic uncertainty does not significantly affect the results. 
The other systematic uncertainty is the CMB effect on the [{\sc Cii}] emission.
As discussed in Section 6.1 of \cite{2020ApJ...896...93H}, 
the [{\sc Cii}] emission may also be affected by the CMB attenuation  
due to the high CMB temperature at $z \sim 6$ 
(see also, \citealt{2018A&A...609A.130L}; \citealt{2014ApJ...784...99G}; \citealt{2019MNRAS.487L..81L}).
Figure 1 of \cite{2019MNRAS.487.3007K} shows 
the CMB suppression effect of the [{\sc Cii}] emission 
with different gas temperatures as a function of gas number density. 
Although only upper limits are derived for $n_{\rm H}$ of J1211--0118 and J0217--0208, 
with a conservative gas density value of $10^4$ cm$^{-3}$, 
we obtain the CMB effect on the [{\sc Cii}] emission line flux of 
$f_{\rm CMB}^{\rm [CII]} = 0.71$--$0.86$ at $30$--$40$~K 
(\citealt{2019MNRAS.487.3007K}; see also, \citealt{2015MNRAS.453.1898P}; \citealt{2015ApJ...813...36V}), 
which is comparable to the dust temperature. 
If the gas density and/or the gas temperature is higher, then the CMB effect is smaller, 
suggesting that the impact of this systematic uncertainty is comparable or smaller than 
that of the $f_{\rm [CII]}^{\rm (ion)}$ scatter.

There is another noticeable difference 
between J0235--0532 and the other two LBGs.  
J0235--0532 has the highest [{\sc Oiii}]/[{\sc Cii}] luminosity ratio 
([{\sc Oiii}]/[{\sc Cii}] $=8.9 \pm 1.7$) among our targets 
as shown in Figure 5 of \cite{2020ApJ...896...93H}. 
In the first place, 
these three $z=6$ LBGs have significantly higher [{\sc Oiii}]/[{\sc Cii}] ratios 
than $z \sim 0$ galaxies with comparable total SFRs.\footnote{It should be noted that 
the [{\sc Oiii}]/[{\sc Cii}] luminosity ratio of some high-$z$ galaxies can be overestimated 
because the [{\sc Cii}] emitting region of ALMA detected high-$z$ galaxies 
is typically about $2$--$3$ times more extended 
than the [{\sc Oiii}] and UV continuum emitting regions 
(\citealt{2020MNRAS.499.5136C}; 
see also, \citealt{2019ApJ...887..107F}; \citealt{2020ApJ...900....1F}; \citealt{2021A&A...649A..31H}).
To capture the extended [{\sc Cii}] emission, 
\cite{2020ApJ...896...93H} have calculated 
the total line fluxes with a large ($2''$ radius) aperture  
(See their Sections 4.1 and 6.1). 
Recently, \cite{2021MNRAS.505.5543V} have proposed that 
the [{\sc Oiii}]/[{\sc Cii}] surface brightness ratio is also useful to overcome this issue. 
We confirm that our $z=6$ luminous LBGs also have high [{\sc Oiii}]/[{\sc Cii}] surface brightness ratios 
and  J0235--0532 shows the highest value among them 
(\citealt{2021MNRAS.505.5543V}; see also, \citealt{2020MNRAS.499.5136C}).}
\cite{2020ApJ...896...93H} have discussed the physical reason for this 
based on comparisons with the results of model calculations 
for both {\sc Hii} regions and PDRs with CLOUDY 
(\citealt{1998PASP..110..761F}; \citealt{2017RMxAA..53..385F}) 
and concluded that high ionization parameters and/or low PDR covering fractions 
can explain high-$z$ galaxy results including the high [{\sc Oiii}]/[{\sc Cii}] ratios 
and low $L_{\rm [CII]}$/SFR ratios. 
\cite{2020ApJ...896...93H} have also found that 
high $n_{\rm H}$, low C/O ratios, and the CMB attenuation effect 
can reproduce a part of the high-$z$ galaxy results.

Because the CO emission originates from different regions 
from [{\sc Oiii}]-/[{\sc Cii}]-emitting regions  
(e.g., Figure 31.2 of \citealt{2011piim.book.....D}), 
it is difficult to make a direct comparison
between our results and the results of \cite{2020ApJ...896...93H}.  
The least we can say is that 
our results suggest a relatively high $n_{\rm H}$ in PDRs of J0235--0532 
compared to the other two LBGs, 
which would be consistent with the results of \cite{2020ApJ...896...93H}, 
although in their study it is not enough to explain the high-$z$ galaxy results. 
In addition, 
$U_{\rm UV}$ of J0235--0532 may be higher than those of the other two LBGs, 
which would be consistent with 
the high [{\sc Oiii}]/[{\sc Cii}] ratio and thus the relatively high ionization parameter, 
although deeper dust continuum observations are required 
for constraining $L_{\rm [CII]}$/$L_{\rm IR}$ to reach a conclusion on this point.

Interestingly, 
this is in line with what is expected from theoretical models. 
The high-$J$ CO lines trace relatively high density regions more directly connected to star formation. 
At such a high density, the self-shielding effect prevents the molecule dissociation 
and at the same time the warm temperature produced by the strong UV radiation 
suggested from the high [{\sc Oiii}]/[{\sc Cii}] ratio is expected to boost the high-$J$ CO emission  
(\citealt{2018MNRAS.473..271V}). 
In this case, the dust temperature is also likely to be high 
(\citealt{2018MNRAS.477..552B}).
In fact, the dust continuum is not detected only for J0235--0532, 
which is consistent with the possibility that the dust temperature of J0235--0532 may be very high. 
To confirm this picture, it would be interesting to carry out deep observations 
to detect high-$J$ CO emission from SFGs 
with high [{\sc Oiii}]/[{\sc Cii}] ratios and/or high $T_{\rm dust}$ 
such as MACS1149-JD1 at $z=9.1096$ 
([{\sc Oiii}]/[{\sc Cii}] $\gtrsim 19$; 
\citealt{2018Natur.557..392H}; \citealt{2019MNRAS.487L..81L}), 
MACS0416-Y1 at $z=8.3118$ 
([{\sc Oiii}]/[{\sc Cii}] $ = 8.6 \pm 2.5$ and $T_{\rm dust} > 80$ K; 
\citealt{2019ApJ...874...27T}; \citealt{2020MNRAS.493.4294B}), 
and 
SXDF-NB1006-2 at $z = 7.2120$
([{\sc Oiii}]/[{\sc Cii}] $\gtrsim 10$; \citealt{2016Sci...352.1559I}).
Note that careful estimates 
on their [{\sc Oiii}]/[{\sc Cii}] luminosity ratios have been provided recently  
by considering the surface brightness dimming effect \citep{2020MNRAS.499.5136C}, 
still showing relatively high [{\sc Oiii}]/[{\sc Cii}] values of 
 [{\sc Oiii}]/[{\sc Cii}] $= 4.2 \pm 1.4$ for MACS1149-JD1, 
 [{\sc Oiii}]/[{\sc Cii}] $= 8 \pm 2$ for MACS0416-Y1, 
 and 
 [{\sc Oiii}]/[{\sc Cii}] $= 4.3 \pm 1.4$ for SXDF-NB1006-2. 
We confirm that 
these sources also have high [{\sc Oiii}]/[{\sc Cii}] surface brightness ratios
\citep{2021MNRAS.505.5543V}.

\subsection{Gas Mass Constraints} \label{subsec:total_gas_mass}

We constrain molecular gas masses in our $z=6$ luminous LBGs 
based on our CO(6--5) results, 
although the systematic uncertainties are not small 
particularly in the CO SLED, 
which has not been investigated well for SFGs at high redshifts. 
Here we present conservative constraints on molecular gas masses in our targets 
by taking account of such uncertainties 
and compare with previous results for lower-$z$ sources.

The total gas mass for molecular clouds, $M_{\rm gas}$, can be estimated from 
the CO(1--0) luminosity in units of K km s$^{-1}$ pc$^2$, $L'_{{\rm CO}(1-0)}$,  
by using Equation (4) of 
\cite{2005ARA&A..43..677S}, 
\begin{equation}
M_{\rm gas} = \alpha_{\rm CO} L'_{{\rm CO}(1-0)}, 
\end{equation}
where $\alpha_{\rm CO}$ is the conversion factor 
from $L'_{{\rm CO}(1-0)}$ to $M_{\rm gas}$. 
We assume a fixed value of 
$\alpha_{\rm CO} = 4.5 \, M_\odot$ (K km s$^{-1}$ pc$^2$)$^{-1}$,\footnote{The 
systematic uncertainties related to this conversion factor is discussed in Section \ref{subsec:systematic_uncertainties}.}
which is consistent with previous results for 
Milky Way 
(\citealt{2013ARA&A..51..207B}), 
$z\sim1$--$2$ SFGs
(\citealt{2010ApJ...714L.118D}; \citealt{2013ARA&A..51..105C}), 
and even an LBG at $z=5.7$
(HZ10; \citealt{2019ApJ...882..168P}).\footnote{\cite{2019ApJ...882..168P} 
have estimated the total gas mass 
by subtracting the contribution of stars and dark matter masses 
from the dynamical mass measured with the significantly detected [{\sc Cii}] line, 
and obtained the $\alpha_{\rm CO}$ value 
from the ratio of the estimated total gas mass 
to the CO luminosity. 
They have found that the obtained $\alpha_{\rm CO}$ value 
is $\alpha_{\rm CO} = 4.2^{+2}_{-1.7} M_\odot$ (K km s$^{-1}$ pc$^2$)$^{-1}$,
which is consistent with that of the Milky Way, 
although the uncertainty is not small.}
The $L'_{{\rm CO}(1-0)}$ values of our $z=6$ luminous LBGs 
can be estimated from $L'_{{\rm CO}(6-5)}$ 
by using the average CO SLED for SFGs. 
Specifically, we assume that the integrated CO(6--5) flux 
is comparable to that of CO(5--4), 
i.e., $I_{\rm CO(6-5)} \simeq I_{\rm CO(5-4)}$, 
and adopt the average integral CO line flux ratio of 
$I_{\rm CO(5-4)}/I_{\rm CO(1-0)} \simeq 5.8 \pm 3.3$, 
which is measured for $z\sim1$--$2$ SFGs 
(\citealt{2015A&A...577A..46D}). 
The large uncertainty of $I_{\rm CO(5-4)}/I_{\rm CO(1-0)}$
is estimated from the standard deviation of the integrated CO flux ratios 
of the $z\sim1$--$2$ SFGs reported in \cite{2015A&A...577A..46D}.\footnote{Although 
here we consider the standard deviation of the previous observation results, 
the uncertainties of the CO SLEDs may be much larger. 
The details of this point are presented in Section \ref{subsec:systematic_uncertainties}.}
We then calculate $L'_{{\rm CO}(1-0)}$\footnote{ 
The CO(1--0) luminosity is calculated 
from 
\begin{equation}
L'_{{\rm CO}(1-0)}
	= \dfrac{I_{\rm CO(1-0)} }{ I_{\rm CO(6-5)} } 
		\left( \dfrac{ \nu_{\rm CO(6-5)}^{\rm (rest)} }{ \nu_{\rm CO(1-0)}^{\rm (rest)} } \right)^{2}
		L'_{{\rm CO}(6-5)}, 
\label{eq:LCO_ratio}
\end{equation}
where $\nu_{\rm CO(1-0)}^{\rm (rest)} = 115.27$ GHz. 
} 
and obtain  $M_{\rm gas}$ constraints 
as summarized in Table \ref{tab:observational_results}.

Note that 
\cite{2018MNRAS.481.1976Z} have reported 
a linear correlation between the [{\sc Cii}] luminosity and the gas mass 
for $z\sim 2$ SFGs  
and obtained a conversion factor from the [{\sc Cii}] luminosity and the gas mass, 
$\alpha_{\rm [CII]} = 31 M_\odot L_\odot^{-1}$. 
By adopting this conversion factor, 
we estimate the gas mass of J0235--0532 
from the [{\sc Cii}] luminosity 
to be only about $1.3 \times 10^{10} M_\odot$, 
which is significantly smaller than that obtained from the CO luminosity.  
This may suggest that 
the conversion factor $\alpha_{\rm [CII]}$ or $\alpha_{\rm CO}$ for high-$z$ luminous LBGs 
is different from those at low redshifts, 
or that the CO SLED is different from those for $z \sim 1$--$2$ SFGs,
although it is difficult to clarify these possibilities with the currently available data. 
In this study, 
we adopt the estimates based on the CO luminosity, 
because it is more commonly used in previous studies 
and would thus be more appropriate for comparisons.

\begin{figure*}[h]
\begin{center}
   \includegraphics[width=0.9\textwidth]{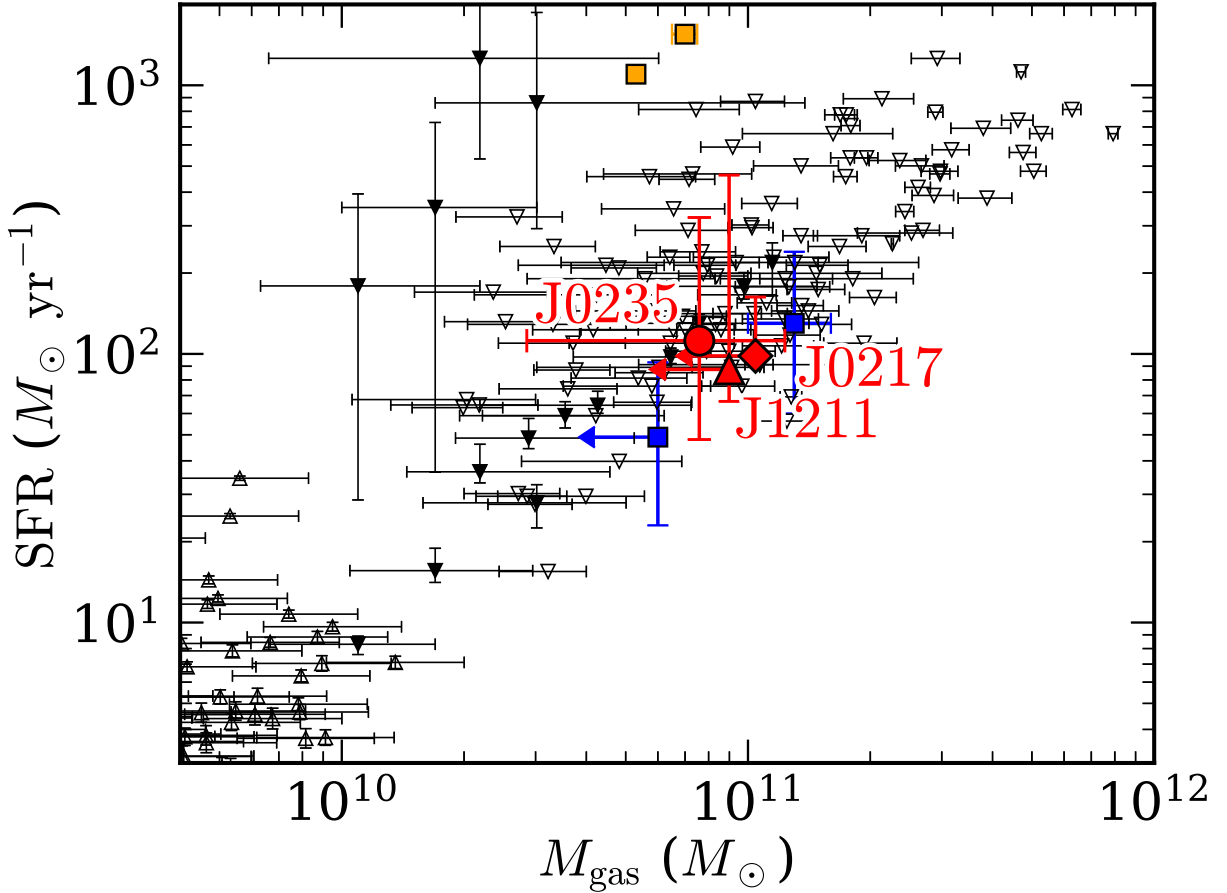}
\caption{SFR vs. $M_{\rm gas}$. 
The red circle is our ALMA result for a luminous LBG at $z=6$ with $\simeq 4 \sigma$ CO(6--5) detection, 
J0235--0532, in the case of  $T_{\rm dust} = 50$ K 
and SFR$_{\rm tot}$ $=$ SFR$_{\rm UV}$ $+$ SFR$_{\rm IR, 2 \sigma}$. 
The error bar along the $y$-axis considers 
the higher dust temperature case of $T_{\rm dust} = 80$ K 
and the minimum SFR case of SFR$_{\rm tot}$ $=$ SFR$_{\rm UV}$. 
Note that, although the intrinsic CO flux and thus the $M_{\rm gas}$ value 
become somewhat larger with a higher dust temperature, 
such a systematic uncertainty is much smaller than 
the uncertainty in the CO flux measurement. 
The red triangle and diamond are also our ALMA results for 
the other luminous LBGs at $z=6$, J1211--0118 and J0217--0208, respectively, 
which shows no significant CO(6--5) detection. 
The red arrows correspond to the $3\sigma$ limits.
The blue squares represent the results of LBG-1 and HZ10 from left to right (\citealt{2019ApJ...882..168P}). 
The blue arrow represents the $3\sigma$ limit. 
The orange squares are the results of AzTEC-3 and CRLE from left to right 
(\citealt{2010ApJ...720L.131R}; \citealt{2014ApJ...796...84R}; 
\citealt{2018ApJ...861...43P}; see also, \citealt{2019ApJ...882..168P}). 
The black filled downward triangles are the results of SFGs with $M_{\rm star} > 3 \times 10^{10} M_\odot$ 
at $z \sim 0$--$4$ using stacked dust SEDs 
(\citealt{2015A&A...573A.113B}). 
The black open downward triangles show the results of SFGs with $M_{\rm star} > 2 \times 10^{10} M_\odot$ 
at $z \sim 1$--$6$ based on sub-millimeter dust continuum measurements 
(\citealt{2016ApJ...820...83S}). 
The black open triangles are the results for low-$z$ galaxies at $z = 0.01$--$0.05$ 
(\citealt{2017ApJS..233...22S}). 
}
\label{fig:Mgas_SFR}
\end{center}
\end{figure*}

In Figure \ref{fig:Mgas_SFR},
we compare total SFRs and $M_{\rm gas}$ estimates of our $z=6$ luminous LBGs 
with dusty starbursts and other SFGs over a wide range of redshifts 
from $z \sim 0$ to $z \sim 6$ 
(\citealt{2015A&A...573A.113B}; \citealt{2016ApJ...820...83S}; 
\citealt{2017ApJS..233...22S})
including HZ10, LBG-1, AzTEC-3, and CRLE 
(\citealt{2010ApJ...720L.131R}; \citealt{2014ApJ...796...84R}; 
\citealt{2018ApJ...861...43P}; see also, \citealt{2019ApJ...882..168P}). 
This figure should be interpreted as an integrated KS relation in a more direct sense 
than the $L_{\rm IR}$ vs. $L'_{\rm CO}$ plot presented as Figure \ref{fig:CO65_LIR} 
(Section \ref{subsec:results_dust}).  
Following \cite{1998ApJ...498..541K}, 
we adopt the molecular gas mass as a proxy for the total gas mass 
for high SFR sources including our $z=6$ luminous LBGs, 
because such sources in the local Universe 
show that the disks are molecular dominated 
(\citealt{1996ARA&A..34..749S}; see also, \citealt{2021ApJ...908...61K}).  
Because we only obtain the upper limit of SFR$_{\rm IR}$ for J0235--0532, 
we present the sum of SFR$_{\rm UV}$ and the $2\sigma$ upper limit of SFR$_{\rm IR}$ 
in the case of $T_{\rm dust} = 50$ K
as its total SFR, 
and consider the higher dust temperature up to $T_{\rm dust} = 80$ K
as well as the minimum SFR case of the SFR$_{\rm UV}$ alone with no SFR$_{\rm IR}$ 
in the relatively large error bars as systematic uncertainties. 
We find that 
our CO-based results for J0235--0532 are in broad agreement with 
SFGs at various redshifts with similar $M_{\rm gas}$ 
including HZ10 and LBG-1. 
J1211--0118 and J0217--0208 are consistent with the previous results 
with similar SFRs, 
although their $M_{\rm gas}$ values are upper limits.

\subsection{Kennicutt-Schmidt Relation} \label{subsec:SKrelation}

Although the currently available data for our $z=6$ luminous LBGs 
do not resolve their internal structures in detail, 
we estimate their sizes to 
calculate their global SFR surface densities and gas surface densities 
for comparisons with the KS relation 
for the average surface densities of SFGs at the local Universe.

The star-forming region sizes of our $z=6$ luminous LBGs 
for calculating global SFR surface densities 
are measured with the HSC $z$-band images, 
which trace the rest UV continuum emission. 
We fit S\'ersic profiles (\citealt{1968adga.book.....S}) 
to the observed surface brightness distributions  
by using GALFIT ver.~3.0.5 
(\citealt{2002AJ....124..266P}; \citealt{2010AJ....139.2097P}),\footnote{\url{https://users.obs.carnegiescience.edu/peng/work/galfit/galfit.html}}
which convolves a galaxy model profile 
with a point-spread function (PSF) profile 
and optimizes the fitting parameters 
based on the Levenberg-Marquardt algorithm 
for $\chi^2$ minimization (e.g., \citealt{1992nrfa.book.....P}). 
We use a PSF image for the position of each of our $z=6$ luminous LBGs 
downloaded from the PSF picker website of the HSC survey.\footnote{\url{https://hsc-release.mtk.nao.ac.jp/psf/pdr2/}}
The output parameters include 
the centroid coordinates of the objects, 
their total magnitude, 
the half-light radius along the semimajor axis, 
the axis ratio, 
and the position angle.  
The S\'ersic index $n$ is fixed at $1.0$.\footnote{We confirm that 
the obtained sizes show little difference if we use $n=1.5$.}
We calculate the circularized half-light radius, 
$r_{\rm e} = \sqrt{q} \, r_{\rm e, maj} $, 
where $q$ is the axis ratio 
and 
$r_{\rm e, maj}$ is the half-light radius along the semimajor axis, 
because it is widely used in size measurements in previous high-$z$ galaxy studies   
(e.g., \citealt{2012ApJ...746..162N}; \citealt{2012ApJ...756L..12M}; \citealt{2013ApJ...777..155O}; 
\citealt{2015ApJS..219...15S}; \citealt{2018ApJ...855....4K}). 
The obtained $r_{\rm e}$ values are presented in Table \ref{tab:targets}.

Note that, for J0235--0532, the output axis ratio 
obtained with GALFIT is enclosed in between star symbols, 
indicating that a numerical convergence issue may have occurred 
in the fitting for this particular source 
(for details, see Section 10 of the GALFIT user's manual). 
As an alternative method, 
for J0235--0532,
we use SExtractor ver. 2.8.6 
(\citealt{1996A&AS..117..393B})\footnote{\url{https://www.astromatic.net/software/sextractor/}}
to calculate the observed half-light radius, $r_{\rm e}^{\rm (obs)}$, 
by using circular apertures that contain a half of the light from a galaxy, 
and correct it for the PSF broadening 
according to 
\begin{equation}
r_{\rm e}
	= \sqrt{ r_{\rm e}^{\rm (obs) \, 2} - r_{\rm PSF}^2}, 
\end{equation}
where $r_{\rm PSF}$ is the half-light radius of the PSF image 
(e.g., \citealt{2010ApJ...709L..21O}; \citealt{2020AJ....160..154H}; 
\citealt{2021MNRAS.502..662B}; \citealt{2022ApJ...927..236R}). 
For this PSF broadening correction, 
we use the PSF images downloaded from the PSF picker website. 
The obtained $r_{\rm e}$ value for J0235--0532 is also presented in Table \ref{tab:targets}. 
We also measure the half-light radii of J1211--0118 and J0217--0208 with SExtractor 
and confirm that the results are consistent with those obtained with GALFIT.

With the obtained $r_{\rm e}$ values, 
we define SFR surface density, $\Sigma_{\rm SFR}$, 
as the average SFR in a circular region whose half-light radius is $r_{\rm e}$ 
(Equation \ref{eq:Sigma_SFR}). 
The obtained $\Sigma_{\rm SFR}$ values are listed in Table \ref{tab:observational_results}.

Because the resolution of our ALMA data is not high enough 
to estimate the sizes of CO-emitting regions,
we calculate their gas surface densities 
by assuming that the sizes of CO-emitting regions are comparable to those of star-forming regions. 
In fact, \cite{2013ApJ...768...74T} have reported that 
molecular gas and UV/optical light distributions 
of $z \sim 1$--$2$ SFGs show comparable sizes, 
in agreement with similar findings in $z \sim 0$ SFGs 
(e.g., \citealt{2001ApJ...561..218R}; \citealt{2008AJ....136.2782L}). 
With the $r_{\rm e}$ values obtained above, 
we define the gas surface density, $\Sigma_{\rm gas}$, as  
in a similar way to $\Sigma_{\rm SFR}$ 
(Equation \ref{eq:Sigma_gas}).\footnote{Note that 
our assumption that these sizes are comparable 
may cause a systematic uncertainty, 
because some previous results indicate that CO-emitting 
region sizes are not comparable to those in the rest UV/optical, 
as described in the last paragraph of this section 
and discussed more quantitatively in Section \ref{subsec:systematic_uncertainties}.  
} 
The obtained $\Sigma_{\rm gas}$ values are also presented 
in Table \ref{tab:observational_results}.

\begin{figure*}[h]
\begin{center}
   \includegraphics[width=0.9\textwidth]{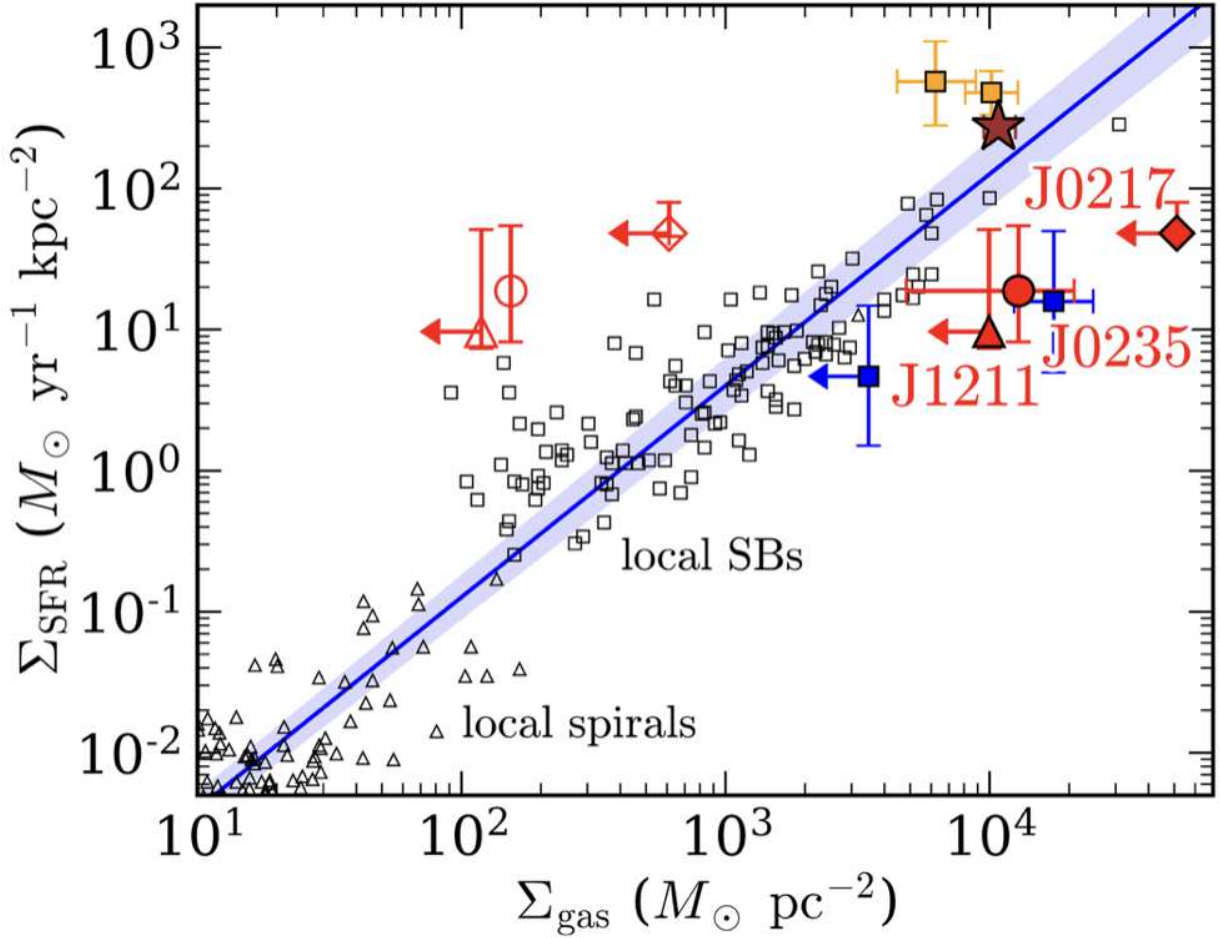}
\caption
{$\Sigma_{\rm SFR}$ vs. $\Sigma_{\rm gas}$. 
The red filled circle is 
our ALMA result for a luminous LBG at $z=6$ with $\simeq 4 \sigma$ CO(6--5) detection, 
J0235--0532, in the case of  $T_{\rm dust} = 50$ K 
and SFR$_{\rm tot}$ $=$ SFR$_{\rm UV}$ $+$ SFR$_{\rm IR, 2 \sigma}$. 
The error bar along the $y$-axis considers 
the higher dust temperature case of $T_{\rm dust} = 80$ K 
and the minimum SFR case of SFR$_{\rm tot}$ $=$ SFR$_{\rm UV}$. 
The red filled triangle and diamond are also our ALMA results for 
the other luminous LBGs at $z=6$, J1211--0118 and J0217--0208, respectively, 
which shows no significant CO(6--5) detection. 
The red arrows correspond to the $3\sigma$ limits. 
We also present the results for our $z=6$ luminous LBGs 
adopting the CO SLED and $\alpha_{\rm CO}$ for Althaea for their total gas mass estimates, 
and the previously obtained [{\sc Cii}] sizes as their gas sizes 
(red open circle: J0235--0532; 
red open triangle: J1211--0118; 
red open diamond: J0217--0208; 
for details, see Section \ref{subsec:systematic_uncertainties}).
The blue squares represent the results of LBG-1 and HZ10 from left to right (\citealt{2019ApJ...882..168P}). 
Note that the data point of HZ10 is shifted by $+0.1$ dex along the $x$-axis for visibility. 
The blue arrow represents the $3\sigma$ limit. 
The orange squares are the results of AzTEC-3 and CRLE from left to right 
(\citealt{2010ApJ...720L.131R}; \citealt{2014ApJ...796...84R}; 
\citealt{2018ApJ...861...43P}; see also, \citealt{2019ApJ...882..168P}). 
The brown star denotes 
the average of the four data points of J0235--0532, HZ10, AzTEC-3, and CRLE 
as the average KS relation at $z=5$--$6$, 
although the number of high-$z$ sources 
whose $\Sigma_{\rm SFR}$ and $\Sigma_{\rm gas}$ are estimated is limited.
The black open triangles and squares denote 
local spiral galaxies and starbursts compiled 
by \cite{2019ApJ...872...16D} and \cite{2021ApJ...908...61K}, respectively. 
The blue solid line corresponds to the KS relation, 
$\log \Sigma_{\rm SFR} = (1.50 \pm 0.02) \log \Sigma_{\rm gas} - 3.87 \pm 0.04$ 
(\citealt{2021ApJ...908...61K}) 
and the blue shaded region represents the $\Sigma_{\rm SFR}$ values 
that can be obtained when the two parameters of the KS relation change 
within the $2\sigma$ uncertainties. 
Note that the $\Sigma_{\rm SFR}$ values 
in \cite{2019ApJ...872...16D} and \cite{2021ApJ...908...61K} 
are corrected by a factor of $\alpha_{\rm KC}$ (Section \ref{sec:introduction}) 
to consider the IMF difference (Table \ref{tab:adopted_IMFs}). 
}
\label{fig:Sigma_gas_Sigma_SFR}
\end{center}
\end{figure*}

Figure \ref{fig:Sigma_gas_Sigma_SFR} plots 
$\Sigma_{\rm SFR}$ of our $z=6$ luminous LBGs 
as a function of $\Sigma_{\rm gas}$. 
For comparison, 
we also present 
normal spiral (disk) galaxies and starbursts at the local Universe 
with the best-fit relation 
between their $\Sigma_{\rm SFR}$ and $\Sigma_{\rm gas}$ 
(the KS relation; \citealt{2019ApJ...872...16D}; \citealt{2021ApJ...908...61K}; 
see also \citealt{1998ApJ...498..541K}),\footnote{Because the Kroupa IMF 
is adopted in \cite{2019ApJ...872...16D} and \cite{2021ApJ...908...61K} 
as listed in Table \ref{tab:adopted_IMFs},  
their $\Sigma_{\rm SFR}$ values are corrected by a factor of $\alpha_{\rm KC}$ 
(Section \ref{sec:introduction}).
}
as well as the $z=5.3$--$5.7$ sources, 
HZ10, LBG-1, AzTEC-3, and CRLE 
(\citealt{2010ApJ...720L.131R}; \citealt{2014ApJ...796...84R}; 
\citealt{2018ApJ...861...43P}; see also, \citealt{2019ApJ...882..168P}). 
We find that 
J0235--0532 and HZ10 are almost at the same position in this plane, 
located below the local KS relation, 
suggesting that 
J0235--0532 and HZ10 have very high gas surface densities 
with relatively low star formation efficiencies. 
We also find that 
J1211--0118 and J0217--0208 are consistent with the local KS relation, 
although their obtained $\Sigma_{\rm gas}$ values 
are upper limits.
Note that in Figure \ref{fig:Mgas_SFR}, 
their data points are consistent with the integrated KS relation, 
while in Figure \ref{fig:Sigma_gas_Sigma_SFR}, 
they are below the local KS relation. 
The reason for this is that 
our $z=6$ luminous LBGs have smaller gas sizes 
and/or larger star-forming region sizes 
than those of local starbursts.
In Figure \ref{fig:Sigma_gas_Sigma_SFR}, we confirm that 
the dusty starbursts at comparable redshifts, AzTEC-3 and CRLE, 
are located above the local KS relation.  
These results may indicate that 
the scatter of the KS relation is larger with increasing  
redshift at least at large $\Sigma_{\rm gas}$ of $\sim 10^4 \, M_\odot$ pc$^{-2}$, 
possibly suggesting that star formation in high-$z$ galaxies with high $\Sigma_{\rm gas}$ 
is diverse, ranging from bursty to slow ones. 
However, the number of high-$z$ data points is still limited; 
this needs to be examined by investigating more objects at high redshifts in the future. 
Averaging the four data points for the $z=5$--$6$ galaxies of 
J0235--0532, HZ10, AzTEC-3, and CRLE, 
we find that the $z=5$--$6$ KS relation at $\Sigma_{\rm gas} \sim 10^4 \, M_\odot$ pc$^{-2}$ 
on average is consistent with the KS relation at the local Universe. 
Again, the number of high-$z$ sources 
whose $\Sigma_{\rm SFR}$ and $\Sigma_{\rm gas}$ are estimated is limited yet. 
It would be interesting to compare the observational results for 
a larger sample of 
high-$z$ galaxies with those of theoretical studies in the future 
(e.g., \citealt{2019MNRAS.489....1F}; \citealt{2021A&A...651A.109D}).

Note that the $r_{\rm e}$ values measured in the rest UV continuum images 
are used in the calculations of both the SFR and gas surface densities 
for our $z=6$ luminous LBGs.  
If their CO sizes are significantly larger than the rest UV sizes, 
the currently presented $\Sigma_{\rm gas}$ values 
corresponds to the upper limits (e.g., \citealt{2020ApJ...899...37K}). 
More quantitative discussion about this point 
is presented in Section \ref{subsec:systematic_uncertainties}.
In order to obtain more accurate $\Sigma_{\rm gas}$ values 
with no such systematic uncertainties, 
high resolution deep CO observations are necessary.

\subsection{Systematic Uncertainties} \label{subsec:systematic_uncertainties}

In Sections \ref{subsec:total_gas_mass}--\ref{subsec:SKrelation}, 
we obtain the total gas mass estimates for our $z=6$ luminous LBGs 
based on our CO(6--5) observation results 
by carefully considering the uncertainties suggested from the previous observation results, 
and discuss the KS relation. 
However, we caution that 
the gas mass estimates have substantial systematic uncertainties.

One is the CO SLED uncertainty. 
In our discussion above, 
we adopt the previously observed CO SLED results for $z \sim 1$--$2$ SFGs 
and consider the significant amount of scatter 
seen in observations of individual objects 
(\citealt{2015A&A...577A..46D}). 
However, our $z=6$ luminous LBGs may be experiencing 
more bursty star formation with higher gas density 
and thus the CO SLED could be more excited. 
For example, previous CO observations for nearby starbursts have revealed that 
their integrated CO flux ratios are about 
$I_{\rm CO(6-5)} / I_{\rm CO(1-0)} \simeq 8$--$20$ 
(Figure 1 of \citealt{2015ApJ...802...81M}).\footnote{In their paper,  
the CO fluxes, $f_{\rm CO}$, have units of W m$^{-2}$; 
for comparison we use the following conversion 
obtained from Equation (\ref{eq:L_CO}), 
\begin{equation}
\dfrac{I_{\rm CO(6-5)}}{I_{\rm CO(1-0)}}
	= \dfrac{f_{\rm CO(6-5)}}{f_{\rm CO(1-0)}} \cdot \left( \dfrac{\nu_{\rm CO(6-5)}^{\rm (rest)}}{\nu_{\rm CO(1-0)}^{\rm (rest)}} \right)^{-1}. 
\end{equation}
Here we do not consider the AGNs and Seyfert galaxies presented in their Figure.} 
Based on the ALMA Spectroscopic Survey in the Hubble Ultra Deep Field (ASPECS), 
\cite{2020ApJ...902..109B} have shown that 
SFGs at $\llangle z \rrangle = 2.5$ have higher CO excitation than those at $\llangle z \rrangle =  1.2$ 
as well as the results of \cite{2015A&A...577A..46D}, 
suggesting the increased CO excitation at higher redshifts 
(e.g., Figure 7 of \citealt{2020ApJ...902..109B}). 
Theoretically, \cite{2018MNRAS.473..271V} have developed a semi-analytical model 
for GMCs where the CO lines are excited, 
and post-processed a state-of-the-art zoom-in cosmological simulation 
of a main-sequence galaxy at $z=6$, Althaea, 
with $M_{\rm star} \approx 10^{10} M_\odot$ and SFR $\approx 100 \, M_\odot$ yr$^{-1}$ 
(\citealt{2017MNRAS.471.4128P}), 
which is in line with the nature of the galaxies discussed in this paper, 
showing that the CO SLED of Althaea has a peak at around  
the upper state rotational quantum number of $J_{\rm up} \simeq 6$.  
Specifically, 
Althaea has CO luminosities of about 
$L'_{\rm CO(1-0)} \simeq 10^{9.2}$ K km s$^{-1}$ pc$^2$ 
and 
$L'_{\rm CO(6-5)} \simeq 10^{8.9}$ K km s$^{-1}$ pc$^2$, 
yielding the CO luminosity ratio of 
$L'_{\rm CO(1-0)} / L'_{\rm CO(6-5)} \simeq 2.0$.
In contrast, 
the CO luminosity ratio that we have adopted in our discussion above is 
$L'_{\rm CO(1-0)} / L'_{\rm CO(6-5)} = 6.2 \pm 3.5$ 
(Equation \ref{eq:LCO_ratio} in Section \ref{subsec:total_gas_mass}).
Although their theoretical result is just for one $z=6$ galaxy, 
their upcoming results with the SERRA simulation 
show that the physical mechanisms exciting the CO SLED 
(i.e., high density and high turbulence) are common in more than 100 high-$z$ galaxies 
(\textcolor{blue}{A. Pallottini in preparation}; 
see also, \citealt{2019MNRAS.487.1689P}). 
If the CO SLEDs of our $z=6$ luminous LBGs are similar to that of Althaea, 
the gas mass estimates become lower by a factor of about $1/3$.  
This should be examined by observing several CO emission lines 
with different excited states from high-$z$ SFGs.

Another systematic uncertainty comes from 
the CO-to-H$_2$ conversion factor, $\alpha_{\rm CO}$. 
In our discussion above, 
we adopt the fixed value of $\alpha_{\rm CO} = 4.5 \, M_\odot$ (K km s$^{-1}$ pc$^2$)$^{-1}$, 
which is consistent with the previous observational results such as 
for Milky Way, $z \sim 1$--$2$ SFGs, and HZ10 at $z=5.7$ 
(\citealt{2013ARA&A..51..207B}; \citealt{2010ApJ...714L.118D}; 
\citealt{2013ARA&A..51..105C}; \citealt{2019ApJ...882..168P}). 
However, it is known that 
the CO-to-H$_2$ conversion factor becomes smaller in galaxies with more active star formation. 
In fact, LIRGs and ULIRGs show low CO-to-H$_2$ conversion factors of 
$\alpha_{\rm CO} \simeq 0.8 \, M_\odot$ (K km s$^{-1}$ pc$^2$)$^{-1}$ 
(\citealt{1998ApJ...507..615D}), 
which is often adopted in previous studies of high-$z$ dusty starbursts 
(e.g., \citealt{2005MNRAS.359.1165G}; \citealt{2010ApJ...720L.131R}; \citealt{2016MNRAS.457.4406A}; 
see also \citealt{2009ApJ...697L..33W}).  
Our $z=6$ luminous LBGs may also have small $\alpha_{\rm CO}$ values compared to the adopted one. 
From a theoretical point of view, \cite{2018MNRAS.473..271V} have shown that 
the simulated $z=6$ galaxy Althaea has a small $\alpha_{\rm CO}$ value of 
$\alpha_{\rm CO} = 1.5 \, M_\odot$ (K km s$^{-1}$ pc$^2$)$^{-1}$ 
(see their Figure 12; 
see also \citealt{2012MNRAS.421.3127N}). 
If our $z=6$ luminous LBGs have small $\alpha_{\rm CO}$ values comparable to Althaea, 
the gas mass estimates based on the CO(6--5) results 
are further reduced by a factor of $1/3$. 
Interestingly, if we adopt the CO SLED and $\alpha_{\rm CO}$ for Althaea, 
the obtained gas mass estimate for J0235--0532 from CO(6--5) is consistent with 
that obtained from the [{\sc Cii}] luminosity 
(Section \ref{subsec:total_gas_mass}).

Furthermore, 
in the calculations of the gas surface densities of our $z=6$ luminous LBGs, 
we assume that 
the sizes of CO-emitting regions are comparable to those of star-forming regions, 
which is also a source of systematic uncertainties. 
As mentioned in Section \ref{subsec:SKrelation}, 
some previous observational studies for $z\sim 0$--$2$ SFGs have shown that this is the case 
(\citealt{2001ApJ...561..218R}; \citealt{2008AJ....136.2782L}; \citealt{2013ApJ...768...74T}), 
while some other studies have shown that the gas sizes are larger. 
For our $z=6$ luminous LBGs, 
\cite{2020MNRAS.499.5136C} have reported their [{\sc Cii}] sizes (FWHMs along the major axis) in their Table A1.  
If the [{\sc Cii}] sizes are comparable to those of CO-emitting regions 
and the CO luminosities are comparable to the current measurements, 
the gas sizes become larger by about a factor of $3$, 
and thus $\Sigma_{\rm gas}$ become smaller by about a factor of $1/10$. 
This can be tested by deep observations for low-$J$ CO emission, 
which better traces the molecular gas distribution and the total gas mass.

If we adopt the CO SLED and $\alpha_{\rm CO}$ for Althaea for total gas mass estimates, 
and use the previously obtained [{\sc Cii}] sizes as their gas sizes, 
then the estimated gas surface densities of our $z=6$ luminous LBGs 
become smaller by about two orders of magnitude than presented in Figure \ref{fig:Sigma_gas_Sigma_SFR}. 
Figure \ref{fig:Sigma_gas_Sigma_SFR}
adds this possibility on the $\Sigma_{\rm SFR}$-$\Sigma_{\rm gas}$ plot. 
In this case, our $z=6$ luminous LBGs are located above the KS relation, 
which means that they are experiencing bursty star formation.
Because their [{\sc Oiii}]/[{\sc Cii}] ratios are relatively high 
compared to local galaxies with similar total SFRs
(Section \ref{subsec:reason_for_luminous_CO}), 
this interpretation may be physically more reasonable 
(\citealt{2021MNRAS.505.5543V}; see also \citealt{2019MNRAS.489....1F}). 
This issue is expected to be clarified by future follow-up observations.

\section{Summary} \label{sec:summary}

In this study, 
we have presented our ALMA observation results 
for the CO(6--5) and dust continuum emission 
from the three luminous  LBGs with $-22.8 < M_{\rm UV} < -23.3$ mag at $z_{\rm spec} = 6.0293$--$6.2037$ 
identified in the Subaru/HSC survey. 
Their [{\sc Oiii}]$88\mu$m and [{\sc Cii}]$158\mu$m emission lines 
have been detected in the previous work (\citealt{2020ApJ...896...93H}). 
Our main results are as follows.

\begin{enumerate}

\item Out of the three $z=6$ luminous LBGs, 
marginal detection 
of the CO(6--5) emission 
at the $\simeq 4\sigma$ significance level 
at the expected frequency 
from the previously detected [{\sc Oiii}]$88\mu$m and [{\sc Cii}]$158\mu$m lines.

\item No dust continuum emission at $\lambda_{\rm rest} \simeq 430 \, \mu$m 
is significantly detected for our $z=6$ luminous LBGs. 
By combining the obtained upper limits 
with the previous results at shorter wavelengths of 
$\lambda_{\rm rest} \simeq 90$--$160 \, \mu$m, 
we have updated the dust continuum SED fitting analyses, 
and confirmed that our obtained constraints on $L_{\rm IR}$ and $T_{\rm dust}$ 
are consistent with the previous results of \cite{2020ApJ...896...93H}.

\item We have compared the CO(6--5) and IR luminosities 
of our $z=6$ luminous LBGs with those of other sources 
over a wide range of redshifts in the literature 
by taking into account the CMB effect and the $T_{\rm dust}$ uncertainty. 
We have found that our $z=6$ luminous LBGs 
are consistent with the previous results 
owing to the relatively large uncertainties.

\item By comparing the $L_{\rm CO} / L_{\rm [CII]}$ and $L_{\rm [CII]} / L_{\rm IR}$ ratios 
of our $z=6$ luminous LBGs with previous observation and PDR model calculation results, 
we have found that J0235--0532 has a relatively high $n_{\rm H}$ value 
comparable to those of low-$z$ LIRGs and ULIRGs, 
as well as those of quasars and Galactic star-forming regions 
with high $n_{\rm H}$ and $U_{\rm UV}$ values. 
We have also found that 
J1211--0118 and J0217--0208 
have lower $n_{\rm H}$ values 
consistent with local SFGs and Galactic star-forming regions 
with relatively low $n_{\rm H}$ values.

\item By carefully taking into account the systematic uncertainties in the CO SLED, 
$M_{\rm gas}$ constraints for our $z=6$ luminous LBGs have been obtained  
based on our CO(6--5) observation results.  
We have found that 
J0235--0532 is in broad agreement with SFGs at various redshifts 
with similar $M_{\rm gas}$ in the literature, 
including the $z=5.3$--$5.7$ SFGs of HZ10 and LBG-1. 
We have also found that 
the $M_{\rm gas}$ upper limits for J1211--0118 and J0217--0208 
are consistent with the previous results with comparable SFRs.

\item We have calculated the global SFR and gas surface densities 
of our $z=6$ luminous LBGs 
based on the total SFR and $M_{\rm gas}$ constraints 
with the star-forming region sizes measured in the HSC images 
capturing the rest UV continuum emission. 
We have found that 
J0235--0532 is almost at the same position as HZ10 
on the $\Sigma_{\rm SFR}$-$\Sigma_{\rm gas}$ plane, 
located slightly below the local KS relation, 
indicating that J0235--0532 and HZ10 have high gas surface densities 
with relatively low star formation efficiencies. 
We have also found that 
the upper limits of $\Sigma_{\rm gas}$ for J1211--0118 and J0217--0208 
are consistent with the local KS relation. 
Because the dusty starbursts at similar redshifts, 
AzTEC-3 and CRLE,  
are located above the local KS relation, 
our results and the previous results may suggest that 
the scatter of the KS relation increases with increasing 
redshift at least at large $\Sigma_{\rm gas}$. 
In addition, the average  $z=5$--$6$ KS relation at $\Sigma_{\rm gas} \sim 10^4 \, M_\odot$ pc$^{-2}$
is in agreement with the local KS relation.
However, the number of high-$z$ sources 
whose $\Sigma_{\rm SFR}$ and $\Sigma_{\rm gas}$ have been estimated is still limited; 
the high-$z$ KS relation needs to be determined with better accuracy to discuss the average and the scatter 
by investigating more objects at high redshifts in the future.

\item We caution that 
the obtained gas mass estimates for our $z=6$ luminous LBGs have substantial systematic uncertainties 
such as the CO SLED, the CO-to-H$_2$ conversion factor $\alpha_{\rm CO}$, and gas sizes. 
If we adopt the CO SLED and the $\alpha_{\rm CO}$ value
suggested by the state-of-the-art zoom-in cosmological simulation 
and the gas sizes measured with [{\sc Cii}] emission, 
the gas surface densities estimated for our $z=6$ luminous LBGs 
can become larger by about two orders of magnitude, 
which opens up two conflicting possibilities regarding the location below or above the KS relation. 
This situation should be clarified by pursuing further CO observations for high-$z$ SFGs.
 
\end{enumerate}

\section*{Acknowledgements}

We acknowledge the constructive comments 
and helpful suggestions from the anonymous referee that helped us to improve the manuscript. 
We appreciate the support of the staff at the ALMA Regional Center, 
especially Kazuya Saigo, for giving us helpful advice on analyzing the ALMA data. 
We are grateful to the staff of the IRAM facilities, 
especially Michael Bremer and Melanie Krips, 
for helping us to reduce the NOEMA data. 
We also thank Daizhong Liu for sharing their data with us, 
and Marc Pound and Mark Wolfire for their helpful advice on using the PDRT calculation results. 

This paper made use of the following ALMA data: 
ADS/JAO.ALMA\#2019.1.00156.S and ADS/JAO.ALMA\#2017.1.00508.S. 
ALMA is a partnership of ESO (representing its member states), NSF (USA) and NINS (Japan), 
together with NRC (Canada), MOST and ASIAA (Taiwan), and KASI (Republic of Korea), 
in cooperation with the Republic of Chile. The Joint ALMA Observatory is operated by ESO, AUI/NRAO and NAOJ.

This work is based on observations 
carried out under project numbers W18FB and S19DK with the IRAM NOEMA Interferometer. 
IRAM is supported by INSU/CNRS (France), MPG (Germany) and IGN (Spain).


The Hyper Suprime-Cam (HSC) collaboration includes the astronomical communities of Japan and Taiwan, and Princeton University.  The HSC instrumentation and software were developed by the National Astronomical Observatory of Japan (NAOJ), the Kavli Institute for the Physics and Mathematics of the Universe (Kavli IPMU), the University of Tokyo, the High Energy Accelerator Research Organization (KEK), the Academia Sinica Institute for Astronomy and Astrophysics in Taiwan (ASIAA), and Princeton University.  Funding was contributed by the FIRST program from the Japanese Cabinet Office, the Ministry of Education, Culture, Sports, Science and Technology (MEXT), the Japan Society for the Promotion of Science (JSPS), Japan Science and Technology Agency  (JST), the Toray Science  Foundation, NAOJ, Kavli IPMU, KEK, ASIAA, and Princeton University.

This paper makes use of software developed for the Large Synoptic Survey Telescope. We thank the LSST Project for making their code available as free software at \url{http://dm.lsst.org}.

This paper is based in part on data collected at the Subaru Telescope and retrieved from the HSC data archive system, which is operated by Subaru Telescope and Astronomy Data Center (ADC) at NAOJ. Data analysis was in part carried out with the cooperation of Center for Computational Astrophysics (CfCA), NAOJ.

The Pan-STARRS1 Surveys (PS1) and the PS1 public science archive have been made possible through contributions by the Institute for Astronomy, the University of Hawaii, the Pan-STARRS Project Office, the Max Planck Society and its participating institutes, the Max Planck Institute for Astronomy, Heidelberg, and the Max Planck Institute for Extraterrestrial Physics, Garching, The Johns Hopkins University, Durham University, the University of Edinburgh, the Queen's University Belfast, the Harvard-Smithsonian Center for Astrophysics, the Las Cumbres Observatory Global Telescope Network Incorporated, the National Central University of Taiwan, the Space Telescope Science Institute, the National Aeronautics and Space Administration under grant No. NNX08AR22G issued through the Planetary Science Division of the NASA Science Mission Directorate, the National Science Foundation grant No. AST-1238877, the University of Maryland, Eotvos Lorand University (ELTE), the Los Alamos National Laboratory, and the Gordon and Betty Moore Foundation.


This work has made use of data 
from the European Space Agency (ESA) mission
{\it Gaia} (\url{https://www.cosmos.esa.int/gaia}), 
processed by the {\it Gaia} Data Processing and Analysis Consortium 
(DPAC, \url{https://www.cosmos.esa.int/web/gaia/dpac/consortium}). 
Funding for the DPAC has been provided by national institutions, 
in particular the institutions participating in the {\it Gaia} Multilateral Agreement.

This work was partially performed using the computer facilities of
the Institute for Cosmic Ray Research, The University of Tokyo. 
This work was supported 
by the World Premier International
Research Center Initiative (WPI Initiative), MEXT, Japan, 
as well as 
KAKENHI Grant Numbers 
15K17602, 
15H02064, 
17H01110, 
17H01114, 
19K14752, 
20H00180, 
and 21H04467 
through the Japan Society for the Promotion of Science (JSPS). 
This work was partially supported by the joint research program of the
Institute for Cosmic Ray Research (ICRR), University of Tokyo. 
AF, AP, and LV acknowledge support from the ERC Advanced Grant INTERSTELLAR H2020/740120. 
AF acknowledges generous support from the Carl Friedrich von Siemens-Forschungspreis 
der Alexander von Humboldt-Stiftung Research Award. 
AKI and YS are supported by NAOJ ALMA Scientific Research Grant Code 2020-16B. 
TH was supported by Leading Initiative for Excellent Young Researchers, 
MEXT, Japan (HJH02007) and KAKENHI (20K22358).

\software{IRAF (\citealt{1986SPIE..627..733T,1993ASPC...52..173T}),\footnote{IRAF is distributed by the National Optical Astronomy Observatory, 
which is operated by the Association of Universities for Research in Astronomy (AURA) 
under a cooperative agreement with the National Science Foundation.} 
SAOImage DS9 \citep{2003ASPC..295..489J},
Numpy \citep{2020Natur.585..357H}, 
Matplotlib \citep{2007CSE.....9...90H}, 
Scipy \citep{2020NatMe..17..261V}, 
Astropy \citep{2013A&A...558A..33A,2018AJ....156..123A},\footnote{\url{http://www.astropy.org}} 
and 
Ned Wright's Javascript Cosmology Calculator \citep{2006PASP..118.1711W},\footnote{\url{http://www.astro.ucla.edu/~wright/CosmoCalc.html}}
CASA (\citealt{2007ASPC..376..127M}),
GILDAS (\citealt{2000ASPC..217..299G}; \citealt{2005sf2a.conf..721P}; \citealt{2013ascl.soft05010G}),  
SExtractor (\citealt{1996A&AS..117..393B}), 
GALFIT (\citealt{2002AJ....124..266P}; \citealt{2010AJ....139.2097P}).
}

\bibliographystyle{aasjournal}
\bibliography{ref}

\appendix

\setcounter{table}{0}
\renewcommand{\thetable}{E.\arabic{table}}

\section{Standard Equations} \label{sec:standard_equations}

In this section, we present the standard equations used in this study for reference. 

In Section \ref{sec:targets}, 
we estimate SFRs for our $z=6$ luminous LBGs  
by using Equation (1) of \cite{1998ARA&A..36..189K}, 
\begin{equation}
{\rm SFR}_{\rm UV}
	=  1.4 \times 10^{-28} \alpha_{\rm SC} \, L_\nu, 
\end{equation}
where $L_\nu$ is the rest UV luminosity density in units of erg s$^{-1}$ Hz$^{-1}$ 
and Equation (4) of \cite{1998ARA&A..36..189K},\footnote{Note that 
this conversion does not take into account 
the contribution from old stars 
whose emission is absorbed by dust and reradiated in the IR spectral range, 
although it would not be significant for high-$z$ LBGs.  
In order to estimate SFRs from IR continuum luminosities 
by appropriately considering the contribution from old stars, 
one needs a more general recipe such as the one derived by \cite{2000PASJ...52..539I} 
(see also, \citealt{2003A&A...410...83H}).} 
\begin{equation}
{\rm SFR}_{\rm IR}
	=  4.5 \times 10^{-44} \alpha_{\rm SC} \, L_{\rm IR}, 
\end{equation}
where $L_{\rm IR}$ is the IR luminosity integrated over the wavelength range of $8$--$1000 \, \mu$m 
in units of erg s$^{-1}$. 
We multiply by $\alpha_{\rm SC}$ 
to convert from the Salpeter IMF to the Chabrier IMF.

In Section \ref{subsec:results_co}, 
from the integrated CO(6--5) emission line flux, 
we obtain the CO(6--5) luminosity in units of $L_\odot$ 
by using Equation (18) of \cite{2014PhR...541...45C},
\begin{equation}
L_{\rm CO}
	= 1.04 \times 10^{-3} I_{\rm CO} \dfrac{\nu^{\rm (rest)}_{\rm CO}}{1+z} D_{\rm L}^2(z), 
\label{eq:L_CO}
\end{equation}
where 
$I_{\rm CO}$ is the integrated CO flux in units of Jy km s$^{-1}$
and 
$D_{\rm L}(z)$ is the luminosity distance in Mpc. 
We also calculate 
the CO(6--5) luminosity in units of K km s$^{-1}$ pc$^2$ defined as 
Equation (19) of \cite{2014PhR...541...45C}, 
\begin{equation}
L'_{\rm CO}
	= 3.25 \times 10^{7} I_{\rm CO} \dfrac{D_{\rm L}^2(z)}{(1+z)^3 \nu^{\rm (obs) 2}_{\rm CO}},
\end{equation}
where 
$\nu^{\rm (obs)}_{\rm CO} = \nu^{\rm (rest)}_{\rm CO}/(1+z)$.

In Section \ref{subsec:results_dust}, 
the intrinsic dust continuum flux densities of a modified blackbody SED 
at a given observed frequency $\nu_{\rm obs}$ are calculated from 
(e.g., \citealt{1999ApJ...517L..19O}; \citealt{2014ApJ...795....5O})
\begin{equation}
f_\nu^{\rm (int)}
	= \dfrac{(1+z) L_{\rm IR}}{4 \pi D_{\rm L}^2(z)} \dfrac{\nu_0^{\beta_{\rm d}} B(\nu_0, \, T_{\rm dust})}{\int \nu^{\beta_{\rm d}} B(\nu, \, T_{\rm dust}) d\nu}, 
\label{eq:modified_blackbody_SED}
\end{equation}
where $B(\nu, \, T)$ is the Planck function, 
and $\nu_0 = \nu_{\rm obs} (1+z)$. 
We assume a spectral index of $\beta_{\rm d} = 1.5$, 
which is consistent with local measurements for SFGs 
(e.g., \citealt{2001MNRAS.327..697D}; \citealt{2010A&A...518L..89G}; \citealt{2012MNRAS.425.3094C}) 
and often adopted in previous high-$z$ studies 
(e.g., \citealt{2014PhR...541...45C}; \citealt{2020A&A...643A..30F}; \citealt{2020ApJ...896...93H}; 
see also \citealt{2021ApJ...923....5S}; \citealt{2022ApJ...928...31S}).
\cite{2020ApJ...896...93H} have confirmed that 
this assumption does not significantly affect the fitting results of the other parameters for our targets.

In Section \ref{subsec:SKrelation}, 
we define SFR surface density, $\Sigma_{\rm SFR}$, 
in units of $M_\odot$ yr$^{-1}$ kpc$^{-2}$ 
as the average SFR in a circular region whose half-light radius is $r_{\rm e}$, 
\begin{equation}
\Sigma_{\rm SFR} 
= \dfrac{\rm SFR}{2 \pi r_{\rm e}^2}.
\label{eq:Sigma_SFR}
\end{equation}
The multiplicative factor of $1/2$ is applied, 
because the SFR is estimated from the total luminosity 
while the area is calculated with the half-light radius 
(e.g., \citealt{2008ApJ...673..686H}; \citealt{2013ApJ...768...74T}; \citealt{2016ApJ...833...70D}). 
In a similar way, we define the gas surface density as 
\begin{equation}
\Sigma_{\rm gas} 
= \dfrac{ M_{\rm gas} }{2 \pi r_{\rm e}^2}.  
\label{eq:Sigma_gas}
\end{equation}

\section{NOEMA Observations} \label{sec:noema_observations}

In addition to the ALMA observations described in Section \ref{sec:observations}, 
two of our targets, 
J1211--0118 and J0217--0208, 
were also observed with NOEMA 
using 9--10 antennas 
between 2019 January 20 and 2019 August 15 
(Proposal IDs: W18FB and S19DK; PI: Y. Ono). 
The antenna configurations were A and D, 
i.e., the most extended and the compact configurations, respectively. 
We used the NOEMA receiver 1 to observe the CO(6--5) emission 
as well as the dust continuum emission from the two LBGs. 
The total observing times were 
5.5 hr for J1211--0118 
and 
13.9 hr for J0217--0208. 
All the NOEMA data are reduced using the GILDAS software.\footnote{\url{https://www.iram.fr/IRAMFR/GILDAS/}} 
The $1\sigma$ flux density levels for the continuum images 
are $13.8 \, \mu$Jy beam$^{-1}$ for J1211--0118 
and $13.9 \, \mu$Jy beam$^{-1}$ for J0217--0208. 
The NOEMA data show no significant detection of either dust continuum emission or CO emission, 
which is consistent with the ALMA results (Section \ref{subsec:results_dust}).

\section{Dust Continuum Emission Maps} \label{sec:dust_continuum_emission_maps}

\begin{figure*}[h]
\begin{center}
   \includegraphics[scale=0.45]{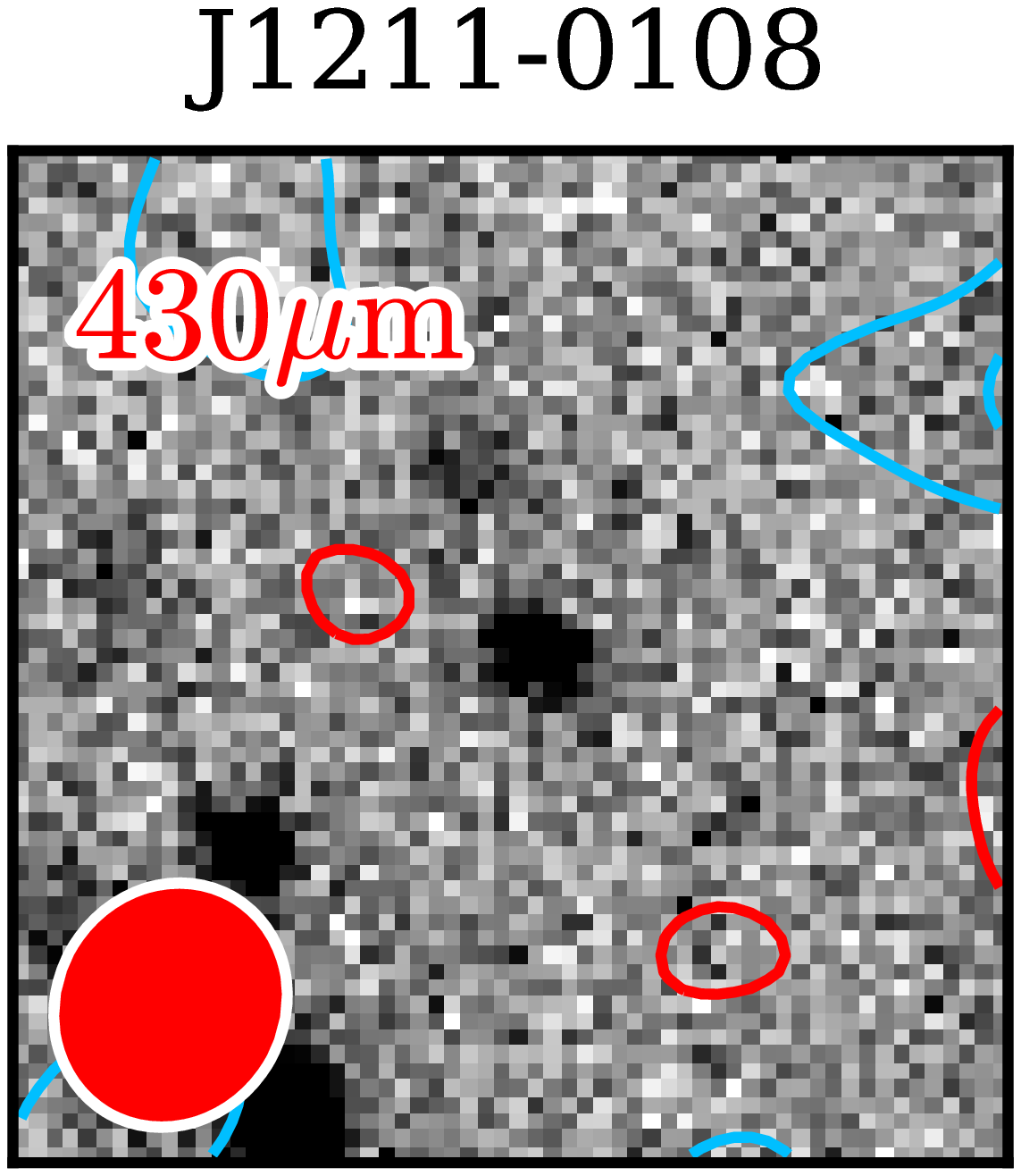}
   \includegraphics[scale=0.45]{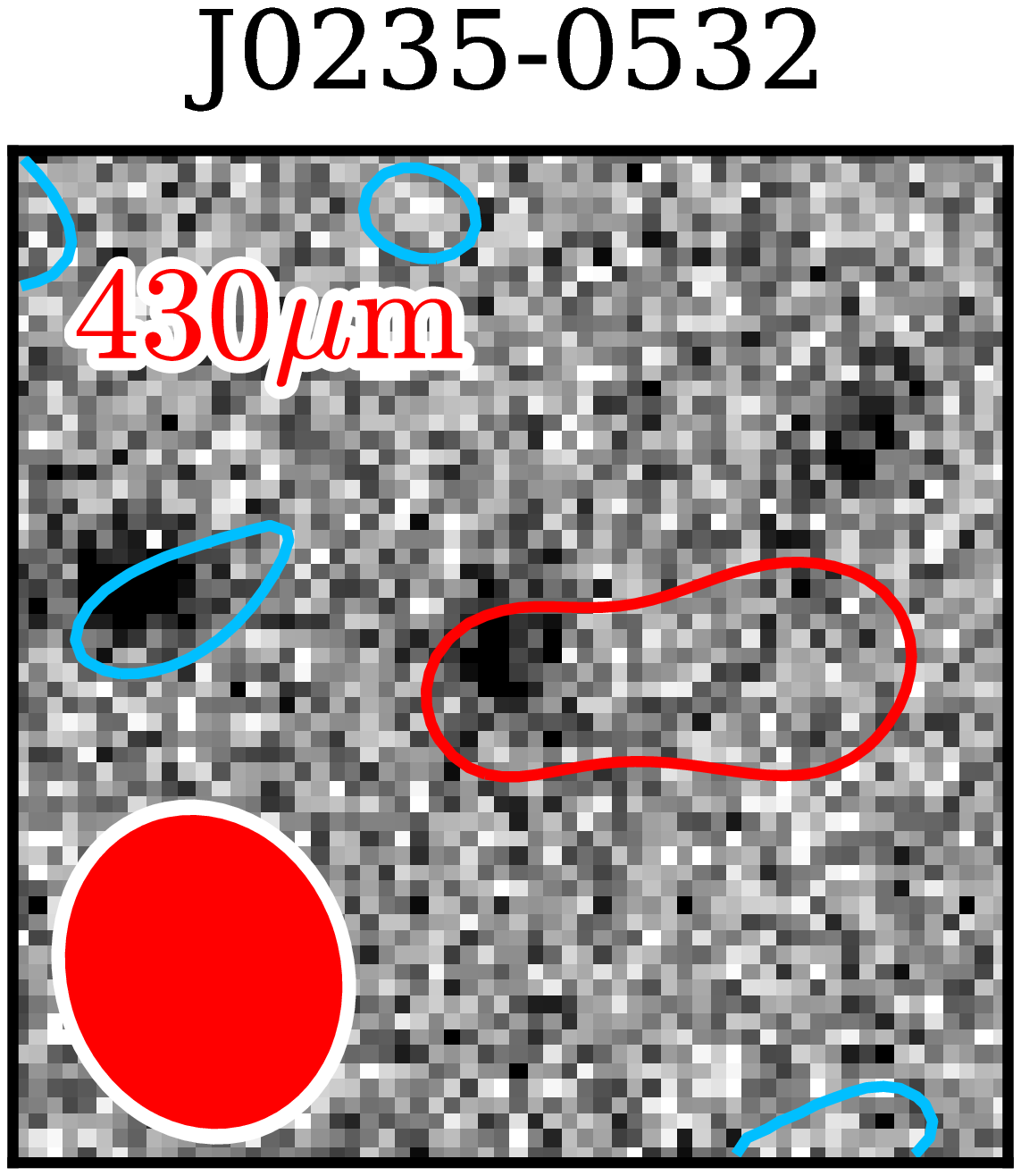}
   \includegraphics[scale=0.45]{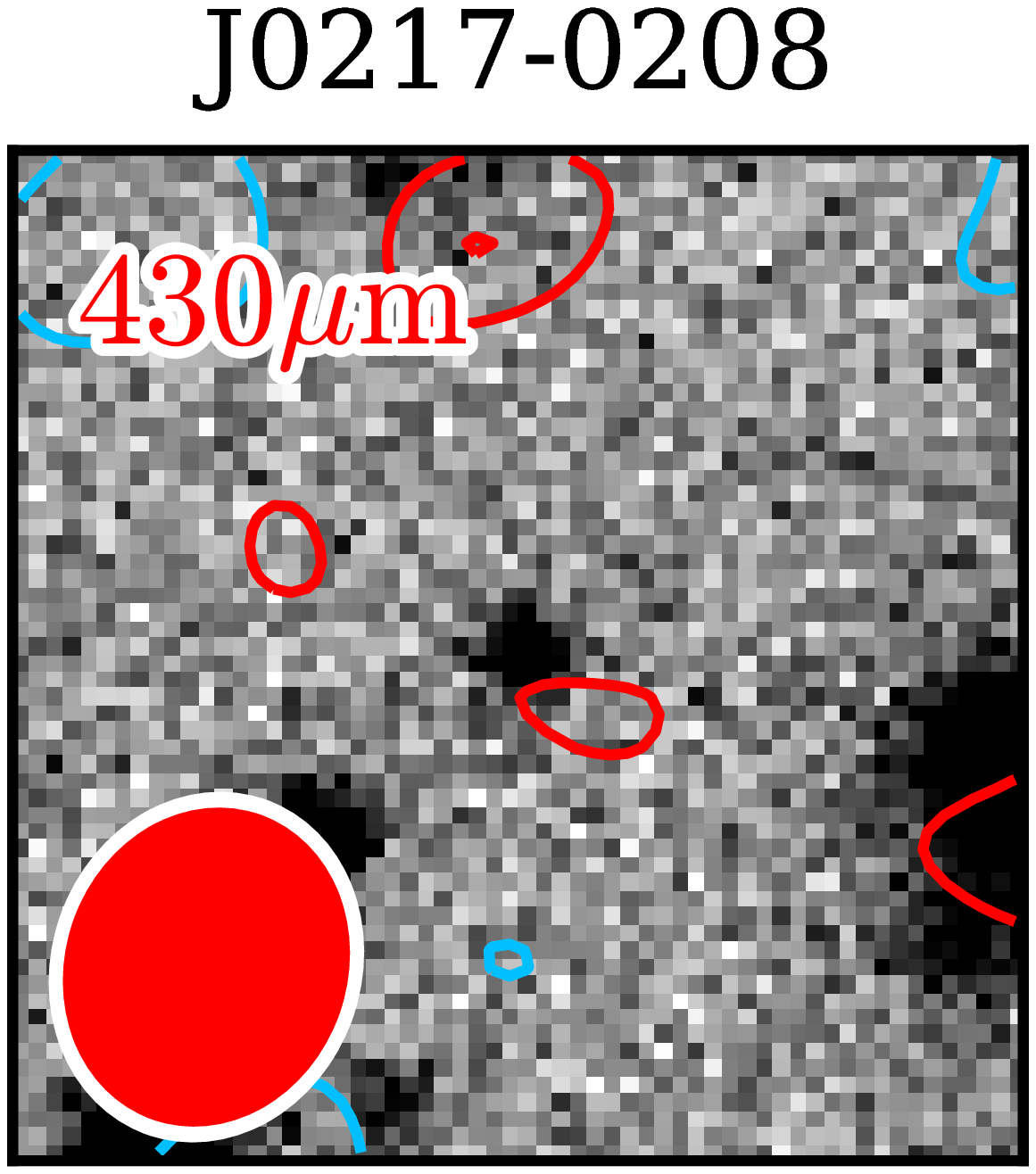}
\caption{
Dust continuum emission maps for our $z=6$ luminous LBGs, 
J1211--0118, J0235--0532, and J0217--0208 from left to right. 
The red contours are continuum emission at $\lambda_{\rm rest} \simeq 430 \, \mu$m 
drawn at $1\sigma$ intervals from $1.5\sigma$. 
Although the dust continuum of J0235 and J0217 may show a $\sim 2\sigma$ signal,
in this paper we conservatively use their $3\sigma$ upper limits. 
The blue lines represent negative contours from $-1.5\sigma$ at $1\sigma$ intervals. 
The red ellipses at the lower left corner denote the ALMA synthesized beams. 
The gray backgrounds are the Subaru HSC $z$-band images 
that capture the rest-UV continuum emission from our targets. 
The size of each image is $10'' \times 10''$. 
}
\label{fig:dust_continuum}
\end{center}
\end{figure*}

In Figure \ref{fig:dust_continuum}, 
we present the dust continuum emission maps 
at $\lambda_{\rm obs} \simeq 3$ mm ($\lambda_{\rm rest} \simeq 430 \, \mu$m) obtained with ALMA 
for our $z=6$ luminous LBGs.  
For J0235--0532, 
the $\pm 300$ km s$^{-1}$ range around the CO(6--5) line is removed.

\section{Extra Results Related to Gas Masses} \label{sec:extra_results}

In addition to the comparisons between gas masses and SFRs 
as well as the Kennicutt-Schmidt relation shown in Section \ref{sec:discussion}, 
in this section, we present the gas fractions and gas depletion timescales 
for comparisons with previous results, 
although the systematic uncertainties on these estimates are also not small 
as discussed in Section \ref{subsec:systematic_uncertainties}.

\subsection{Gas Fraction} \label{subsec:gas_fraction}

We constrain gas fractions for our $z=6$ luminous LBGs. 
The gas fraction is defined as  
\begin{equation}
f_{\rm gas} 
	= \dfrac{M_{\rm gas}}{M_{\rm gas} + M_{\rm star}}, 
\end{equation}
where $M_{\rm star}$ is the stellar mass. 
For our $z=6$ luminous LBGs, 
$M_{\rm star}$ can be roughly estimated from $M_{\rm UV}$ 
by using the relation between $M_{\rm star}$ and $M_{\rm UV}$ for SFGs at similar redshifts, 
e.g., Equation (2) of \cite{2015ApJS..219...15S}, 
\begin{equation}
\log M_{\rm star} = - 2.45 - 0.59 M_{\rm UV} + \log \beta_{\rm SC},  
\end{equation}
where $\beta_{\rm SC} = 1/1.64 \simeq 0.61$ is the factor to convert 
from $M_{\rm star}$ with the \cite{1955ApJ...121..161S} IMF 
to that with the \cite{2003PASP..115..763C} IMF 
(\citealt{2014ARA&A..52..415M}; see also Table \ref{tab:adopted_IMFs}). 
We present the obtained $f_{\rm gas}$ constraints in Table \ref{tab:observational_results}.
Note that our $f_{\rm gas}$ constraints do not include the systematic uncertainties 
in the stellar mass estimates from the UV luminosity, 
which is about $\pm 0.5$ dex 
due to differences in stellar population properties such as star formation history 
(\citealt{2015ApJS..219...15S}). 
For more robust discussion, 
deep rest-frame optical data that can probe the stellar continuum emission are required.

\begin{figure*}[h]
\begin{center}
   \includegraphics[width=0.9\textwidth]{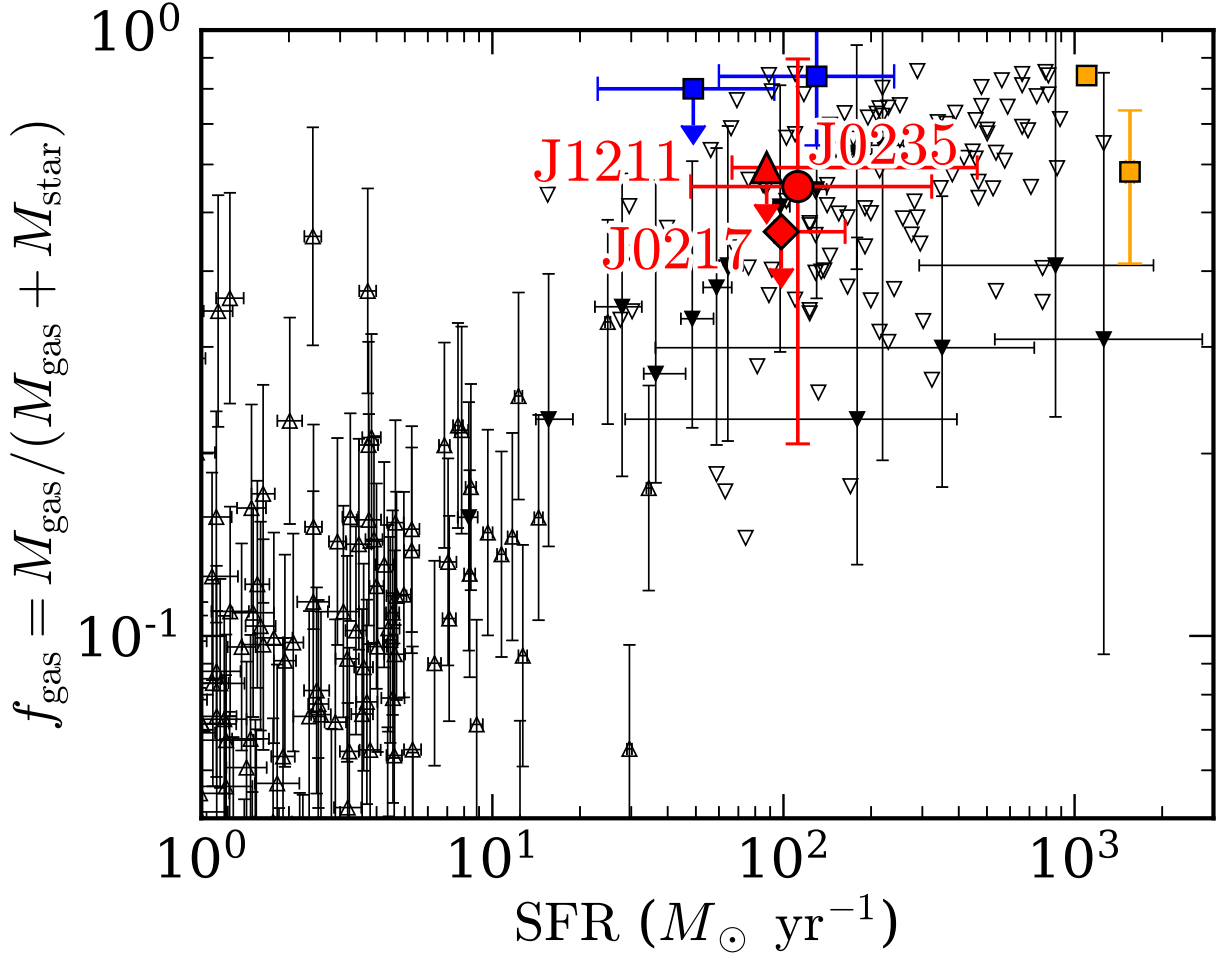}
\caption
{Gas fraction, $f_{\rm gas} = M_{\rm gas} / (M_{\rm gas} + M_{\rm star})$, 
as a function of SFR. 
The red circle is our ALMA result for a luminous LBG at $z=6$ with $\simeq 4 \sigma$ CO(6--5) detection, 
J0235--0532, in the case of  $T_{\rm dust} = 50$ K 
and SFR$_{\rm tot}$ $=$ SFR$_{\rm UV}$ $+$ SFR$_{\rm IR, 2 \sigma}$. 
The error bar along the $y$-axis considers 
the higher dust temperature case of $T_{\rm dust} = 80$ K 
and the minimum SFR case of SFR$_{\rm tot}$ $=$ SFR$_{\rm UV}$. 
The red triangle and diamond are also our ALMA results for 
the other luminous LBGs at $z=6$, J1211--0118 and J0217--0208, respectively, 
which shows no significant CO(6--5) detection. 
The red arrows correspond to the $3\sigma$ limits. 
The blue squares represent the results of LBG-1 and HZ10 from left to right (\citealt{2019ApJ...882..168P}). 
The blue arrow represents the $3\sigma$ limit. 
The orange squares are the results of AzTEC-3 and CRLE from left to right 
(\citealt{2010ApJ...720L.131R}; \citealt{2014ApJ...796...84R}; 
\citealt{2018ApJ...861...43P}; see also, \citealt{2019ApJ...882..168P}). 
The black filled downward triangles are the results of SFGs with $M_{\rm star} > 3 \times 10^{10} M_\odot$ 
at $z \sim 0$--$4$ using stacked dust SEDs 
(\citealt{2015A&A...573A.113B}). 
The black open downward triangles show the results of SFGs with $M_{\rm star} > 2 \times 10^{10} M_\odot$ 
at $z \sim 1$--$6$ based on sub-millimeter dust continuum measurements 
(\citealt{2016ApJ...820...83S}). 
The black open triangles are the results for low-$z$ galaxies at $z = 0.01$--$0.05$ 
(\citealt{2017ApJS..233...22S}). 
}
\label{fig:fgas_SFR}
\end{center}
\end{figure*}

In Figure \ref{fig:fgas_SFR}, 
we present $f_{\rm gas}$ of our $z=6$ luminous LBGs 
with those of lower-$z$ SFGs 
(\citealt{2015A&A...573A.113B}; \citealt{2016ApJ...820...83S}; 
\citealt{2017ApJS..233...22S})  
as well as the $z=5.3$--$5.7$ sources, 
HZ10, LBG-1, AzTEC-3, and CRLE 
(\citealt{2010ApJ...720L.131R}; \citealt{2014ApJ...796...84R}; 
\citealt{2018ApJ...861...43P}; see also, \citealt{2019ApJ...882..168P}) 
as a function of total SFR. 
We find that J0235--0532 has a 
comparable gas fraction to lower-$z$ SFGs with similar SFRs. 
We also find that 
the obtained upper limits on the gas fractions 
of J1211--0118 and J0217--0208 
are consistent with lower-$z$ SFGs with similar SFRs. 
Compared to HZ10, 
the gas fraction of J0235--0532 is consistent owing to the large uncertainties, 
while those of J1211--0118 and J0217--0208 are significantly lower, 
although their total SFRs are comparable.

\subsection{Evolution of the Gas Depletion Timescale} \label{subsec:tdep_evolution}

\begin{figure*}[h]
\begin{center}
   \includegraphics[width=1.0\textwidth]{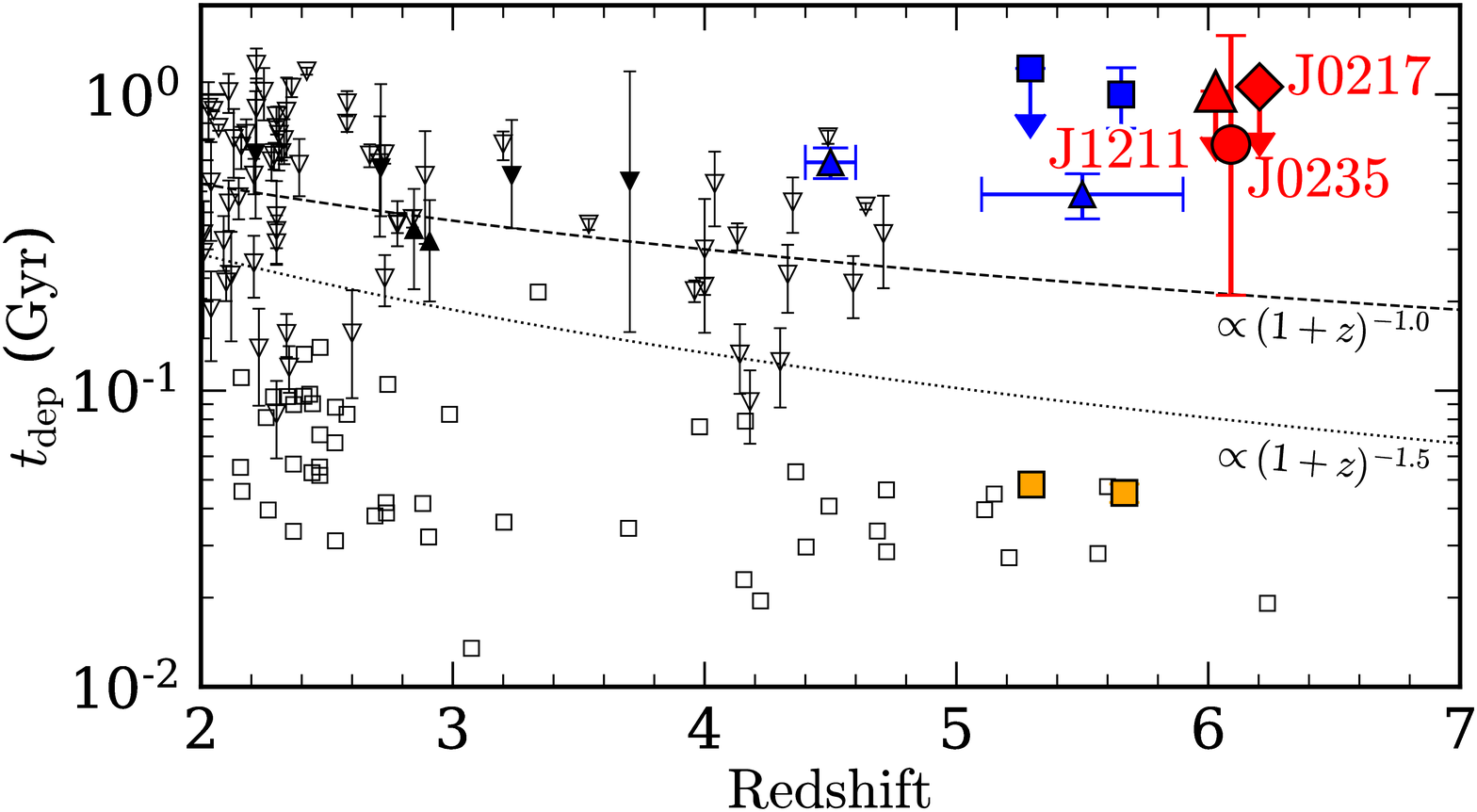}
\caption
{Redshift evolution of the gas depletion time, 
$t_{\rm dep} = M_{\rm gas} /$SFR. 
The red circle is our ALMA result for a luminous LBG at $z=6$ with $\simeq 4 \sigma$ CO(6--5) detection, 
J0235--0532, in the case of  $T_{\rm dust} = 50$ K 
and SFR$_{\rm tot}$ $=$ SFR$_{\rm UV}$ $+$ SFR$_{\rm IR, 2 \sigma}$. 
The error bar along the $y$-axis for J0235--0532 considers 
the higher dust temperature case of $T_{\rm dust} = 80$ K 
and the minimum SFR case of SFR$_{\rm tot}$ $=$ SFR$_{\rm UV}$. 
The red triangle and diamond are also our ALMA results for 
the other luminous LBGs at $z=6$, J1211--0118 and J0217--0208, respectively, 
which shows no significant CO(6--5) detection. 
The red arrows correspond to the $3\sigma$ limits. 
The blue squares represent the results of LBG-1 and HZ10 from left to right (\citealt{2019ApJ...882..168P}). 
The blue arrow represents the $3\sigma$ limit. 
The orange squares are the results of AzTEC-3 and CRLE from left to right 
(\citealt{2010ApJ...720L.131R}; \citealt{2014ApJ...796...84R}; 
\citealt{2018ApJ...861...43P}; see also, \citealt{2019ApJ...882..168P}). 
The blue filled triangles show the average results of $z=4.4$--$5.9$ LBGs 
with $M_{\rm star} = 10^{8.4 - 11} M_\odot$ 
obtained in the ALPINE survey 
(\citealt{2020A&A...643A...5D}), 
where $M_{\rm gas}$ are estimated from the [{\sc Cii}] luminosities. 
The black filled triangles denote the results based on CO observations for $z\sim3$ LBGs 
(\citealt{2017A&A...603A..93M}).
The black filled downward triangles are the results of SFGs with $M_{\rm star} > 3 \times 10^{10} M_\odot$ 
using stacked dust SEDs 
(\citealt{2015A&A...573A.113B}). 
The black open downward triangles show the results of SFGs with $M_{\rm star} > 2 \times 10^{10} M_\odot$ 
based on sub-millimeter dust continuum measurements 
(\citealt{2016ApJ...820...83S}; \citealt{2016ApJ...833..112S}). 
The black open squares are the results for 
lensed/unlensed dusty starburst sources compiled by 
\cite{2016MNRAS.457.4406A}. 
The dashed and dotted lines correspond to the curves of 
$t_{\rm dep} \propto (1+z)^{-1.0}$ and $t_{\rm dep} \propto (1+z)^{-1.5}$ 
(e.g., \citealt{2013ApJ...768...74T}; \citealt{2012MNRAS.421...98D}),
which are normalized to the typical gas depletion time of $1.5$ Gyr 
observed for local galaxies 
(\citealt{2008AJ....136.2782L}; \citealt{2011ApJ...730L..13B}; 
\citealt{2011MNRAS.415...61S}; \citealt{2012ApJ...758...73S}; 
see also, \citealt{2013ApJ...778....2S}). 
}
\label{fig:tdep_redshift}
\end{center}
\end{figure*}

Figure \ref{fig:tdep_redshift} shows 
the gas depletion timescale, $t_{\rm dep} = M_{\rm gas} /$SFR, 
as a function of redshift.  
For comparison, we also present 
the results for the four $z=5.3$--$5.7$ sources of HZ10, LBG-1, AzTEC-3, and CRLE 
(\citealt{2010ApJ...720L.131R}; \citealt{2014ApJ...796...84R}; 
\citealt{2018ApJ...861...43P}; see also, \citealt{2019ApJ...882..168P}),  
the average results for $z = 4.4$--$5.9$ LBGs with $M_{\rm star} = 10^{8.4-11} M_\odot$ 
obtained in the ALMA Large Program to INvestigate [{\sc Cii}] at Early times 
(ALPINE; \citealt{2020A&A...643A...5D}), 
and other SFGs including dusty starbursts over a wide range of redshifts 
(\citealt{2015A&A...573A.113B}; \citealt{2016ApJ...820...83S}; 
\citealt{2016MNRAS.457.4406A}; \citealt{2016ApJ...833..112S}; 
\citealt{2017A&A...603A..93M}; \citealt{2017ApJS..233...22S}).

As expected from Figure \ref{fig:Mgas_SFR}, 
$t_{\rm dep}$ of J0235--0532 is comparable to those of lower-$z$ SFGs at $z \sim 2$--$3$. 
Based on previous results for lower-$z$ SFGs, 
\cite{2013ApJ...768...74T} have suggested a redshift dependence of 
the gas depletion timescale in the form of 
$t_{\rm dep} \propto (1+z)^{-1.0}$, 
which is shallower than what is expected 
if $t_{\rm dep}$ is proportional to the dynamical timescale, 
$t_{\rm dep} \propto (1+z)^{-1.5}$  
(\citealt{2011MNRAS.416.1354D}; \citealt{2012MNRAS.421...98D}; see also, \citealt{2013ApJ...778....2S}). 
Our results for J0235--0532 show that 
the $t_{\rm dep}$ value is likely to be larger than
expected from the previously reported redshift dependencies. 
For the other two $z=6$ luminous LBGs, 
J1211--0118 and J0217--0208, 
we have obtained upper limits on their $t_{\rm dep}$, 
indicating that their $t_{\rm dep}$ values can be significantly shorter than J0235--0532.
In other words, 
there is a possibility that $t_{\rm dep}$ values of high-$z$ SFGs are not necessarily as large as 
those of lower-$z$ SFGs, 
suggesting that J0235--0532 may be an outlier with large $t_{\rm dep}$.
Similar arguments can be made at slightly lower redshifts 
based on the results of HZ10 and LBG-1 as well as the ALPINE results. 
Because the previous results for lower-$z$ SFGs 
show a large scatter of $t_{\rm dep}$, 
it would be interesting to investigate a typical $t_{\rm dep}$ value 
by observing more high-$z$ SFGs with better sensitivities 
in future studies 
to characterize the typical star formation properties in high-$z$ SFGs.

\section{Adopted IMFs in the literature} 

In this paper, we have adopted the \cite{2003PASP..115..763C} IMF 
with lower and upper mass cutoffs of $0.1 M_\odot$ and $100 M_\odot$, respectively, 
as described in Section \ref{sec:introduction}. 
However, some previous studies have adopted different IMFs, 
and corrections for IMF differences are required 
when comparing physical quantities related to IMFs such as SFR and $M_{\rm star}$. 
For convenience in such purposes, 
Table \ref{tab:adopted_IMFs} summarizes the IMFs adopted in the previous studies 
whose SFR or $M_{\rm star}$ estimates are used for comparisons with our results. 
Where necessary to convert SFR and $M_{\rm star}$ values in the literature,
we use constant factors of  $\alpha_{\rm SC}$, $\alpha_{\rm KC}$, and $\beta_{\rm SC}$, 
as described in Section \ref{sec:introduction} and Appendix \ref{subsec:gas_fraction}. 

\newcommand\Bethermin{{\cite{2015A&A...573A.113B}}} 
\newcommand\Magdis{{\cite{2017A&A...603A..93M}}} 
\newcommand\Dessauges{{\cite{2020A&A...643A...5D}}} 

\begin{deluxetable}{cc}[h] 
\tablecolumns{2} 
\tablewidth{0pt} 
\tablecaption{IMFs Adopted in the Previous Studies
\label{tab:adopted_IMFs}}
\tablehead{
    \colhead{Previous Study} 
    &  \colhead{IMF}
}
\startdata
\cite{2014ApJ...796...84R} & \cite{2003PASP..115..763C} \\ 
\Bethermin & \cite{2003PASP..115..763C} \\ 
\cite{2015ApJS..219...15S} & \cite{1955ApJ...121..161S} \\ 
\cite{2016ApJ...820...83S} & \cite{2003PASP..115..763C} \\ 
\cite{2016MNRAS.457.4406A} & \cite{2003PASP..115..763C} \\ 
\cite{2016ApJ...833..112S} & \cite{2003PASP..115..763C} \\ 
\Magdis & \cite{2003PASP..115..763C} \\ 
\cite{2017ApJS..233...22S} & \cite{2003PASP..115..763C} \\ 
\cite{2018ApJ...861...43P} & \cite{2003PASP..115..763C} \\ 
\cite{2019ApJ...872...16D} & \cite{2001MNRAS.322..231K} \\ 
\cite{2019ApJ...882..168P} & \cite{2003PASP..115..763C} \\ 
\cite{2020ApJ...896...93H} & \cite{2003PASP..115..763C} \\ 
\Dessauges & \cite{2003PASP..115..763C} \\ 
\cite{2021ApJ...908...61K} & \cite{2001MNRAS.322..231K}  
\enddata 
\end{deluxetable} 

\end{document}